\documentclass[namedreferences,hyperref,optionalrh]{spr-sola}
\usepackage{graphicx}        
\usepackage{color}           

\renewcommand{\vec}[1]{{\mathbfit #1}}

\chardef\us=`\_
\usepackage{lineno}
\begin{document}

\begin{frontmatter}
\title{Observation and modeling of a geo-effective event observed on 2011 May 28 from the solar surface to 1\,au}

\author[addressref={aff1},email={nishu.karna@cfa.harvard.edu}]{\inits{N.}\fnm{Nishu}~\snm{Karna}\orcid{0000-0002-1314-4690}}
\author[addressref=aff1,email={tniembro@cfa.harvard.edu }]{\inits{T.}\fnm{Tatiana}~\snm{Niembro}\orcid{0000-0001-6692-9187}}

\address[id=aff1]{Center for Astrophysics $|$ Harvard $\&$ Smithsonian, Cambridge, MA, USA}

\runningauthor{Karna and Niembro}
\runningtitle{Geomagnetic storm}

\begin{abstract}
In this study, we present a comprehensive observational and modeling study of a geo-effective event with D$_{ST}$ index of -80\,nT observed on 2011\,May\,28 when a coronal hole was bordering an active region.  {We analyze HMI and EUV images and found that this event involved two filament eruptions ~8 hours apart from two different active region closed to each other.} We produce 3D magnetic field configurations  {for the active regions} that are consistent with the observations and employ numerical models to track the CME/ICME propagation up to 1\,au. From our, magnetic models we found that the nearby coronal hole reduced the stability threshold of the flux ropes,  { with axial flux values approximately three times lower than in comparable cases without coronal holes. A derivative analysis applied to STEREO coronagraph and OMNI database in situ data revealed no evidence of CME–CME interaction during the early stages of their evolution and identified distinct signatures of two CMEs, along with the interacting flow associated with the nearby coronal hole at 1\,au. Moreover, we used hydrodynamic simulations constrained by remote sensing and in situ data to track the different structures in the solar wind. We found a good agreement between data and the models.}  {Additionally, we found that the presence of the coronal hole may have suppressed interactions between CMEs, with the transients subsequently propagating along the solar wind streams emerging from the coronal hole.}


\end{abstract}
\end{frontmatter}


\section{Introduction}
     \label{S-Introduction} 

 {Geomagnetic storms \citep[GS, ][]{1994JGR....99.5771G, Zhanget07, Echeret08}} are disturbances in Earth's magnetosphere driven by solar wind variations which can be measured in terms of the Disturbance Storm-Time index \citep[D$_{ST}$, ][]{osti_4554034}. In general, these variations are plasma flows with speeds and magnetic field strengths that are unusually large, which may lead to severe geomagnetic consequences such as the malfunction of space vehicle operation, the interruption of radio communication, and the disruption of the power grids. Most of these solar wind perturbations are related to coronal mass ejections \citep[CMEs, ][]{1986SSRv...44..139S, 1988JGR....93.8519T, 2002GeoRL..29.1287S, 2023EP&S...75...90W} and high-speed streams \citep[HSSs, ][]{Cranmer09} from coronal holes \citep[CHs, ][]{Cranmer09} with both considered as major drivers of space weather \citep{2006PCE....31...81G}. GSs are categorized as weak with -50\,nT$<D_{ST}<$-30\,nT, moderate with -100\,nT$<D_{ST}<$-50\,nT and, intense geomagnetic storms (IGSs, \citet{Borovsky17}) with $D_{ST}<$-100\,nT values. Statistical studies have shown that moderate and intense GSs are related to CMEs while moderate and weak GSs are related to HSSs \citep{2019InJPh..93.1103W}. 

Moreover, a strong GS-CME correlation has been noticed when CMEs are fast, with large angular widths and, with source regions close to the center of the disk \citep{2009JGRA..114.0A22G, 2016GSL.....3....5L} particularly when these CME are tracked and seen on coronagraph images (from spacecraft near Earth's location) as halo or partial-halo CMEs \citep{1982ApJ...263L.101H, 2010SunGe...5....7G}. However, these reports are far from complete, as they treat GS events as solely correlated to CMEs or HSSs and lack discussions of the  {event complexity \citep{Shenetal13, Shenetal14, Liuetal15, Kilpuaetal17, Preteetal24}}.

Other studies found halo CMEs unrelated to GS events and strong relationships between non-Halo CMEs and GSs which led to false space weather predictions \citep{2015ApJ...812..145M}. A possible explanation of these exemptions suggests that  {CMEs are influenced by HSSs \citep{2002SSRv..101..229C, Cranmer09, 2020JGRA..12527530W, Heinemannetal19, Kayetal22}} implying that the global magnetic field surrounding the CME source location plays an important role by deflecting CMEs \citep{2000AdSpR..26...43S, 2004JGRA..109.7105Y, 2020JGRA..12527530W} and causing them to be more geo-effective when propagating or deflected towards the ecliptic plane \citep{2011JGRA..116.4104W}. It was also found that the IGS occurrence is generally associated with an eruptive event in an  {active region (AR, \citet{Asaietal09, Lugazetal11})}, an area with a strong closed magnetic field located near an equatorial CH \citep{Gonzalezet96, bravo98}.

 {Some studies have shown lack of deflection e.g. \citet{Chertoket18} reported that two ICMEs launched from AR arrived at Earth earlier than expected due to their propagation within a high-speed, low-density stream originating from an extended CH. Rather than being deflected, the Earth-directed components of both CMEs appeared to be channeled along the HSS. Similarly, \citet{Abuninet20} found that two relatively weak transients propagated entirely within and between fast wind streams from different sectors of CH, which acted to suppress their expansion. These findings underscore the critical role of coronal hole–associated HSSs in modifying ICME kinematics, structure, and arrival timing at Earth.}

There has been several interesting results tying open-field coronal structures near close-field structures. \citet{Habbalet08} found that ARs can increase plasma outflows in neighboring CHs. \citet{Karnaetal2020} studied the connection between the CH location to the AR in the maximum phase of four solar cycles (21-24) finding a strong odd-even trend between the wind properties and the number of ARs close to CHs suggesting a possible link between solar wind properties and the interaction between open and closed magnetic field structures. \citet{vanet2012} studied the magnetic topology of an AR bordering a CH and found that the proximity of an AR to a CH has significant implications on coronal outflows and solar wind. They found that ARs bordering CHs can contribute to the slow solar wind. However, it remains unclear how CH-AR interactions could influence geo-effective CMEs. Moreover, the nature of the formation and evolution of the magnetic field at the boundary of open and closed field regions is not well understood. To understand how the influence of a coronal hole plays a major role in the CME geo-effectiveness, we need to study the three-dimensional structure of the coronal magnetic configuration prior to the eruption in the source region at the surface of the Sun and we need to propagate the eruption including coronal hole fast solar wind.

\begin{figure}[ht!]
\centering\includegraphics[width=0.5\textwidth, trim={0 0.3in 0 0}]{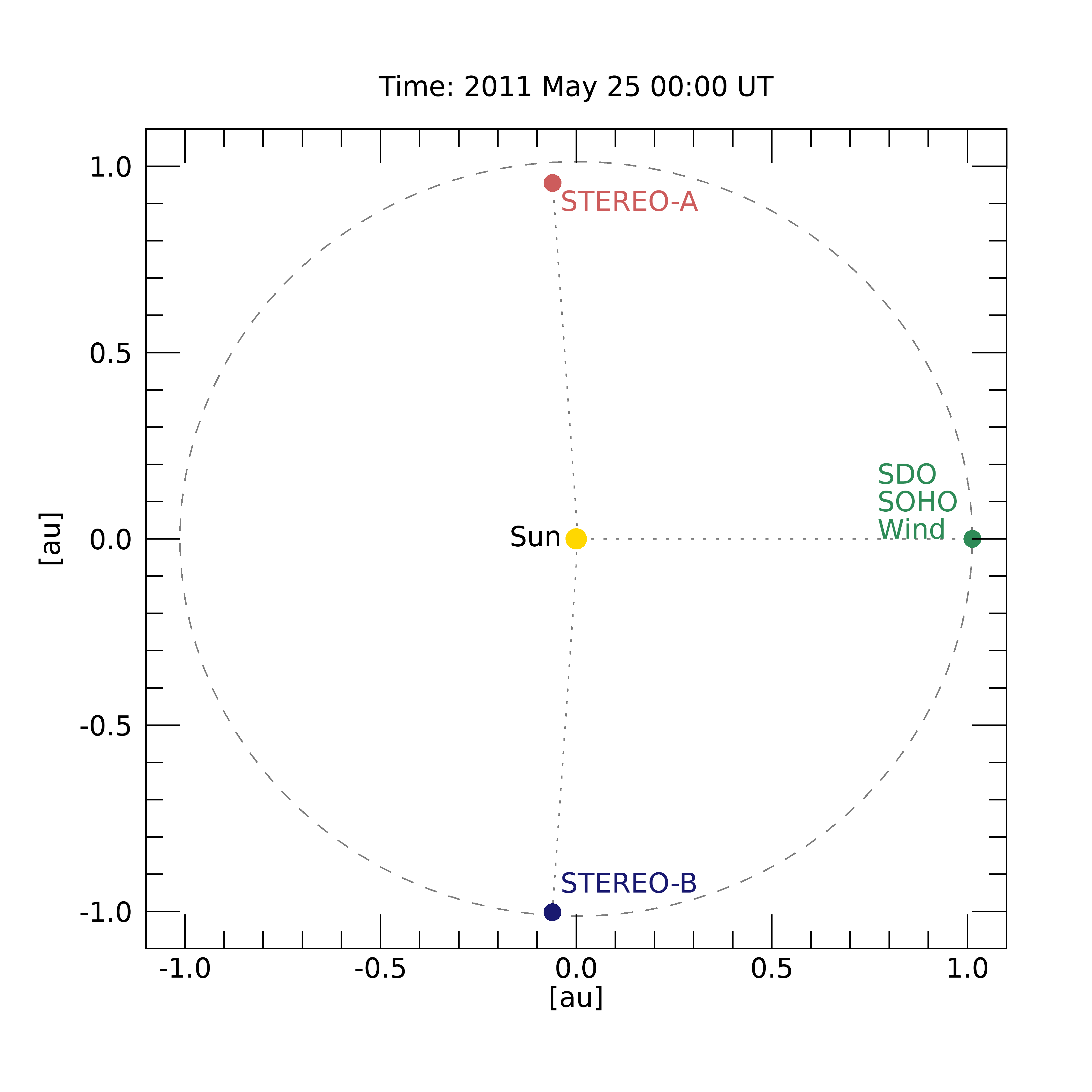}
\caption{Spacecraft location on 2011\,May\,25\,00:00\,UT. The Sun's location is shown in {\em yellow}. STEREO-A direction and location are shown in dash line and {\em red} circle, STEREO-B is marked in {\em blue} and Earth (SDO, SOHO and Wind) in {\em green}. STEREO-A/B were separated by about 90$^{\circ}$ from Earth.}
\label{fig:orbit} 
\end{figure}

In this work, we present a comprehensive observation from multiple spacecraft (Figure\,\ref{fig:orbit}) and modeling of a geo-effective event with D$_{ST}$ index -80\,nT observed on 2011\,May\,28 driven by two interplanetary coronal mass ejections (ICMEs) embedded in a HSS. Hereinafter, we use CME\,1 and CME\,2 to refer to the ICME counterparts. We would like to note that the sources of these CMEs are ARs located nearby a CH. This GS has been the subject of  {previous} studies \citep{Wood_2017, Nitta_2017, Chi_2018}. For instance, \citet{Wood_2017} performed full 3D reconstructions of the CME structure and kinematics using the {\it{Solar TErrestrial RElations Observatory}} \citep[STEREO, ][]{kaiseret08} and the {\it{Solar and Heliospheric Observatory}} \citep[SOHO, ][]{1995SoPh..162....1D} imaging constraints. The author considered the CME\,1 reaching Earth as GS responsible while the CME\,2 just barely missed to the west. Dissimilar, \citet{Nitta_2017} suggested the CME\,2 as a stealth CME \citep{2021SSRv..217...82N, 2021SSRv..217...84N} and as the only one causing the GS. In contrast, \citet{Chi_2018} applied a graduated cylindrical shell model to reconstruct the 3D\,geometry, propagating direction, velocity, and brightness of the two CMEs and concluded that both ICMEs caused the GS event.  {The in situ structure comprised two distinct ICME ejecta that were closely connected in both time and space.} We also note that, this event is reported as a single magnetic obstacle in the Wind ICME list \citep{Nieves2016, Nieves2018, Nieves2019}. None of these reports discussed: a) near the Sun, the effects of the nearby AR-CH on the CMEs; b) in the heliosphere, the corresponding HSS effect in both ICME propagation and; c) on Earth, their role in the GS effectiveness. The lack of these discussions motivated us to revisit the event.

Our focus is to identify and determine the impact of the AR nearby CH on the CMEs dynamics originated by two filament eruptions and their role in the GS effectiveness. In the low corona where magnetic pressure dominates over the dynamic pressure, we produce 3D magnetic field configurations consistent with multi-spacecraft observations to study the CH presence on the filament eruptions. Meanwhile, in the inner heliosphere where the solar wind is ruled by the flow dynamic pressure, we employ hydrodynamic (HD) analytical and numerical models to track the CMEs/ICMEs propagation embedded in the HSS originated from a CH. 
  
The manuscript is outlined as follows. In Section\,2, we present the observations and data analysis techniques used to identify and track the CMEs on multiple spacecraft data sets and at different distances from the Sun. In Section\,3, we explain our magnetic models and results. In Section\,4, we describe the eruption propagation using our HD models. Then, we conclude with a summary of our findings and discussion in Section\,5.

\section{Observations}
\label{sec:obs}

\subsection{Remote sensing}
\label{sec:remote}

\begin{figure}[hbt]
\centering\includegraphics[width=0.8\textwidth]{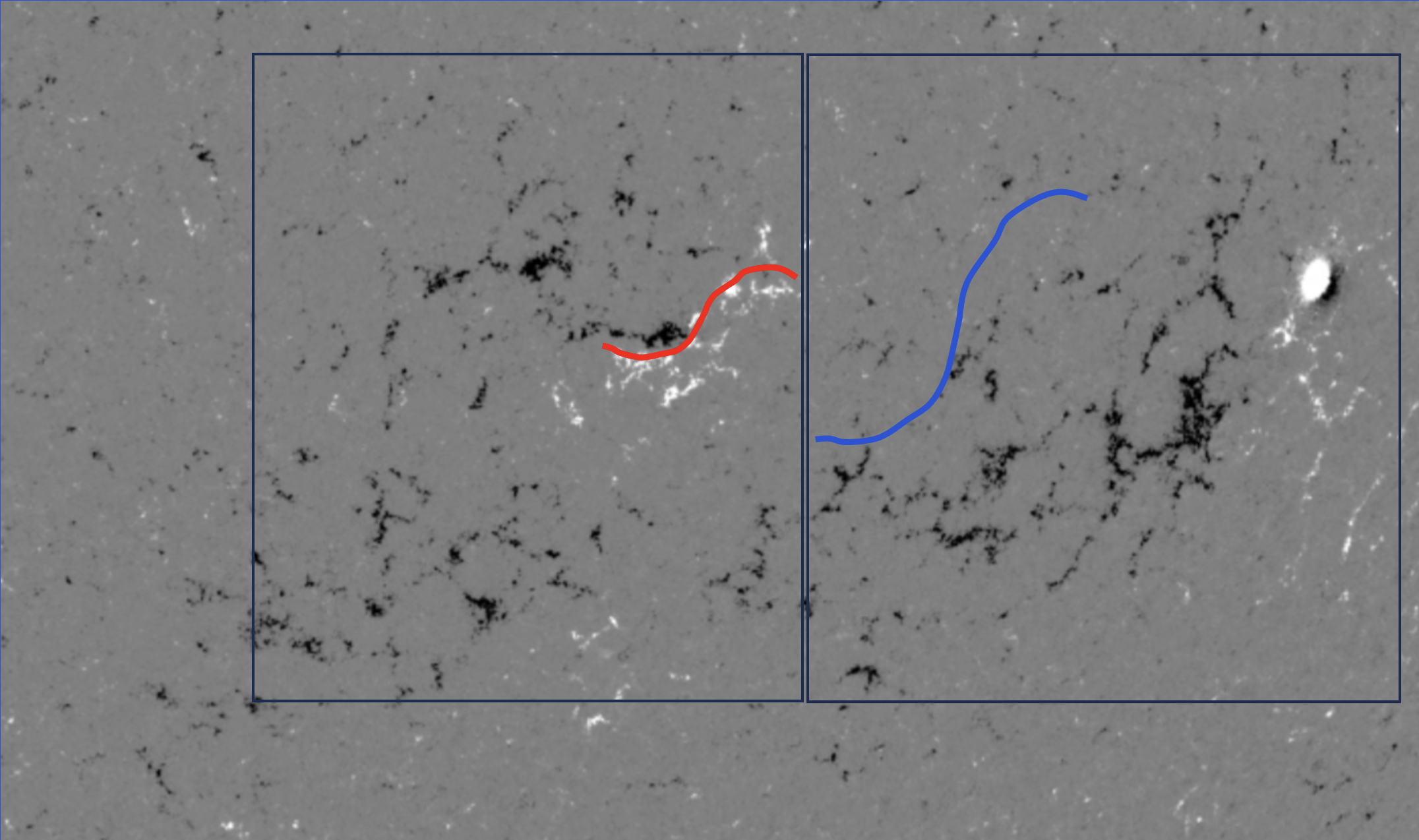}
\caption{ {HMI magnetogram showing the eruption locations of two filaments. The rectangular boxes enclose the two active regions: NOAA\,11218 ({\it left}) and NOAA\,11216 ({\it right}). The red and blue S-shaped lines indicate the locations of the filament eruptions. The grayscale intensity is scaled between $+$100\,G ({\it white}) and $-$100\,G ({\it black}).}}
\label{fig:ar} 
\end{figure}

\begin{figure}[hbt]
\centering\includegraphics[width=1.0\textwidth]{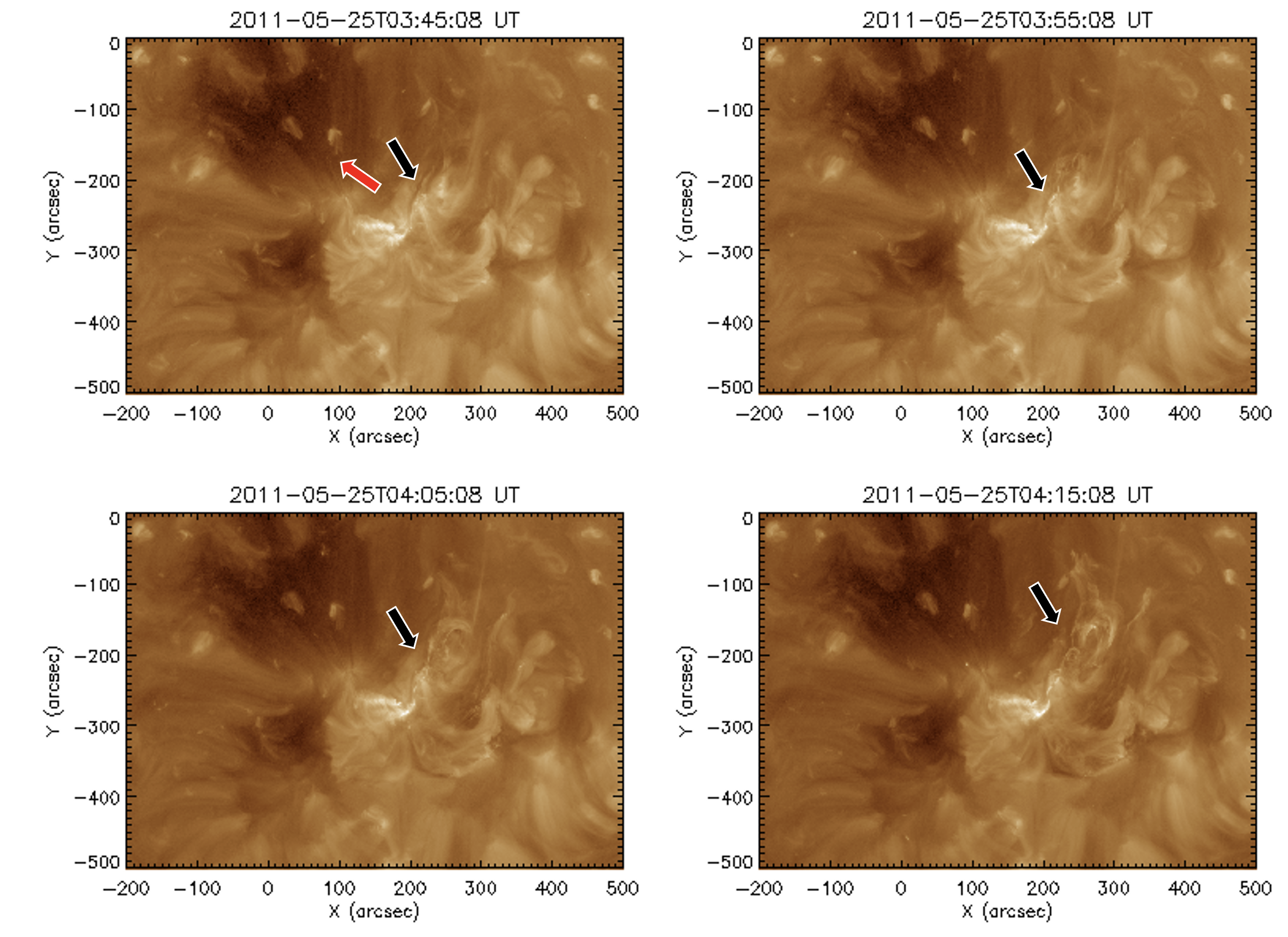}
\caption{AIA\,193\,\AA\ observation snapshots of the filament eruption on 2011\,May\,25\,03:45 and 04:15\,UT related to CME\,1. In the \textit{top left} panel, the {\em black} arrow is pointing to the filament before launch, and in panels \textit{top right, bottom left, and bottom right} the eruption evolution. The {\em left} dark regions (pointed by the {\em red} arrow in \textit{top left} panel) indicates the coronal hole.}
\label{fig:aia193} 
\end{figure}

\begin{figure}[hbt]
\centering\includegraphics[width=1.0\textwidth]{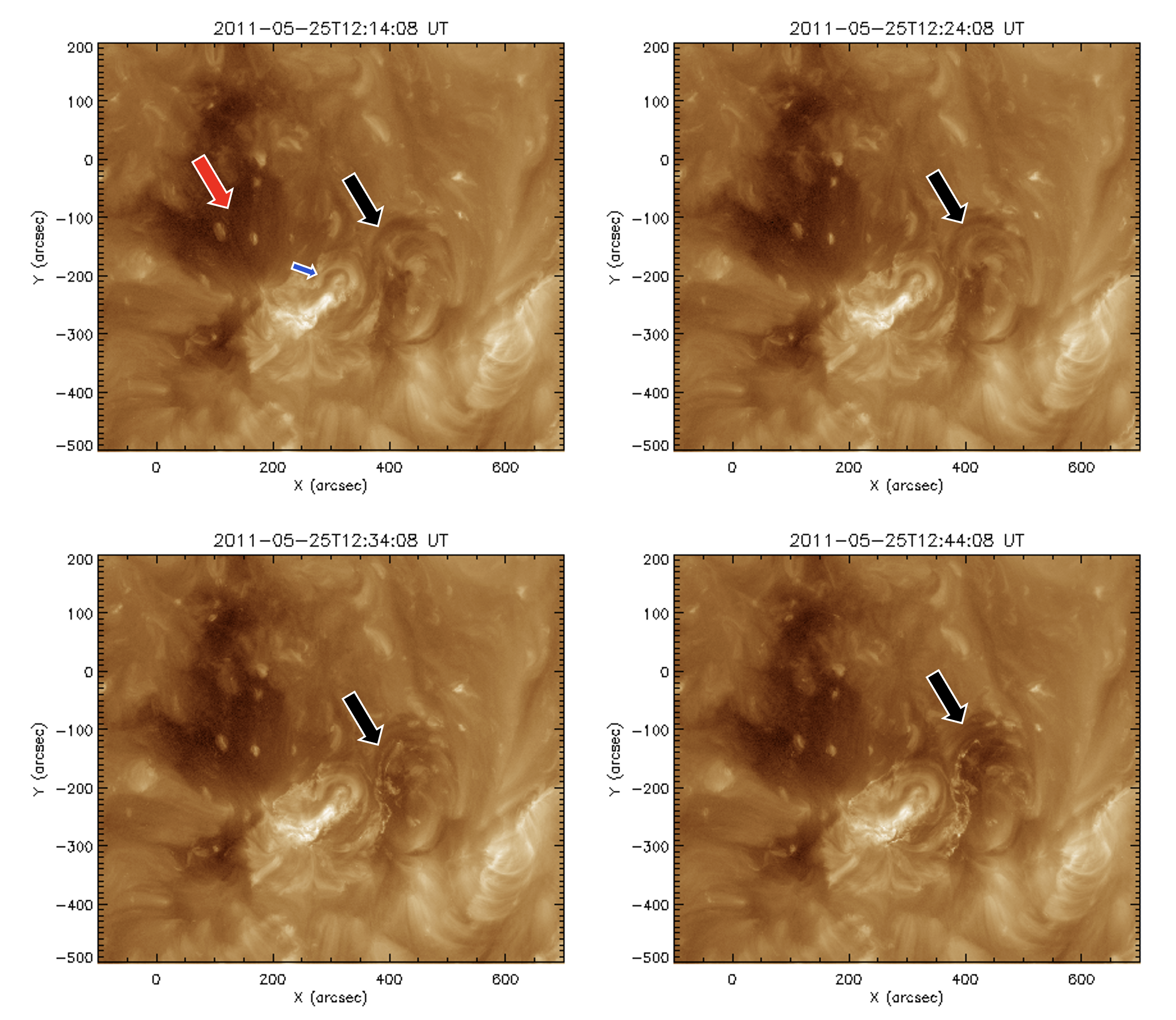}
\caption{Same as Figure\,\ref{fig:aia193} but for CME\,2. We show AIA\,193\,\AA\ observation snapshots of the second filament eruption on 2011\,May\,25 from 12:14\,UT to 12:44\,UT which is close to the location of the first eruption (pointed by the small blue arrow). In 
 the \textit{top left} panel, the {\em black} arrow points to the S-shaped filament before the eruption and panels (\textit{top right, bottom left}, and \textit{bottom right}) show the eruption evolution. The left big regions (pointed by the red arrow in the \textit{top left} panel) correspond to the coronal hole.}
\label{fig:event2} 
\end{figure}

On 2011\,May\,25, extreme ultra-violet (EUV) observations from the Atmospheric Imager Assembly \citep[AIA, ][]{lemenet12} on board the Solar Dynamics Observatory \citep[SDO,][]{Pesnell:2012rr} showed that CME\,1 (Figure\,\ref{fig:aia193}) and CME\,2 (Figure\,\ref{fig:event2}) were associated with two filament eruptions from two different ARs close to each other. The CME\,1 filament eruption onset was registered at 03:50\,UT from NOAA\,11218 active region at S16W13 (Figure\,\ref{fig:ar} left rectangular box) while the second filament erupted at 12:24\,UT from NOAA\,11216 active region at S18W25 (Figure\,\ref{fig:ar} right rectangular box).

In Figure\,\ref{fig:aia193}, we show EUV observations of the CME\,1 filament eruption configuration. The filament is pointed by the {\em black} arrow, before (\textit{top left}) and during the eruption evolution (\textit{top right, bottom left}, and \textit{bottom right}). The corresponding AR was bordering a large CH pointed by the {\em red} arrow in the \textit{top left} panel. The CH\,region was first observed on the eastern limb on 2011\,May\,21 and rotated to the backside on 2011\,June\,07. It was centered on the central meridian on 2011\,May\,25 extending in both north and south directions. In association with this eruptive event, the C2 and C3\,white imagers \citep[from the SOHO Large Angle and Spectrometric COronagraph Experiment, LASCO, ][]{1995SoPh..162..357B} observed a partial CME with $38^\circ$ width. CME\,1 entered the C2\,field of view (2--6\,R$_\odot$) on 2011\,May\,25 at 05:24\,UT and the C3\,field of view (6--30\,R$_\odot$) at 09:06\,UT. According to the SOHO/LASCO\,CME\,catalog \footnote{\url{https://cdaw.gsfc.nasa.gov/CME_list/}}, CME\,1 had a velocity of 276\,km\,s$^{-1}$. CME\,1 was also observed by both COR\,2\ from both STEREO-A/B Sun-Earth Connection Coronal and Heliospheric Investigation instrument suite \citep[SECCHI, ][]{2008SSRv..136...67H} on 2011\,May\,25 at 05:39\,UT \citep{Chi_2018}.

The panels in Figure\,\ref{fig:event2} show observations of the second filament eruption i.e., CME\,2,
(also pointed by the {\em black arrow} in the \textit{top left} panel configuration before and during the eruption evolution (\textit{top right, bottom left}, and \textit{bottom right} panels). The filament had an S-shaped morphology and lay in the periphery of the eruption region of CME\,1 (pointed by the small blue arrow in the \textit{top left} panel). The elapse time \citep[or waiting time, ][]{2020A&A...635A.112L} between the two eruptions was $\sim$8\,hours. CME\,2 entered C2 field of view around 2011\,May\,25 at 13:25\,UT and C3 starting at 14:06\,UT with a speed of $561$\,km\,s$^{-1}$. CME\,2 appeared in the COR2 field of view from both STEREO-A/B on 2011\,May\,25 at 13:39\,UT.

{ In \cite{Chi2018}, the speed of the CMEs was computed by fitting the Graduated Cylindrical Shell \citep{Thernisien_etal_2006ApJ, Thernisien_etal_2009SoPh, 2011ApJS..194...33T} model to fit a 3D croissant-shape to the ejecta over the coronagraph images at different times to estimate the speed of the CMEs. However, the method relies on locating the legs of the CME at the center of the Sun. The locations of the eruptions are considerably far away (roughly for CME\,1 in (200 arcsec, -250 arcsec) and (400 arcsec, -250 arcsec) for CME\,2) from the (0,0) which may end in additional uncertainties to consider when trying to describe the CME evolution. Additionally, the purpose of this particular investigation \citep{Chi2018} is to show that when the CMEs are not clearly identified in the coronagraphs, particularly from Earth's point of view, we are missing information about the event (e.g., the number of CMEs) that compromises our understanding of Sun-Earth connection and forecasting efforts.}

Although the implementation of the GCS is a first good approximation, it is missing the importance of the location of the source as well as the role-play of the coronal hole in the evolution of the CMEs. Some of the limitations are discussed by \cite{2013JGRA..118.6866C, Thernisien_etal_2009SoPh, 2012JGRA..117.6106N}. We decided to implement an empirical analysis over the coronagraph images, not only to track the CMEs and estimate their speed and density but also to get relative values to estimate the speed and density of the ambient solar wind conditions. Moreover, the analysis enables the identification of all possible transients related to the event. It also provides time series of the speed and density that we used to simulate the event from 10--20\,R$_{\odot}$ up to 1\,au. With the analysis, we can also determine if CME interaction took place or not in the first stages of their evolution. All these are based on the observations and not on the shape fitting which forces assumptions over the event such as the direction of the CMEs. We analyzed coronagraph images covering from 2011 May 23 to 27. The time range starts two days before the first eruption and ends at the CMEs Earth's arrival time.
 
We used the method described by \citet{Lara2004} to estimate the relative brightness ($\Delta SB$) time series along the ecliptic plane (and towards the Earth) from the COR\,2 images at different heliospheric distances (one hundred distances covering a range of 4--12\,R$_{\odot}$). {Please note that the method is similar to other common methods used by the solar science community, for example, the one described by \cite{2009ApJ...698..852C}}. 

{ From the brightness time series ($\Delta SB$), we computed the first and second derivatives, calculated the second derivative histograms, and fit Gaussian distributions. The average $\sigma$ of the fitting is used as a filter to smooth the $\Delta SB$ time series. This works as a low pass-filter in which only values of the second derivative lower than $\sigma$ are used. Consecutive series elements with a lower value than $\sigma$ are averaged resulting in the change of the $\Delta SB$ time series to step-function shapes}. Each step (k) of the input time series $\{x_i, x_{i+1}, x_{i+2}, ..., x_n\}$ corresponds to:
\begin{equation}\label{eq:step}
    f_k(x_i, x_{i+1}, x_{i+2},...., x_{i+j}) = \overline{(x_i, x_{i+1}, x_{i+2}, ..., x_{i+j})},
\end{equation}  
where the bar represents the median values of the subset formed by the continuous points $x_i, ...,x_j$ such that $x_i < \sigma$. The index $k$ changes automatically when the condition $x_i < \sigma$ does not apply (is false), defining the next step of the resulting series. 
\begin{figure}[htbp]
    \centering
    \includegraphics[width=1\textwidth, trim={0in 0.0in 0in 0.45in},clip]{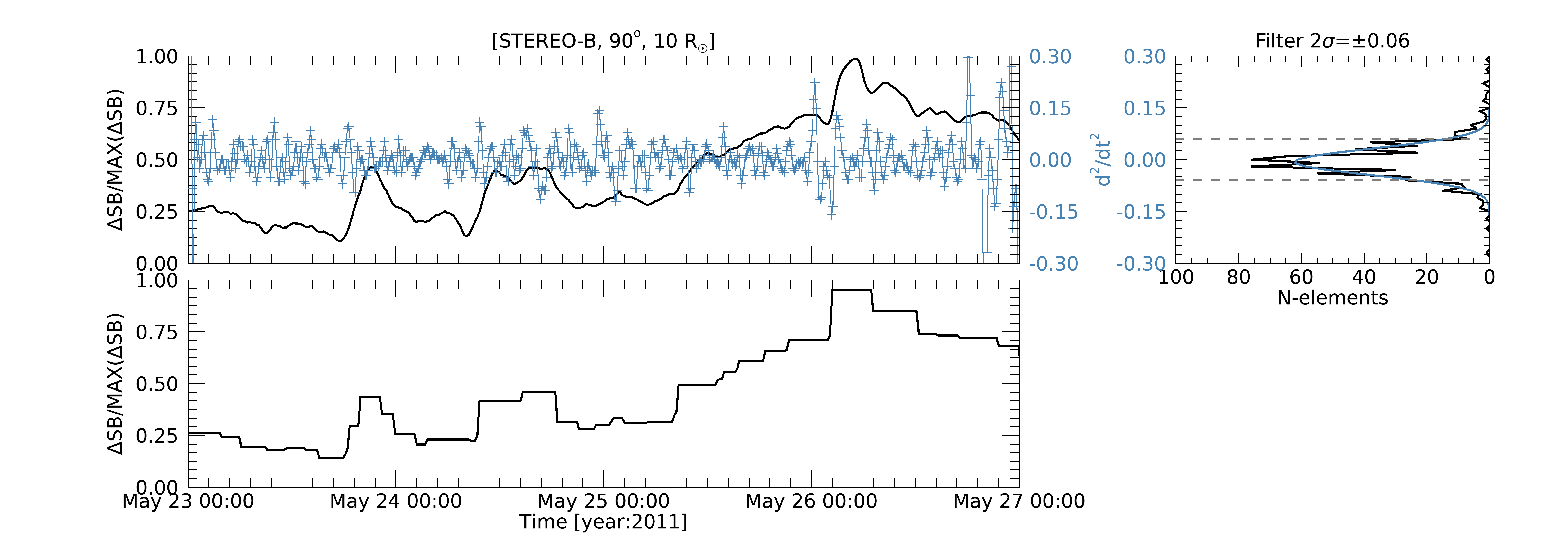}
 \caption{{Example of the construction of the step-function $\Delta SB$ time series from the COR\,2 images. \textit{Top left:} Original $\Delta SB$ time series (shown in {\em black}) and its second derivative (in {\em blue}). \textit{Top right:} Histogram of the second derivative with the $\sigma$ value used as a filter to obtain step-functions of the $\Delta SB$) time series. \textit{Bottom left:} step-function of the $\Delta SB$ time series obtained when consecutive series elements with a lower value than $\sigma$ are averaged.}}
    \label{fig:example01}
\end{figure}

\begin{figure}[htbp]
\centering\includegraphics[width=0.40\textwidth, trim={0in 0.0in 0in 0.7in},clip]{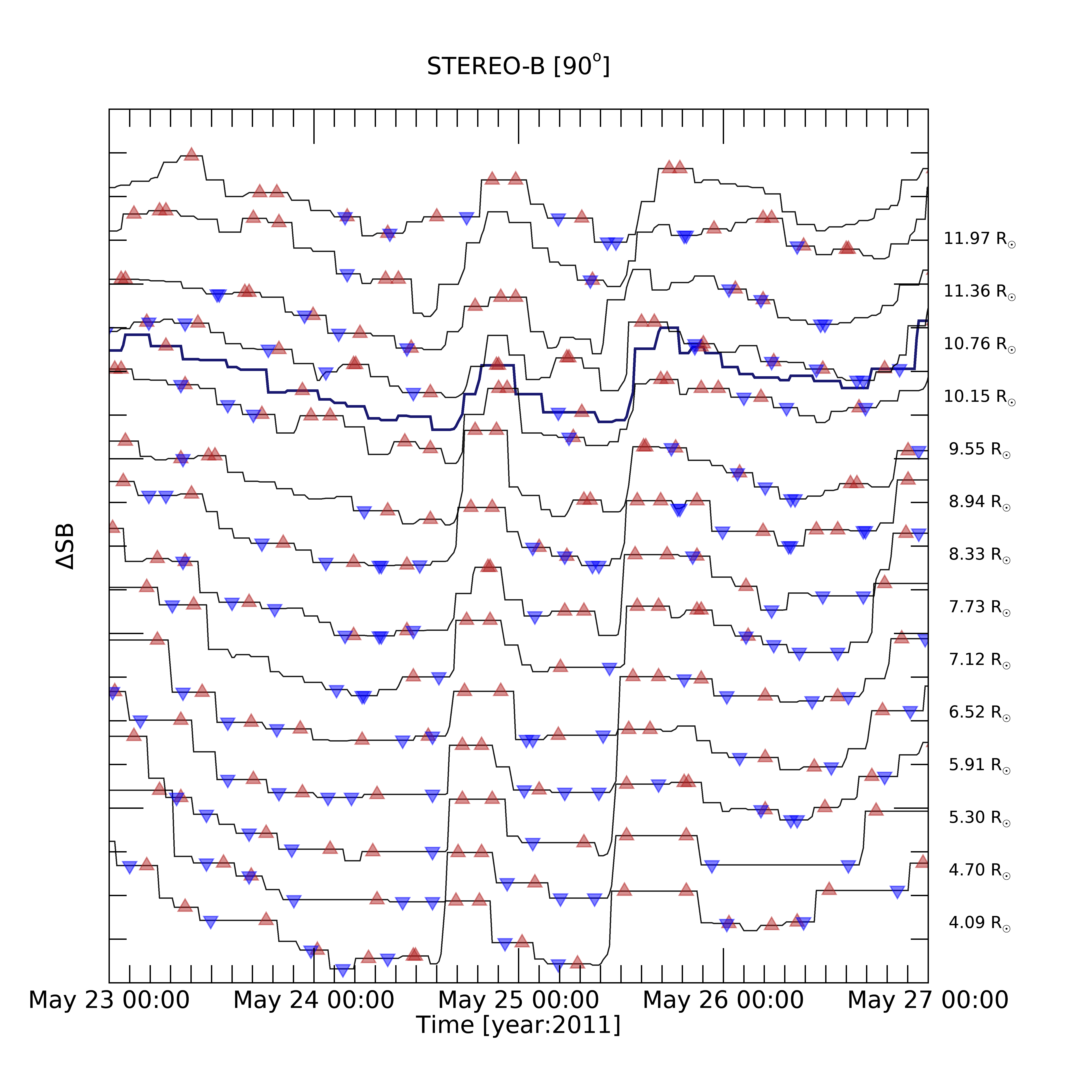}
\centering\includegraphics[width=0.59\textwidth, trim={0in 0.0in 0in 0.7in},clip]{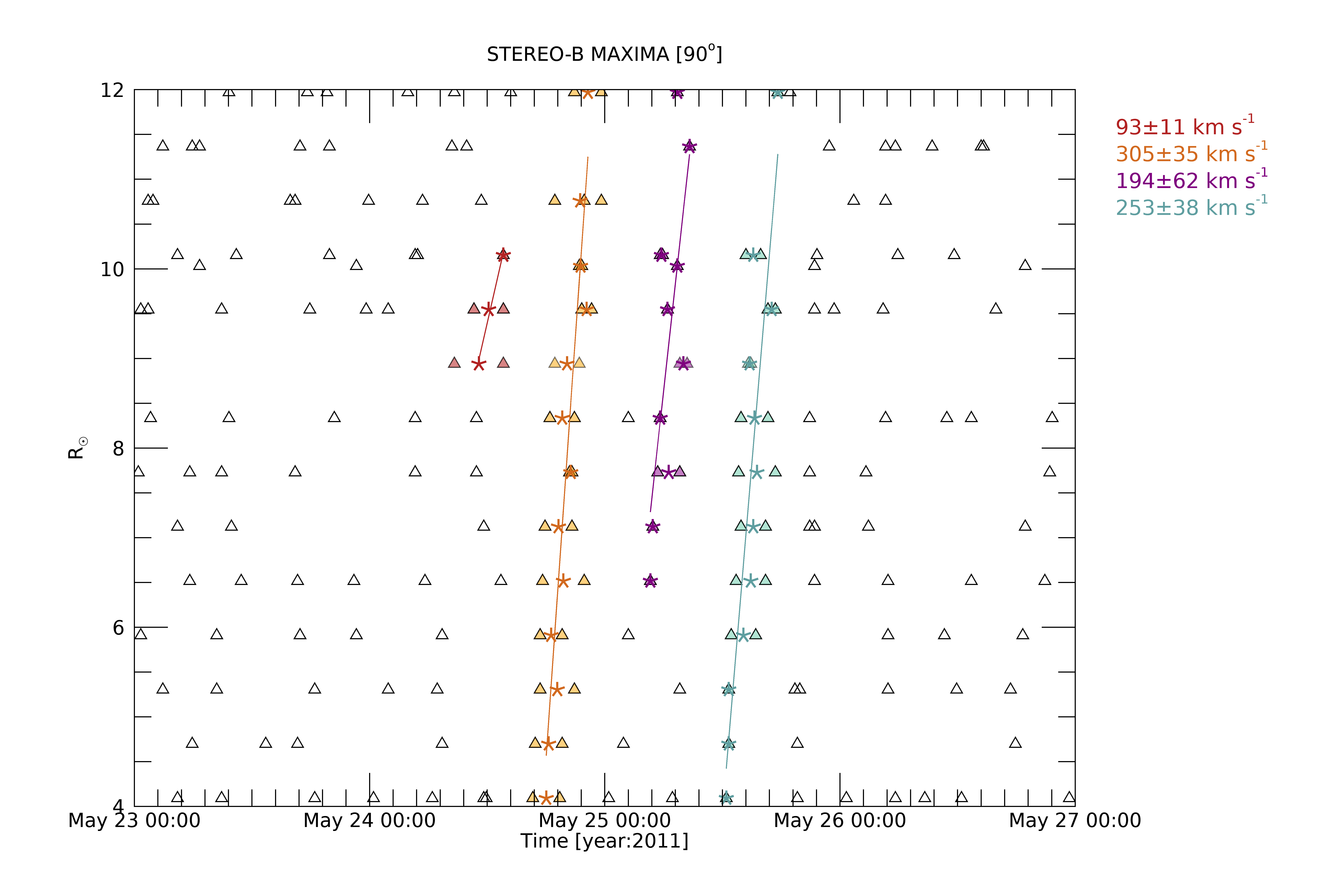}
\caption{STEREO-B COR2 Brightness analysis from 4--12\,R$_{\odot}$. In the {\em left} panel, the {\em blue triangle down} symbols and {\em red triangle up} symbols are the minima and maxima of the second derivative. The maxima and minima trace and constrain both transients. The {\em right} panel shows the maxima. The {\em purple} represents CME\,1 and the {\em aqua}, CME\,2. With linear fitting, we estimated speeds of $194 \pm 62$\,km\,s$^{-1}$ and $253 \pm 38$\,km\,s$^{-1}$, respectively. With the model, we also identified two other structures denoted by {\em red} and {\em yellow} triangles.}
\label{fig:sb_STB} 
\end{figure}

\begin{figure}[htbp]
\centering\includegraphics[width=0.40\textwidth, trim={0in 0.0in 0in 0.7in},clip]{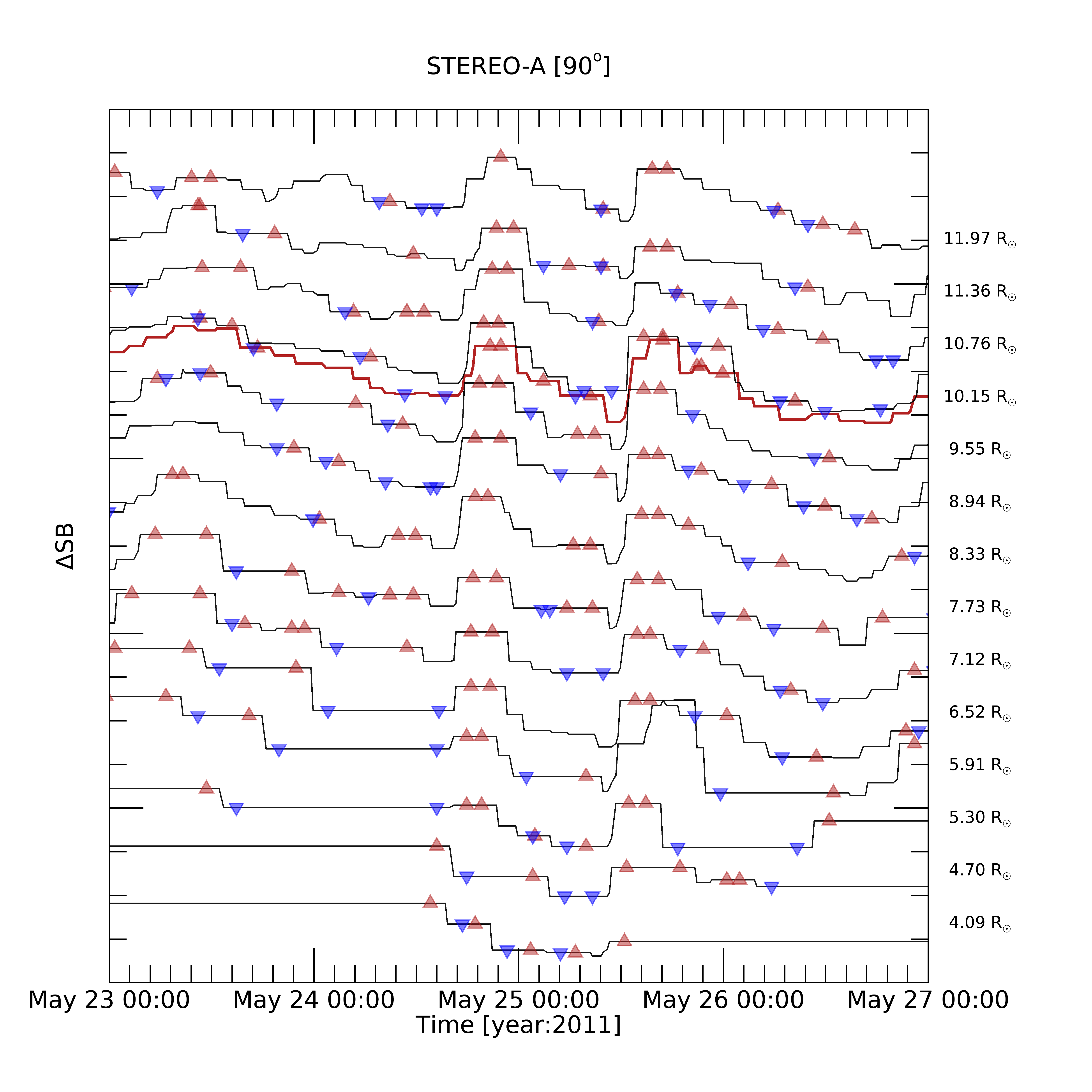}
\centering\includegraphics[width=0.59\textwidth, trim={0in 0.0in 0in 0.7in}, clip]{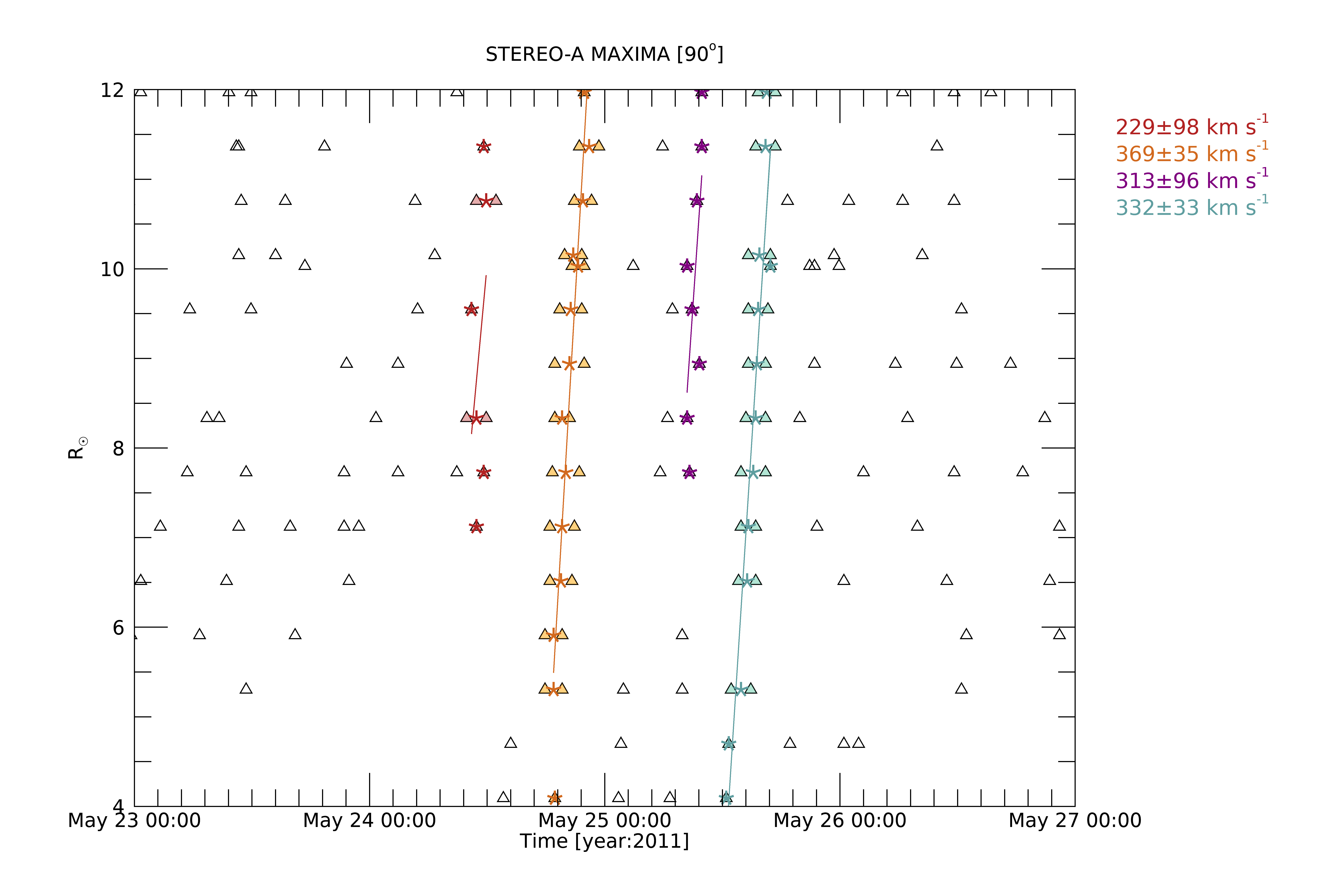}
\caption{Same as Figure \ref{fig:sb_STB} for STEREO-A COR2.}
\label{fig:sb_STA} 
\end{figure}

{ In Figure\,\ref{fig:example01}, we show the step-function time series construction. The original $\Delta SB$ time series is shown in \textit{top left} panel in \textit{black} with its second derivative overplot (shown in \textit{blue}). We obtained the second derivative histogram (\textit{top right} panel) and fitted a Gaussian distribution to estimate $\sigma$. Using $\sigma$ as a low-pass band filter, we computed the step-function. We repeated the analysis for different angles around the direction of propagation of the CME and also for 100 distances from the Sun (in the left panels of Figures\,\ref{fig:sb_STB} and \ref{fig:sb_STA}, one can see at least 15 of them).}

{ Moreover, we improved our CME identification with the first and second derivatives (see left panels of Figures\,\ref{fig:sb_STB} and \ref{fig:sb_STA}), we located different transients (in time and space) using the maxima (the first derivative equal to zero and the second derivative being negative, shown in {\it red}). The minima (the first derivative equal to zero and the second derivative being positive, in {\it blue}) were used as transient constraints. This is consistent with the idea that CMEs are noticed as perturbations propagating in the solar wind. CMEs or any other transient will be related to maxima as they will have, at least in this case, speeds faster than the ambient solar wind. }

{ We also used the maxima location to compute the speed $V_{CME}$ of both CMEs through linear fitting (see right panels of Figures\,\ref{fig:sb_STB} and \ref{fig:sb_STA}). The transient densities $N_{CME}$ were estimated following \cite{2021JSWSC..11...11H}, that is, by computing baseline signal-to-noise ratios (SNR with $SNR = (image-background)/\sqrt{image}$) from the COR2 images and estimating $N_{CME} = 1/10\,N_{corona}$ with $N_{corona} = 110 \times SNR^2$. The background is computed when applying the method over the images as described by \cite{Lara2004}.} 

Figure\,\ref{fig:sb_STB} (for STEREO-B) and Figure\,\ref{fig:sb_STA} (for STEREO-A), we show the brightness time series when choosing $\sigma = 0.005$. We marked the maxima (shown in {\em upward red triangle} symbols) and the minima (in {\em downward blue triangle}) in the {\em right} panels. In the {\em left} panel, CME\,1 is marked in {\em purple} and CME\,2 in {\em teal}. We found that CME\,1 and CME\,2 propagated with the speeds of $194 \pm 62$\,km\,s$^{-1}$ and $253 \pm 38$\,km\,s$^{-1}$ from STEREO-B point of view (Figure\,\ref{fig:sb_STB}). Similarly, from STEREO-A (Figure\,\ref{fig:sb_STA}) the speeds were $313 \pm 96$\,km\,s$^{-1}$ and $332 \pm 33$\,km\,s$^{-1}$, respectively. The corresponding points used for the linear fits are shown in {\em star} symbols in {\em left} panels. We identified two other structures that we used as reference when comparing STEREO-A with STEREO-B results shown in {\em red} and {\em yellow} lines.  

Alternatively, following the methodology described by \cite{Lara2004} we computed the speed (V$_{CME}$) of CME\,1 between 482 and 581\,km\,s$^{-1}$ and the CME\,2 ranging from 486 to 511\,km\,s$^{-1}$. In comparison, though we found larger uncertainties between the two {approaches}, the latter method provided accurate speed measurements whereas the maxima method discerned solar wind structures very well.

We also computed the densities (N$_{CME}$) of the transients following \cite{2021JSWSC..11...11H}. We obtained 2625-3300\,cm$^{-3}$ for the CME\,1 and 2500-3300\,cm$^{-3}$ for the CME\,2.

 {Analysis from these remote sensing observations especially COR2 indicates the presence of two clearly separated structures traveling up to 20~R$_\odot$ without interacting, with no signs of potential interaction detected at larger heliocentric distances.}


\subsection{In situ}
\label{sec:insitu}

The counterpart of the eruption event arrived to Earth, three days later on 2011\,May\,28 at 04:30\,UT. We analyzed its plasma and magnetic field characteristics based on data from OMNI \citep{2005JGRA..110.2104K} that records near-Earth solar wind magnetic field and plasma parameter data from several currently-operating missions, the Advanced Composition Explorer \citep[ACE, ][]{1998SSRv...86....1S}, Deep Space Climate Observatory \citep[DISCOVR,][]{Burt2012} and Wind \citep{Ogilvie1995}. From the OMNI\,database, we also obtained the geomagnetic and solar activity index, Disturbance Storm Time D$_{ST}$ \citep{2006JGRA..111.2202W, dstlink}, registered by the World Data Center for Geomagnetism, Kyoto \citep{dstlink}.  {The OMNI time series were obtained in one-minute while the D$_{ST}$ in one-hour time resolutions.}

\begin{figure}[htbp]
\centering\includegraphics[width=1\textwidth, trim={0 0.95in 0.2in 0}]{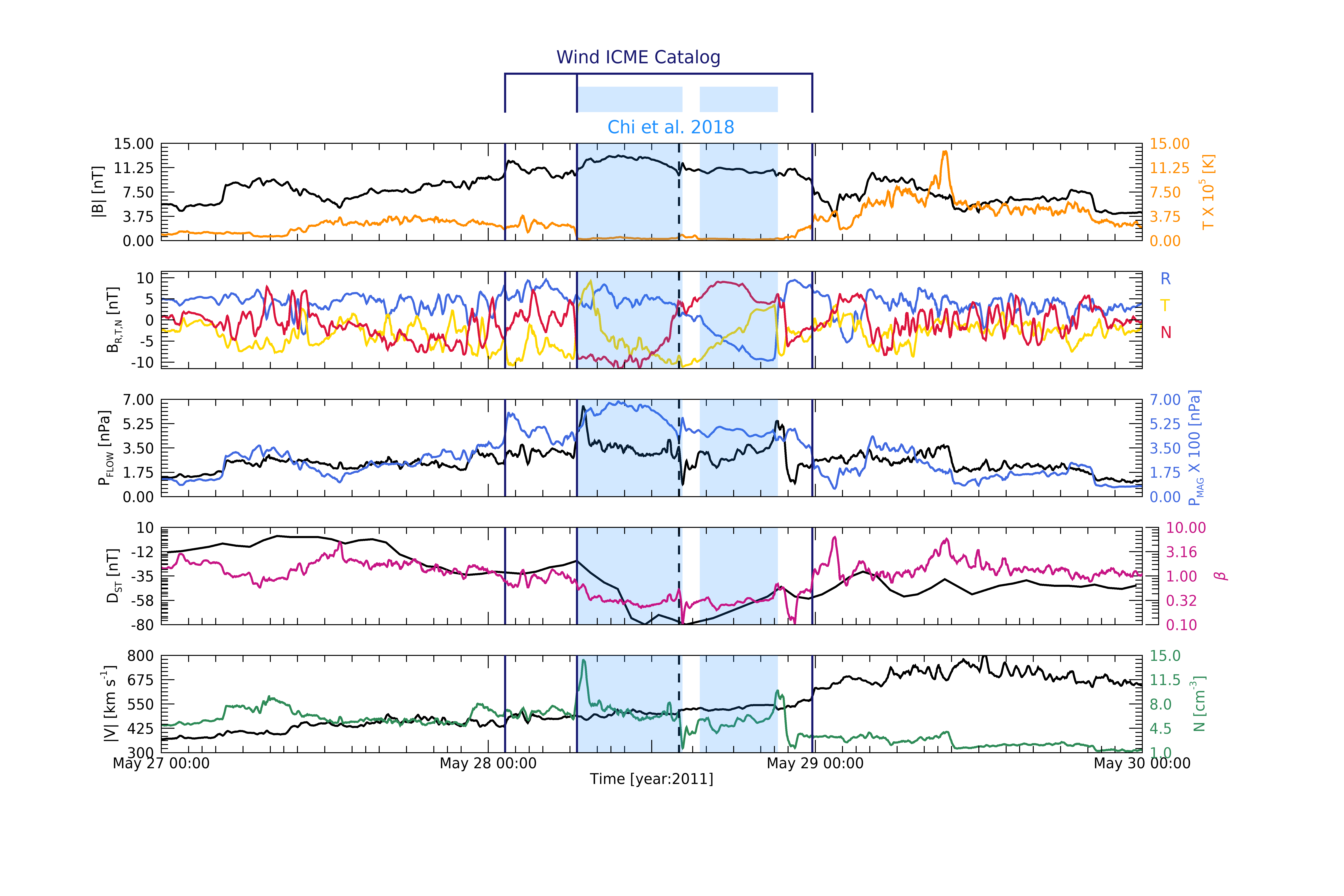}
\caption{Solar wind parameters measured at 1\,au, from {\em top} to {\em bottom}: magnitude of the magnetic field ({\bf B}, shown in {\em black}), temperature (T, in {\em orange}); ({\em second} panel) the radial (B$_R$, {\em blue}), tangential (B$_T$, {\em yellow} and normal (B$_N$, {\em red}) components of the magnetic field; ({\em third} panel) the computed flow (P$_{FLOW}$, in {\em black}) and magnetic (P$_{MAG}$, in {\em blue}) pressures; ({\em fourth} panel)the D$_{ST}$ index (in {\em black}) and beta parameter ($\beta$ in {\em pink}); ({\em fifth panel}) the magnitude of the speed ({\bf V} in {\em black}) and density (N in {\em green}). The {\em navy} vertical lines represent the CME arrival time and start/end times of the CME reported in the Wind ICME Catalog  {shifted an hour to fit the OMNI time frame}. We note that the {\em black} dashed line indicates the arrival of CME\,2 according to our analysis.}
\label{fig:param} 
\end{figure}

The {\em first} panel of Figure \ref{fig:param} shows the magnetic field ({\bf B}) in {\em black} and the temperature (T) in {\em orange}. In the {\em second} panel, we present the radial (B$_R$, {\em blue}), tangential (B$_T$, {\em yellow}) and normal (B$_N$, {\em red}) components of the magnetic field. In the third panel, we computed the flow (P$_{FLOW} \, [nPa] = 2 \times 10^{-6} N [\,cm^{-3}] \, V [km \, s^{-1}]^2 $, in {\em black}) and magnetic (P$_{MAG}\, [nPa] = (10^{-2}/ 8 \pi) \, B [nT]^2$, in {\em blue}) pressures. In the fourth panel, the D$_{ST}$ index (in {\em black}) and beta parameter ($\beta$ in {\em pink}) are shown. Finally, in last panel, we present the velocity ({\bf V} in {\em black}) and density (N in {\em green}). Last, in {\em blue} shadow areas, we denoted the time when the spacecraft passed both CMEs according to \citet{Chi_2018} and in {\em navy} the the CME arrival time (2011\,May\,28\,00:14\,UT) and start (2011\,May\,28\,05:31 UT) and end (2011\,May\,28\,22:47\,UT) times related to the CME reported in the Wind ICME Catalog.  {These times where shift an hour to be comparable with the OMNI time frame}. We also included a {\em black} dashed line indicating the CME\,2 arrival time according to our analysis discussed in the following sections. In the figure, we also indicate the HSS arrival time ( {around 2011\,May\,27\,07:00\,UT}) where {\bf V} started rising  {and reached values above} 450\,km\,s$^{-1}$ \citep{2003JGRA..108.1156C} while N decreases after spiking and later increasing again due to the ICMEs arrival. The HSS is also identified because of both, P$_{MAG}$ and P$_{FLOW}$ increasing. The  temperature slightly rose between the HSS start and the ICMEs arrival times and eventually decreased within the ICMEs. The ICMEs were followed by the fast flow forming the HSS. The D$_{ST}$ index (bottom panel) declines to -80\,nT within the ICMEs.  {Please note that we are plotting the 15-minute average of the OMNI data. This time resolution is sufficient to see the differences between both CMEs.}

\begin{figure}[hbt]
\centering\includegraphics[width=1\textwidth, trim={0.4in 0.91in 1in 0.2in}]{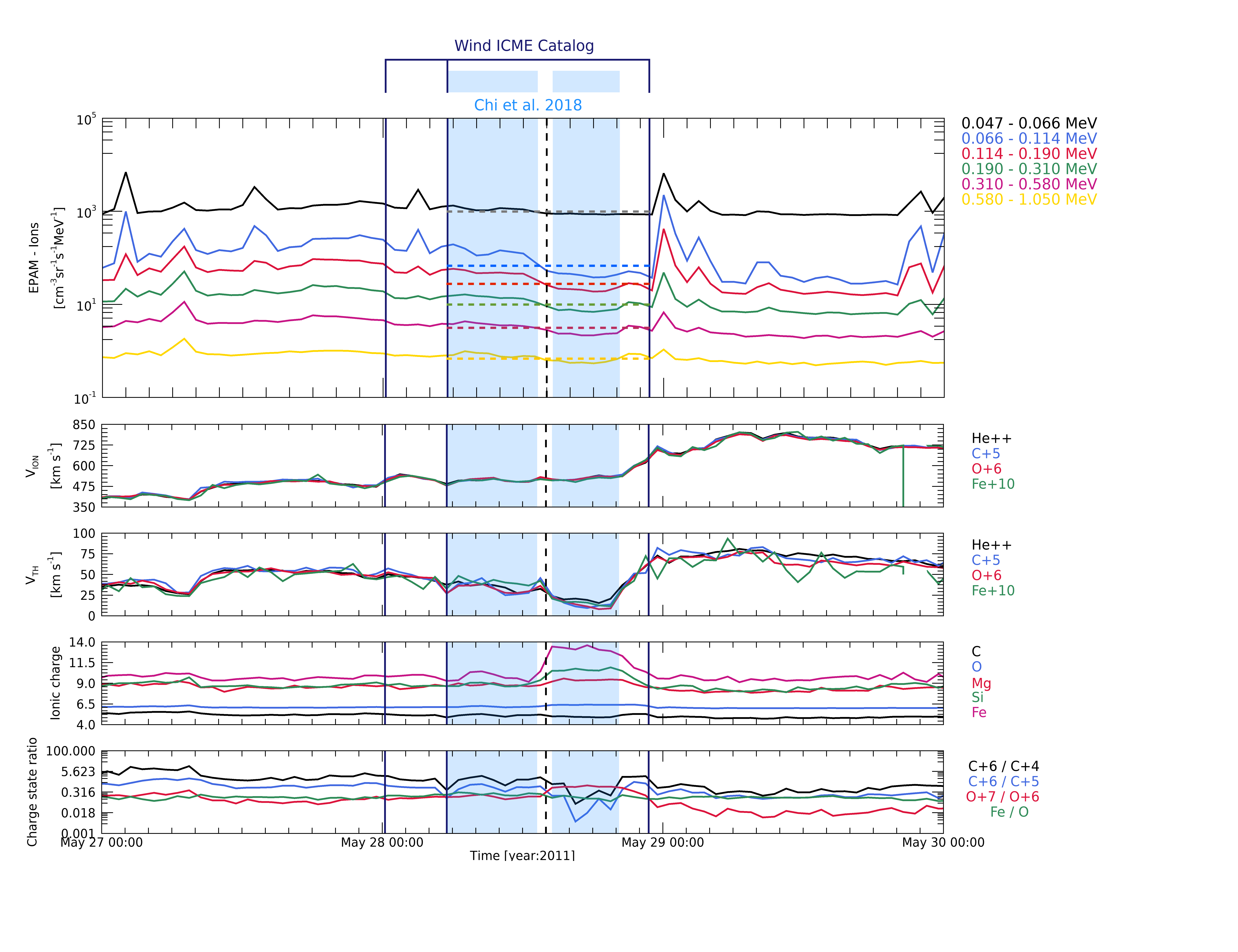}
\caption{EPAM and SWICS observations during the event of 2011\,May\,28. We show: the EPAM ion observations ({\it first panel}) for energies between 0.066 and 1.05\,MeV (with colors indicating the energy channel in MeV), proton speed ({\it second}), thermal speed ({\it third}), ionic charge ({\it fourth}) and charge state ratio ({\it bottom}). The {\em navy} vertical lines represent the CME arrival time and start/end times of the CME reported in the Wind ICME Catalog. We note that the {\em black} dashed line indicates the arrival of CME\,2 according to our analysis.  {In the {\it first} panel, we have included base lines for comparison purposes. All time series show differences between the CMEs.}}
\label{fig:osir} 
\end{figure}

 {Next, we analyzed the ion, chemical and isotopic composition measurements of the solar wind at 1\,au in Figure\,\ref{fig:osir}. We used the Electron, Proton, and Alpha Monitor (EPAM) instrument \citep{Gold98} and the Solar Wind Ion Composition Spectrometer \citep[SWICS,][]{Gloeckler98} on ACE.  {The EPAM and SWICS time series are shown in one-hour temporal resolution.} Figure\,\ref{fig:osir} shows the ion observations for energies between 0.047--1.05\,MeV ({\it first} panel), the ion  velocity ({\it second} panel), thermal speed ({\it third} panel), ionic charges ({\it fourth} panel), and solar wind charge state ratio and averages of some heavy-ion density ratios ({\it bottom} panel). A bump in the energies was observed before and during the speed rising due to the HSS arrival ({\it first} panel). Between the HSS start and ICMEs arrival times, we observed an increase in energies and thermal speed, a decrease in solar wind average ionic charge and charge state ratios like O7+/O6+, C6+/C5+ and C6+/C4+. We also noticed slight decrease in solar wind elemental abundance ratios i.e., Fe/O ({\em bottom} panel) while the speed increases ({\em second} panel). These decreases correlate to the transitions from slow to fast solar wind from HSSs.}

 { Within the CMEs, we observed a significant decrease in thermal speeds and ions energies, along with enhancements in ionic charge states and charge state ratios, along with increases in solar wind speed.
After the ICMEs, we observed similar signatures of the solar wind as those at the HSS arrival.}

In addition, to characterize HSS and transients boundaries, we analyzed the behavior of the speed and density (time) rates through the statistical distribution of their derivative and using the distribution standard deviation. Same as the brightness distribution analysis, we computed the first and second derivatives and used Equation\,\ref{eq:step} to smooth and construct step-function shape of V and N time series. After fitting the Gaussian distribution to the histogram of the V and N second derivative, we select two different filters multiple (N$_{TR}$) of $\sigma$, one ($\sigma=$0.45) that shown to be related to the HSS and the other ($\sigma=$0.85) to structures comparable is size to ICMEs.
\begin{figure}[htbp]
    \centering
    \includegraphics[width=1\textwidth, trim={0in 0.95in 0.2in 0in}]{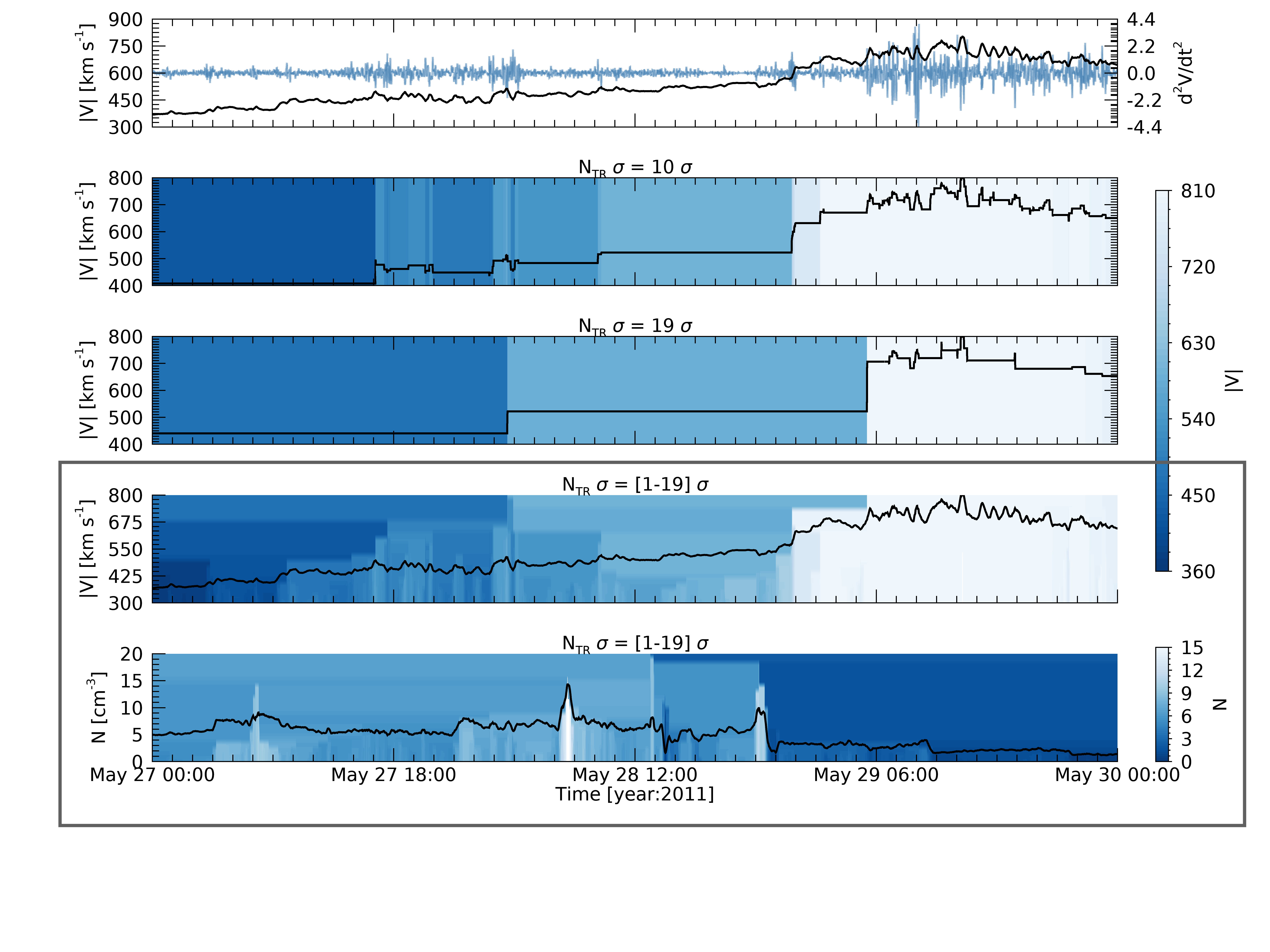}
    \caption{Construction of the V step-function time series based on OMNI data. From {\em top} to {\em bottom}: {\em first} panel show the original 15-minute V time series (shown in {\em black}) and its second derivative (in {\em blue}). In the {\em second} and {\em third} panels, we present the V step-function time series when choosing N$_{TR}\sigma = $10\,$\sigma$ and 19\,$\sigma$, respectively. With the {\em blue}--{\em white} color scale, we emphasize the solar wind separation in patches of different speeds. The separation is different depending on N$_{TR}$. The last two panels, inside the {\em gray} rectangle summarize the results of implementing the method for 1$<$N$_{TR}<$19 to V and N time series in graphic maps. We combine the V and N results and found that with N$_{TR} \sigma = $10\,$\sigma$, we identified the ICMEs and N$_{TR} \sigma = $19\,$\sigma$ the HSS.}
    \label{fig:dev}
\end{figure}

The first three panels in Figure\,\ref{fig:dev}, shows an example of the construction of V step-functions choosing two different N$_{TR} \sigma$. Depending on the value of N$_{TR} \sigma$, the solar wind is split in patches at different speeds. We use a {\em blue}--{\em white} color scale to emphasize these patches. In the {\em third} panel (N$_{TR}=$19 over V), the solar wind is split in three different V regions with the {\em dark blue} patch at 450\,km\,s$^{-1}$, the {\em blue} one at 540\,km\,s$^{-1}$ and the {\em white} patch at 810\,km\,s$^{-1}$. Using the same N$_{TR} \sigma$ but over N ({\em top blue} gradients in the {\em bottom} panel), the solar wind is divided in two regions where the first one is denser than the second one. The transition between these two N patches correlates to the {\em blue} V region shown in the third panel. The filter N$_{TR}=$19 over V and N worked to identify the HSS and its respective slow and fast flows as well as a transition region between these two. Similarly, when choosing N$_{TR}=$10 over V and T, we were capable of identifying at least to ICMEs reaching the spacecraft. 

To corroborate our results, we repeat the process for N$_{TR}=$1,2,...,19 over V and N. With each of the {\em blue}--{\em white} color scale profiles, we constructed graphic maps. These maps are shown in the two {\em bottom} panels in Figure\,\ref{fig:dev}. Please notice that several values of N$_{TR}$ result in very similar {\em blue}--{\em white} color scale profiles indicating that we can use this method for the identification process. We will explore this response in a future work.

 {So far, the method has demonstrated its ability to distinguish and improve the structural characterization of CMEs and HSSs, using simple statistical analysis applied to spacecraft measurements. We found that the boundaries identified by this method are consistent with the observed characteristics of HSSs and CMEs in the in situ data as seen in Figures\,\ref{fig:param} and \ref{fig:osir}). Moreover, our method identifies two CMEs.}



\section{Magnetic modeling 1-3 \texorpdfstring{R$_{\odot}$}{rsun}:} 
\label{sec:mag}
\subsection {Flux~rope Insertion Method}

In the corona, filaments lie in regions  where the plasma pressure is small compared to the magnetic pressure $\beta << 1$ i.e. the plasma is magnetically dominated. Moreover, the magnetic field plays a fundamental role in formation, stability and eruptions of the filaments \cite{Priestet89}. Therefore, it is important to understand the 3D magnetic configuration.

We utilize the Coronal Modeling System (CMS) software, a Non-linear Force Free Field {({NLFFF})} implementation software with a flux insertion capability developed by \citet{vanBallegooijen04}, to reconstruct the magnetic field in the solar corona using the {\em HMI} line of sight (LoS) magnetograms and global magnetic synoptic maps. The method
involves inserting a magnetic flux~rope into a potential field model based on an observed photospheric magnetogram and then uses magneto-frictional relaxation  to drive the magnetic field toward a force-free state while preserving the topology of the magnetic field lines \citep{Bobra08}. The technique has been successful to model sigmoids, filaments/prominences and pseudostreamers \citep{savchevaet16, Bobra08, asgari12,karnaetal24, yingnaet12, karnaet2019, karnaet2021}. 

At first, a wedge-shaped domain covering a large area surrounding the active region and coronal hole is selected. The domain extends from the photosphere to a ``source surface" at a radial distance of about 3 $R_\odot$ from the Sun center. The magnetic field in the domain is described in terms of the vector potential $(\vec{B}=\nabla \times \vec{A})$, so that axiomatically, the solenoid condition, ${\nabla}\cdot \vec{B} =0$, is satisfied. Next, a flux~rope is inserted following the path of an observed filament that appears dark in EUV images. The flux~rope path starts in a region of positive polarity near the PIL, then follows the observed filament path, and ends in a region with negative polarity on the PIL`s opposite side. The HMI\,LoS\,magnetogram serves as a lower boundary of the radial component $\vec{B}(r)$ of the magnetic field as a function of longitude and latitude whereas the HMI\,magnetic field synoptic map serves as the side boundary conditions. We apply open boundary conditions at the top, where the field is assumed to be radial. 

A set of models are constructed with different combinations of axial {f}lux, $\phi_{axi}$, and poloidal {f}lux, $F_{pol}$ per unit length along the flux rope. The axial flux runs horizontally and the poloidal flux wraps around the flux rope.
The models are then relaxed to a force-free state using magneto-frictional relaxation, which consists of evolving the coronal field via the induction equation expressed in terms of the vector potential \citep{Bobra08}. For an NLFFF model, the induction equation is iterated until the magneto-frictional velocity vanishes as per the assumption that the configuration is in equilibrium, and an unstable model keeps evolving with a subsequent iteration of the magneto-frictional equation. During the relaxation process, magnetic fields are allowed to vary throughout the wedge volume, while keeping the radial field in the photosphere fixed.{{ The end result of the magneto-frictional relaxation is a 3D NLFFF model with a magnetic flux~rope located at the location of the observed filament. The advantage of using this method is that it uses observed line-of-sight  photospheric magnetogram as a boundary condition and  EUV and Xray images to constrain coronal magnetic field. Moreover, this technique is much more economic in numerical resources.

From all the 3D NLFFF models we try to find the model that best fits the observed coronal structure in EUV  observations. The best-fit selection of stable and unstable methods involves tracing field lines through the 3D models and comparing these field lines with loops observed in the EUV images \citep{Savcheva09}. The fitting uses a 3D visualization tool that overlies and displays field lines in projection on the AIA plane of the sky view. We first select a few prominent loops in the AIA image and save their coordinates relative to the Sun center and align them with a model. We then measure the distance between the observed loops and the model field lines and calculate the quadratic average. The model with the lowest quadratic average value is considered the best-fit model.}} The NLFFF stable solutions are reasonable representations of the equilibrium coronal magnetic field and the best fit unstable models are suitable representations of an eruption morphology, if not the dynamics \citep{savchevaet16, karnaet2021}.

\subsection{Application to CME\,1}

We ran a set of {18} models with different choices of axial {f}lux, $\Phi_{axi}$ and poloidal {f}lux, $F_{pol}$ ranging from $0.01-5\times10^{21}$\,Mx  {and $0.5-5 \times10^{10}$\,Mx\,cm$^{-1} $}, respectively. We used the filament path observed in the EUV images as the length of the flux~rope. We tested the flux~rope height parameter by inserting flux~rope at different heights and found that the height had a significant effect on the FR's stability. The range in axial and poloidal fluxes and FR initial heights are appropriate for the general properties of the observed filaments.

\begin{figure}[htbp]
\centering\includegraphics[width=1\textwidth]{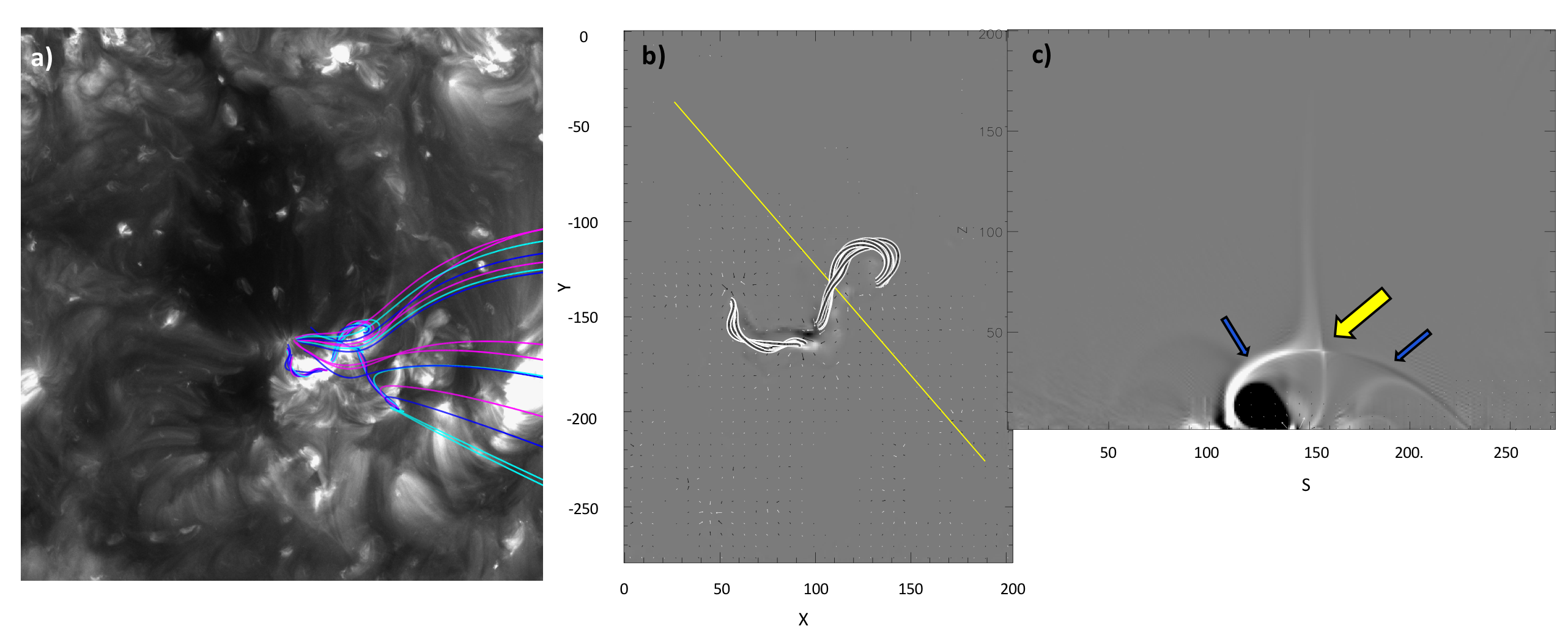}
\caption{{\em (a)} Some field lines traced from the best-fit stable model (axial flux $1 \times 10 ^{20}$\,Mx, poloidal flux  $1 \times 10^{10}$\,Mx\,cm$^{-1}$) before the eruption (CME\,1, {Figure~\ref{fig:aia193} {\em top left panel})}, representing the 3D magnetic field structure and are plotted over AIA\,193\,\AA\ image taken at 2011\,May\,25\,03:29\,UT. {\em (b)} A horizontal map of the best stable model showing the current density distribution in the computational domain at 8\,Mm height with few magnetic field lines for 30,000 iterations. {\em (c)} Distribution of the current density in a vertical cross-section of the flux~rope. The location of the vertical plane is shown by the {\em yellow} lines in {\em (b)}. The two {\em blue} arrow points to two dome and the {\em yellow} arrow points to the center null point.}
\label{fig:stable1} 
\end{figure}

\begin{figure}[htbp]
\centering\includegraphics[width=1\textwidth]{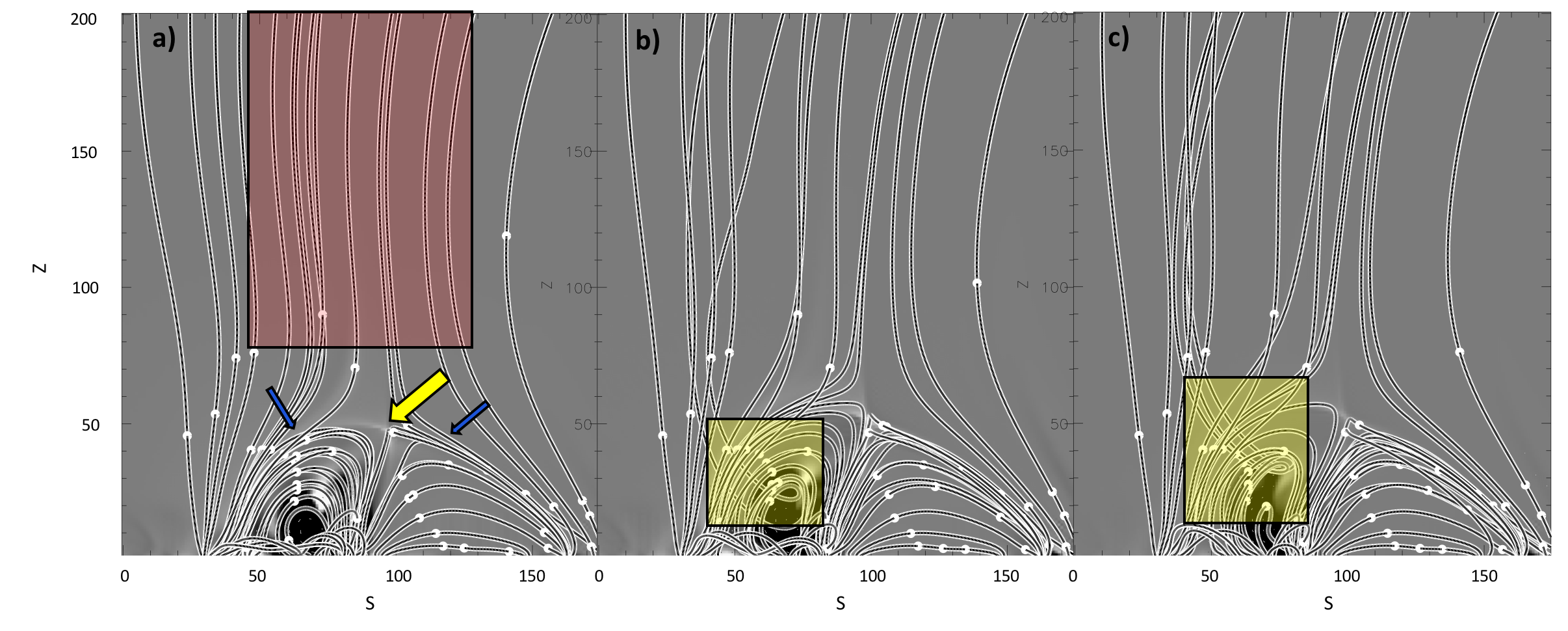}
\caption{Best fit unstable model (axial flux $3 \times 10 ^{20}$\,Mx, poloidal flux  $1 \times 10^{10}$\,Mx\,cm$^{-1}$) evolution with few magnetic field lines in SZ plane for selected iterations: 10,000 {\em (a)}, 40,000 {\em (b)}  and  70,000 {\em (c)}. The location of the vertical plane is shown by the {\em yellow} lines in Figure\,\ref{fig:stable1} {\em (b)}. Two domes (pointed by two blue arrows) with center null points (pointed by the yellow arrow) and streamer lines enclosed by the  red rectangular box are observed. The reconnection region of the field lines around the left cusp is enclosed by the
yellow shaded rectangular box.}
\label{fig:unstable1} 
\end{figure}

From our different combinations of axial and poloidal fluxes, the model with axial flux $1\times10^{20}$\,Mx and poloidal flux $1 \times10^{10}$ \,Mx\,cm$^{-1}$ is the best stable model (Figure\,\ref{fig:stable1}). 
We also found that the models with axial flux larger than $2\times10^{20}$\,Mx were unstable.
The flux~rope in these unstable models continues to expand and rise in height with continued iterations. We ran the relaxation code for our models until 90,000 iterations. Although the iterations do not provide the exact time of the eruption evolution, a sense of time can be obtained by interpreting the sequence of magneto-frictional steps as an evolution of the CME propagation. Previous work from \citet{Savchevaet15} shows that the structure of the unstable flux rope remains consistent with the erupting field. Out of {7} unstable models, the model with axial flux $3\times10^{20}$\,Mx and poloidal flux $1 \times10^{10}$  \,Mx\,cm$^{-1}$ is the best unstable model (Figure\,\ref{fig:unstable1}). 

Figure\,\ref{fig:stable1} {\em (a)} is a 3D view of the best stable model. Figure\,\ref{fig:stable1} {\em (b)} shows the horizontal map of the best stable model. Magnetic field lines are shown on a background of the radial component of the current density at 8\,Mm height with few magnetic field lines for 30,000\,iterations and Figure\,\ref{fig:stable1} {\em (c)} shows vertical cross sections through the flux~ropes at the location of the yellow lines in {\em (b)}. {{A very sharp boundary is observed with two lobes (pointed by two blue arrows), and an X-point or a magnetic null (pointed by the yellow arrow) exists at a height of 70\,Mm in the corona.}}

Figure\,\ref{fig:unstable1} is the best fit unstable model evolution for selected iterations: 10,000 {\em (a)}, 40,000 {\em (b)} and 70,000 {\em (c)} in a vertical cross-section of the flux~rope at the location of the yellow line in Figure\,\ref{fig:stable1} {\em (a)}. { {Two domes (pointed by two blue arrows in {\em (a)} panel) with center null points (pointed by the yellow arrow in {\em (a)} panel) and streamer lines (red-shaded rectangular box in the panel {\em (a)}) are observed in the SZ plane. This configuration resembles the x-point topology. The reconnection region of the open and closed-field lines around the left dome is enclosed by the yellow-shaded rectangular box in {\em (b)} and {\em (c)} panels.  {The magnetic field lines reconnect over the left dome in the panels {\em (a) and (b)}. Additionally, the current density associated with the erupting flux rope increases and rises to greater heights with successive iterations.} The eruption propagates towards the null point making a curve. Though after the eruption, the x-point topology still existed. }}

\subsection{Application to CME\,2}

\begin{figure}[htbp]
\centering\includegraphics[width=1\textwidth]{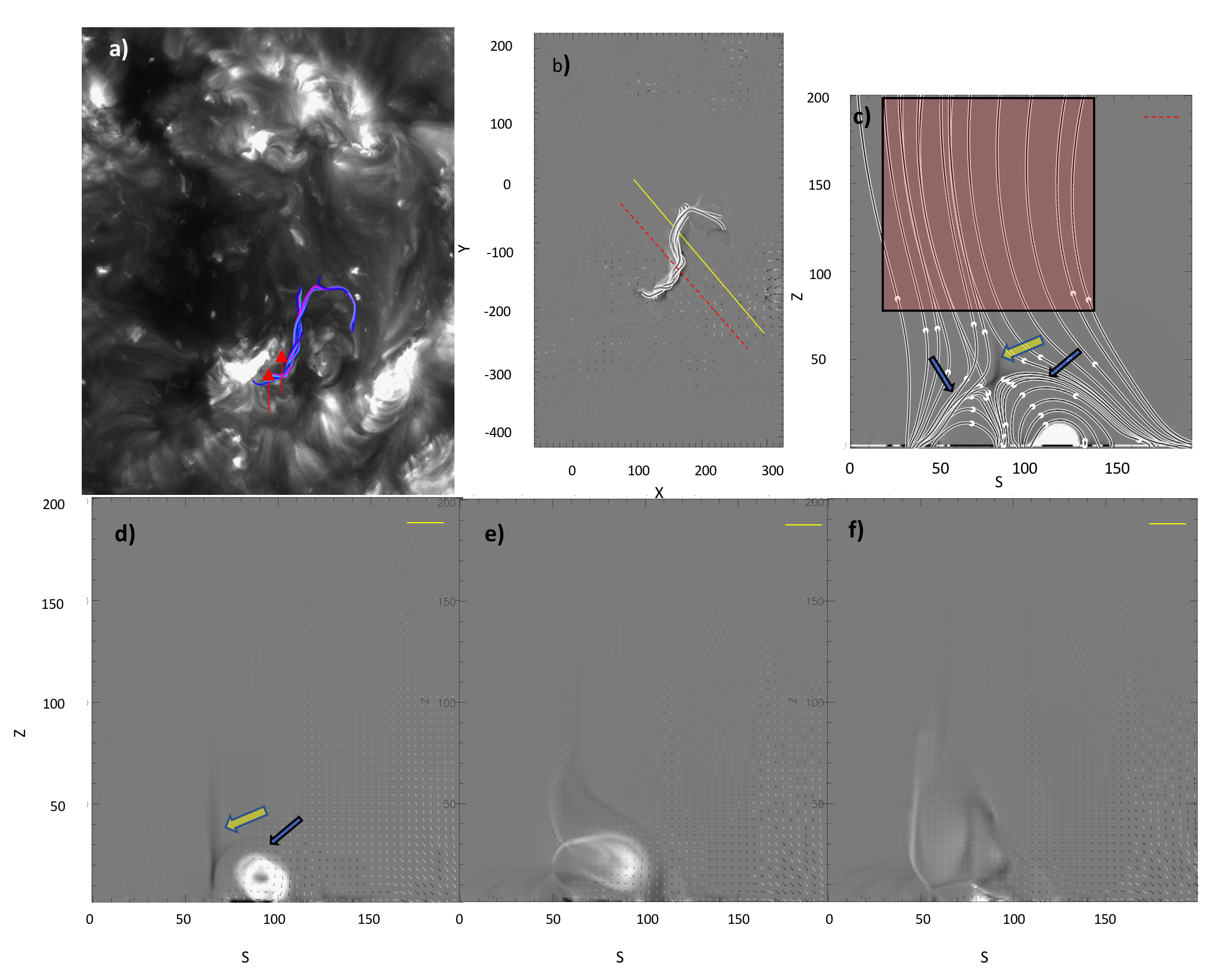}
\caption{Panel\,{\em a} presents some field lines traced from the best-fit stable model (axial flux $5 \times 10 ^{19}$\,Mx, poloidal flux $7 \times 10^{9}$\,Mx\,cm$^{-1}$) before the second eruption (CME\,2, {Figure~\ref{fig:event2} {\em top left panel})}, representing the 3D magnetic field structure and are plotted over AIA\,193\,\AA\ image taken on 2011\,May\,25\,10:30\,UT. The two {\em red} arrows point to the location of a null  {point}. Panel\,{\em b} shows a horizontal map of the best stable model showing the current density distribution in the computational domain at 8\,Mm height with few magnetic field lines for 30,000. Panels\,{\em (c-f)}, the best fit unstable model (axial flux $7 \times 10 ^{19}$\,Mx, poloidal flux  $9 \times 10^{9}$\,Mx\,cm$^{-1}$) evolution in SZ space. Panel\,{\em c} is the evolution for 10,000 in a vertical cross-section of the flux~rope at the location of the {\em red} dashed line in panel\,{\em b}. Two domes (pointed by two {\em blue} arrows) with center null points (pointed by the {\em yellow} arrow) and streamer lines enclosed by the {\em red} rectangular box are observed. Panels\,{\em (d-f)} show best fit unstable model evolution in SZ space for selected iterations: 10,000 {\em (d)}, 40,000 {\em (e)} and, 70,000 {\em (f)}. The location of the vertical plane is shown by the {\em yellow} lines in panel\,{\em b}. The separatrix and a null  {point} are pointed by {\em blue} and {\em yellow} arrows in panel\,{\em d}.} 
\label{fig:case2} 
\end{figure}

Similar to the first event model, we ran a set of 20 models with different choices of axial {f}lux ranging from $0.01-5\times10^{21}$\,Mx and poloidal {f}lux, ranging  {from $0.5-5 \times10^{10}$\,Mx\,cm$^{-1} $}, respectively. There was a clearly observed filament in the EUV images. We found that the model with $5 \times 10 ^{19}$\,Mx axial flux and  $7 \times 10^{9}$\,Mx\,cm$^{-1}$ poloidal flux was the best fit stable model and the model with  $7 \times 10 ^{19}$\,Mx axial flux and $9 \times 10^{9}$\,Mx\,cm$^{-1}$ poloidal flux is the best unstable model.
Figure\,\ref{fig:case2} shows the best stable and unstable model for the second event. {\em (a)} is the 3D structure of the filament observed before eruption plotted over AIA\,193\,\AA\ image taken on 2011\,May\,25\,10:30\,UT. Panel\,{\em b} shows the 2D horizontal map of the best stable model showing the current density distribution in the computational domain at 8\,Mm height with few magnetic field lines for 30,000. Panel\,{\em c} is the best fit unstable model evolution for 10,000 in a vertical cross-section of the flux~rope at the location of the {\em red} dashed line in panel\,{\em b}. Two domes (pointed by two {\em blue} arrows) with center null points (pointed by the {\em yellow} arrow) and streamer lines enclosed by the {\em red} rectangular box are observed. The null point is more like a line of nulls oriented along the southeast direction pointed by the two red arrows in panel\,{\em a}, passes in between the two eruptions’ locations. Panels\,{\em d-f} is the best fit unstable model evolution for selected iterations: 10,000 {\em (d)}, 40,000 {\em (e)} and 70,000 {\em (f)} in a vertical cross-section of the flux~rope at the location of the {\em yellow} line in panel\,{\em (b)}. The separatrix and a null  {point} are pointed by blue and yellow arrows in panel\,{\em (d)}. At the initial iterations, the eruption is propagated non-radially making a curve towards the null point as seen in {\em (e)} similar to CME\,1. The non-radial propagation toward the null point is due to deflection from nearby CHs. Moreover, CMEs motion is guided toward the null point which has the weaker field regions at different heights in the overlying configuration.

\subsection{Magnetic Models result}

Table\,\ref{tab:b} gives a summary of the parameters for the two events. For each model, we computed a number of parameters: total poloidal flux, a ratio of axial to poloidal flux, free energy (difference between the total magnetic energy and potential energy), and relative magnetic helicity \citep{Bobra08}. { { The absolute twist angles for unstable models for the Event~\,1 and Event\,2 were $1.3 \pi$ and $7.8 \pi$ estimated using $\phi= 2 \pi F_{pol} L/\Phi_{axi}$ \citep{Savcheva09}.}}

Comparing both events, we notice that CME\,1 had larger axial and poloidal flux compared to CME\,2. Both events result in a positive relative helicity but Event\,1 had higher helicity than Event\,2. Free energy is slightly higher in the stable model from Event\,1 compared to the stable model from Event\,2, but both events' unstable models have higher free energy than the stable models.  {Event\,2, twist angle exceeded the the critical twist ($2.5\pi$) derived by \citet{hood81}).}

{ {Using the NLFFF model, we were able to reconstruct the pre-eruptive magnetic configurations of two filaments that eventually gave rise to two CMEs. The reconstructed fields revealed critical magnetic structures, including magnetic null points, separatrices, and regions conducive to magnetic reconnection. 
}}

\begin{table}
\centering
{
 \begin{tabular}{|c c c c c|} 
 \hline

Parameters & \multicolumn{2}{c}{Event\,1} & \multicolumn{2}{c|}{Event\,2} \\[2ex]
 & stable &  unstable &  stable & unstable \\[2ex] 
 \hline 
   Axial Flux ($10^{19}$\, Mx) & 10 & 30 & 5 & 7 \\[2ex] 
   Poloidal Flux ($10^{09}$\, Mx\,cm$^{-1})$ & 10 & 10 & 7 & 9 \\[2ex] 
   Total Poloidal ($10^{19}$\, Mx) & 9.8 & 9.8 & 10.6 & 13.7  \\[2ex] 
   Axial Flux/Poloidal Flux ratio & 1.02& 3.06 & 0.47 & 0.51  \\[2ex] 
   Free energy ($10^{30}$ Mx$^2$) & 5.98 & 6.06 & 4.94 & 5.56 \\[2ex] 
   Helicity ($10^{40}$ erg)& 17.9 & 30.6 &6.8 &10.7 \\[2ex]
    flux~rope length (Mm) & 98 & 98 & 152 & 152 \\[2ex] 
 
 \hline
 \end{tabular}}
 \caption{{Representative Model Grid Parameters for Event\,1 (CME\,1) and Event\,2 (CME\,2).}}
 \label{tab:b}
\end{table}


\section{HD modeling 4-214 \texorpdfstring{R$_{\odot}$}{rsun}:} 
\label{sec:hd}

{ {While the NLFFF analysis provides valuable insight into the coronal magnetic field configuration and eruption near the Sun, it is equally important to investigate how the resulting CMEs and HSS propagate through the heliosphere and interact with the Earth’s environment. In this section, we present a complementary study to trace the evolution of the CMEs and HSS from the Sun to Earth by performing hydrodynamic (HD) numerical simulations based on two approaches: (1) forward modeling using STEREO-A and STEREO-B white-light observations, and (2) backward analytic reconstructions using in,situ measurements.}}

On one side, we performed forward reconstructions starting from remote sensing observations which are limited by the scattered light and because they only cover few tens of solar radii above the limb. On the other side, we performed backward reconstructions which are known to be limited as they rely on the plasma stationary state unabling to resolve stream-line crossings. The limitations of this approach have been discussed by \cite{Burlaga_1982, 1981JGR....86.6685P, 2021JSWSC..11....7B}.

With this, we obtained three different simulations, two of them, from the remote sensing observations from both STEREO-A and STEREO-B and one from the OMNI (in\,situ) dataset. { {These simulations, which do not require high-performance computing, allowed us to bridge remote and in,situ observations and identify both CMEs and the HSS as the main drivers of the GS event.}}

For the backward reconstructions, we solved the equations describing the polytropic case of a solar wind expanding adiabatically as described by \cite{2022PhPl...29l2901S} \citep[see also][]{Parker1958, 1987flme.book.....L}.  

For the simulations, we used the numerical HD code YGUAZÚ-A \citep{Raga2000} which integrates the HD equations for a plasma (with adopted abundances by number H=0.9 and He= 0.1) over a five-level binary adaptive 1024$\times $1024  point grid (computational domain) and time steps of 15\,min. This code has previously been used to study CMEs evolution \citep{Niembro2019}. 

\subsection{Setting the initial conditions 4--12\,R$_{\odot}$}
\label{sec:init}

In the numerical simulations, first, the computational domain is filled by an isotropic ambient solar wind (with a speed V$_{SW}$ and density N$_{\odot}$). Then, sudden changes of the flow are imposed at an injection radius R$_{inj}$. This means that, the HD code inputs are V and N\,time series at R$_{inj}$. Each sudden change (for an interval of time) may represent CMEs and flows at different speeds (including HSSs). 

We selected R$_{inj}=$10\,R$_{\odot}$ and obtained V and N\,time series from the remote sensing observations from (1) STEREO-A and (2) STEREO-B and also from (3) OMNI data.

\begin{figure}[htbp]
\centering\includegraphics[width=0.71\textwidth]{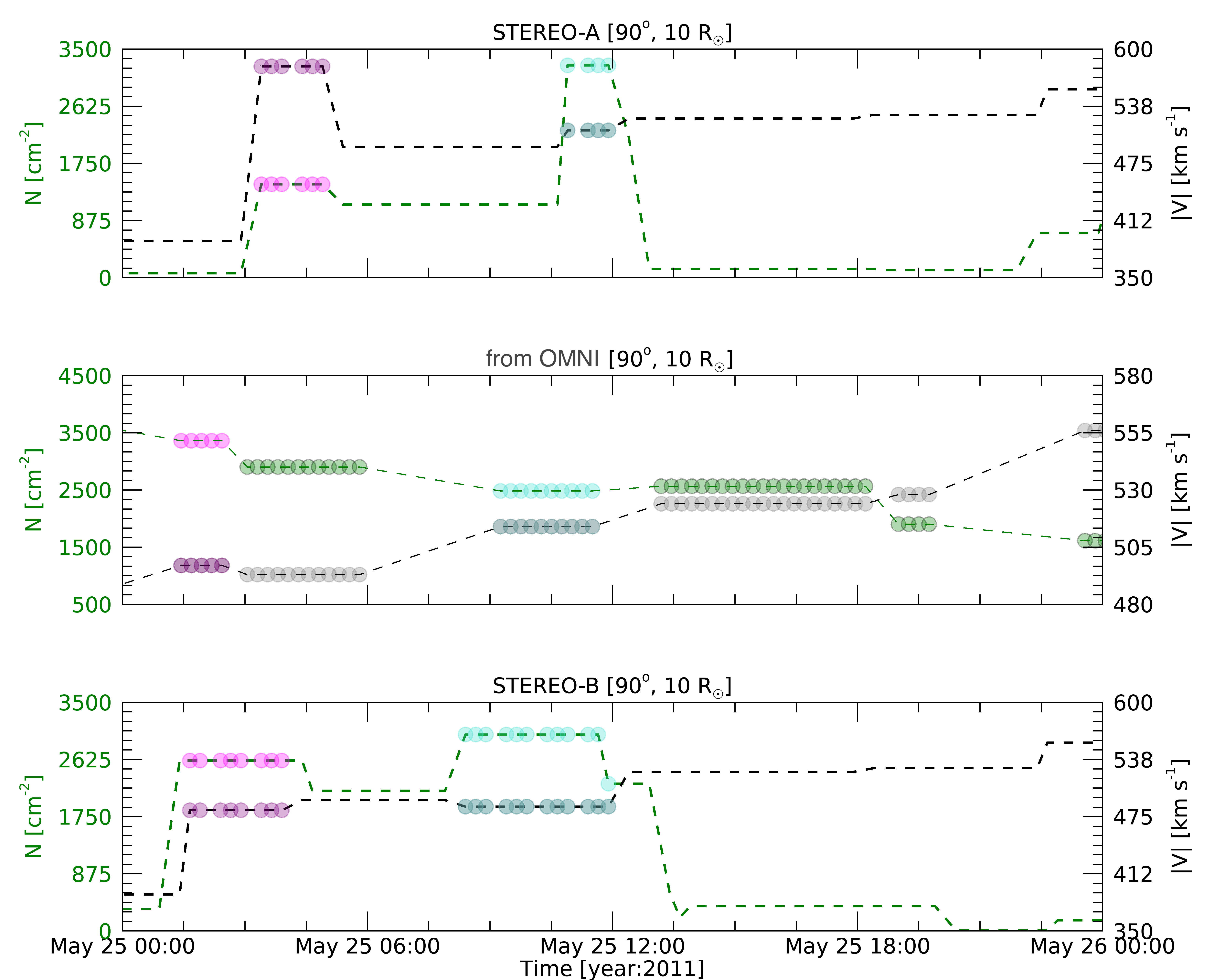}
\caption{Speed and density profiles obtained from STEREO-A ({\em top}) and STEREO-B ({\em bottom})\,COR2 remote sensing observations and the analytical hydrodynamical reconstruction based on OMNI in\,situ measurements ({\em middle} panel). We matched, in time and space, the parcels of plasma related to both transients. CME-1 was traced in {\em purple} and characterized at 10 R$_{\odot}$ along the ecliptic plane moving away from the Sun with $V \sim$482--581\,km\,s$^{-1}$ and $N \sim$2625--3300\,cm$^{-3}$ within a time window of 1.5-2.5\,h. CME\,2 moved with $V \sim$486--511\,km\,s$^{-1}$ and $N \sim$2500--3300\,cm$^{-3}$ within 1.5--3\,h.}
\label{fig:bright} 
\end{figure}

In Section\,\ref{sec:remote}, we described a statistical analysis performed over the STEREO-A/B COR\,2 brightness time series ({\em left} panels of Figure\,\ref{fig:sb_STB} for STEREO-B and Figure\,\ref{fig:sb_STA} for STEREO-A) and the respective methodology to obtain the N\,time series and also the CMEs speed. These profiles are shown in Figure\,\ref{fig:bright}. 

Additionally, in Section\,\ref{sec:insitu}, we obtained 15-minute V and N series at 1\,au. From these series, we backward reconstructed the solar wind conditions to R$_{inj}$ (show in the {\em middle} panel of Figure\,\ref{fig:bright}).

The three perspectives (STEREO-A/B and OMNI) provide estimations of the speed and density of the CME\,1 (shown {\em pink}) and CME\,2 (in {\em aqua}) while only from the OMNI backward reconstruction, we obtained the speed and density of different patches of the solar wind (shown in {\em green} in the {\em middle} panel). 

From this figure, we found that the three spacecraft predict the passage of the first eruption on 2011\,May\,25 between 01:30 and 04:30\,UT propagating over the ecliptic plane at V ranging from 482 to 581\,km\,s$^{-1}$ with N between 1500 and 3300\,cm$^{-3}$, while the second was observed on 2011\,May\,25 between 08:00 and 12:00\,UT at $486<V<511$\,km\,s$^{-1}$ and $2600<N<3300$\,cm$^{-3}$.  

\subsection{Numerical models}

{ The three sets of V and N\,time series, were used as inputs in the HD code to obtain V and N\,time series at different distances. In Figure\,\ref{fig:sim0}, we show the simulation results (V and N time series) at different heliospheric distances: STEREO-B (show in {\em navy}), STEREO-A (in {\em red}) and OMNI backward reconstruction (in {\em green}) overlapped. With a {\it yellow} square, we denote the presence of solar wind conditions between the two CMEs dissipating at distances near the Earth. Therefore, the simulations predicted no interaction between the CMEs. This flow is inhibiting their interaction. 

\begin{figure}[htbp]
    \centering
    \includegraphics[width=1\textwidth, trim={0in 0.97in 0.2in 0in}]{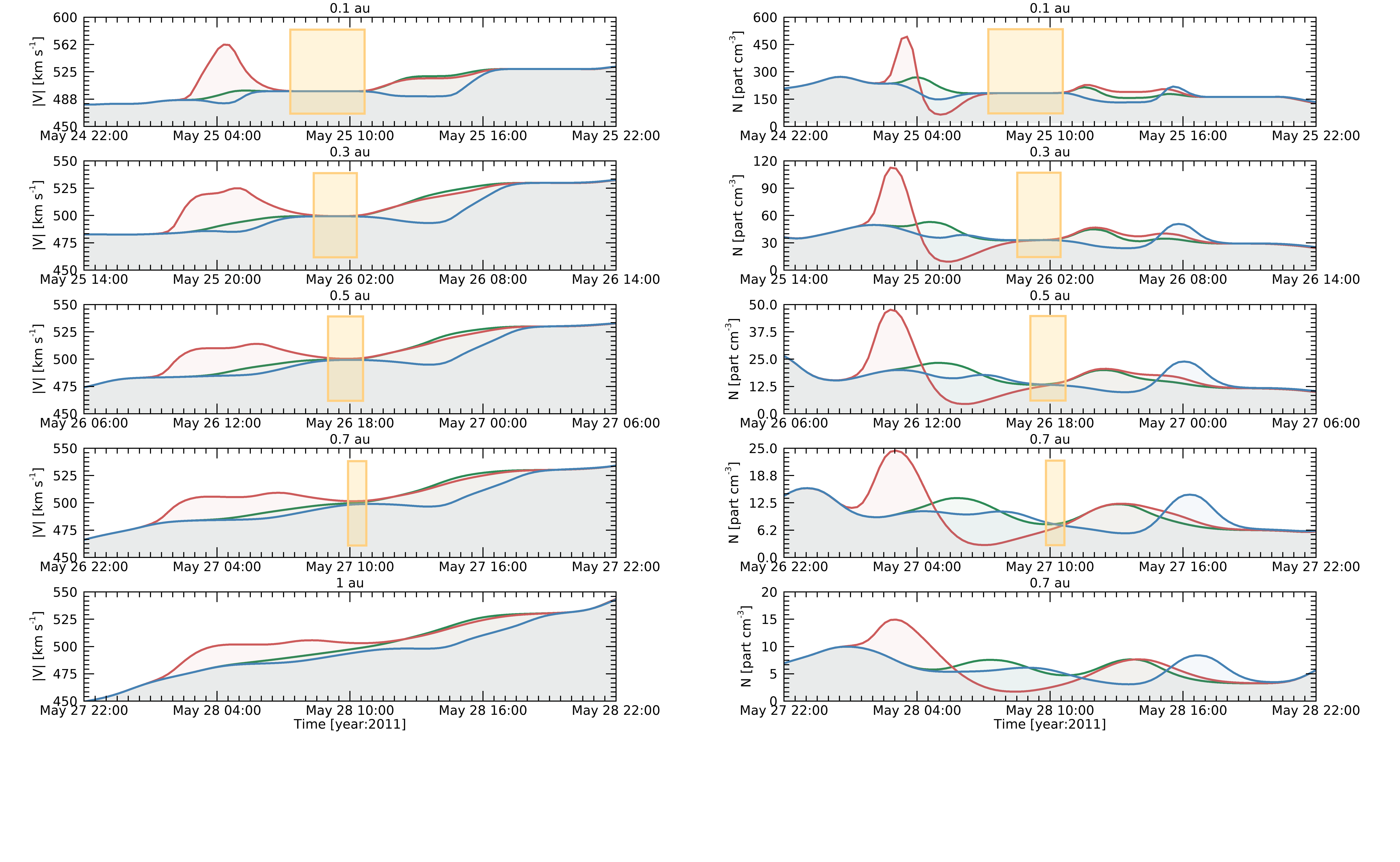}
    \caption{Simulation results (V and N time series) at different heliospheric distances: STEREO-B (show in {\em navy}), STEREO-A (in {\em red}) and OMNI backward reconstruction (in {\em green}) overlapped. The yellow square shows that between the two CMEs there is a patch of solar wind related to the coronal hole evolution. The numerical models predict that the CMEs started to interact near 1\,au.}
    \label{fig:sim0}
\end{figure}

The flow between the CMEs is characterized with a speed of 530--550\,km\,s$^{-1}$ (see Figure\,\ref{fig:bright}), that is, a speed faster than the flow before CME\,1 (375--400\,km\,s$^{-1}$) but slower than the one from the CH flow ($>700$\,km\,s$^{-1}$, after CME\,2).} 

\begin{figure}[htbp]
    \centering
    \includegraphics[width=1\textwidth, trim={0in 0.95in 0.2in 0in}]{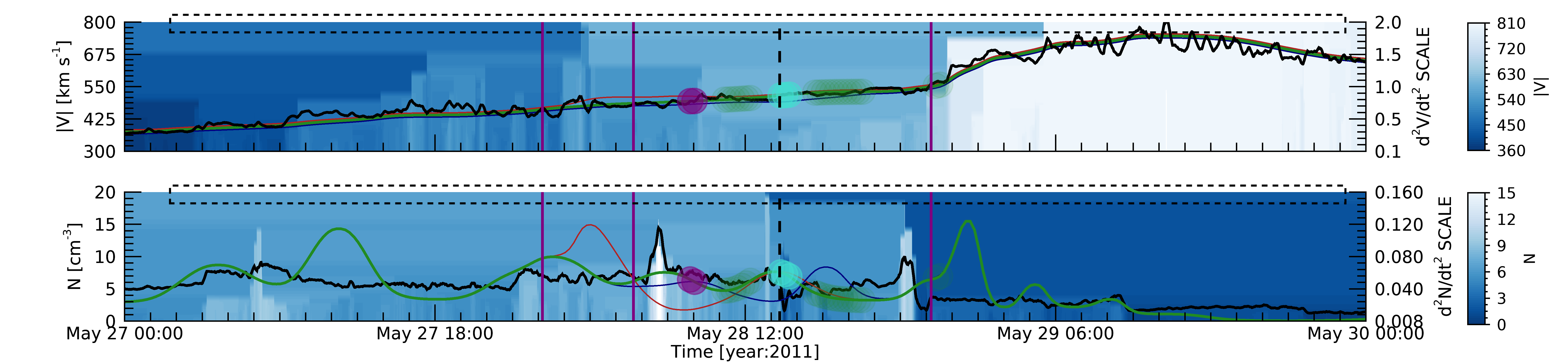}
    \caption{Last two panels of Figure\,\ref{fig:dev} with the V and N synthetic time series from: STEREO-B (show in {\em navy}), STEREO-A (in {\em red}) and OMNI backward reconstruction (in {\em green}) overlapped. Plasma from CME\,1 is marked with {\em purple circles}, CME\,2 in {\em aqua} and solar wind between both CMEs in {\em green}. The {black} dashed line denotes the predicted CME\,2 arrival time.}
    \label{fig:sim}
\end{figure}

In Figure\,\ref{fig:sim}, we show the last two panels from Figure\,\ref{fig:dev} overlapping the V and N synthetic profiles (obtained from the simulations) from: STEREO-B (in {\em navy}), STEREO-A (in {\em red}) and OMNI backward reconstruction (in {\em green}). Because of the conservative conditions of the HD model, we tag the CME flows. The {\em purple circles} correspond to the first eruption while the {\em aqua} to the second. We also colored {\em circles} in {\em green} that belong to solar wind flow between both transients. Gaps between {\em circles} are regions in which the solar wind patches have interacted (resulting from the stream-line crossings). The backward reconstruction predicts solar wind flows between both transients, { the model predicts that the CMEs were not interacting between them until they reached Earth}.

\begin{figure}[htbp]
\centering\includegraphics[width=0.8\textwidth]{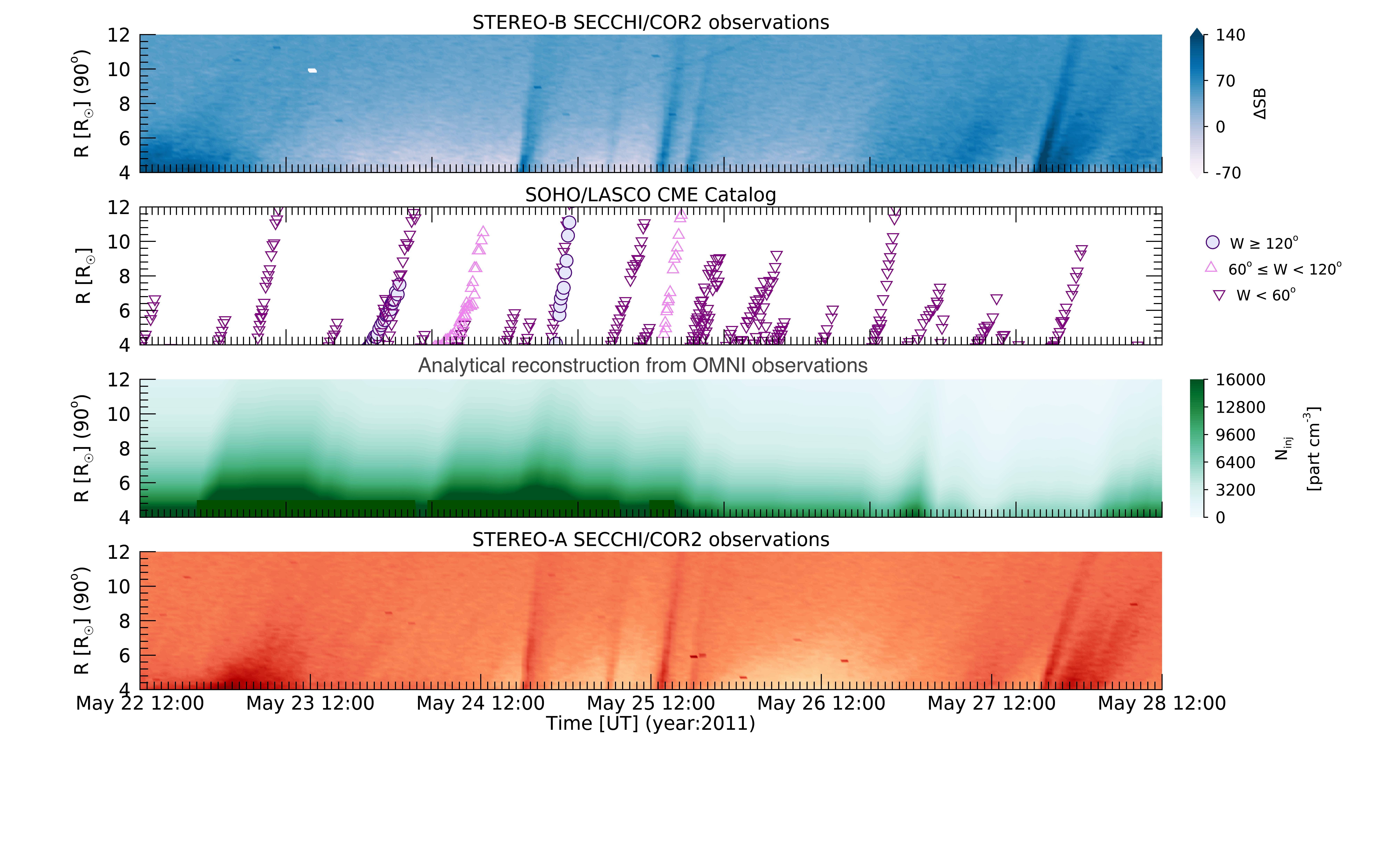}
\caption{Height-time plots (4-12\,R$_{\odot}$) from: remote sensing observations (STEREO-B ({\em first} panel) and STEREO-A ({\em fourth} panel) COR2 images, SOHO/LASCO\,CME\,Catalog ({\em second} panel) and OMNI\,backward reconstruction ({\em third} panel). All CMEs listed in the catalog are shown with different symbols representing their angular width (W).}
\label{fig:ht} 
\end{figure}

As we can compute V and N\,time series at any other heliospheric distance. We constructed height-time plots from 4 to 12\,R$_{\odot}$ for each spacecraft to compare with the SOHO/LASCO\,CME\,Catalog
\citep{2009EM&P..104..295G}. In Figure \ref{fig:ht}, we present the STEREO-B\,COR2 height-time plot in {\em blue} shades) along the ecliptic plane. In the {\em bottom} panel, we show the corresponding one to STEREO-A\,COR2 in {\em red} shades). The CMEs reported in the SOHO/LASCO\,CME\,Catalog in the {\em second} panel. Last, in the {\em third} panel, the height-time plot obtained from OMNI backward reconstruction. 

In addition, Figure\,6 of \citet{Chi_2018} shows height time plots which covers up to 30 solar radii with the presence of two CMEs.


\section{Discussion}

Here, we present a study of a geomagnetic storm driven by two ICMEs, both associated with two active region filaments eruptions occuring near a coronal hole. Using combination of multi-spacecraft observations and both magnetic and hydrodynamic modeling, we explore how the proximity of the coronal hole influenced the magnetic field configuration and the subsequent propagation of the CMEs.

The first goal of this study was to examine the impact of nearby coronal hole on the the pre and post eruptive magnetic configuration of filaments. From our magnetic models, we found that both our stable and unstable models had a lower axial and poloidal value compared to other active region case studies performed by other researchers using the same modeling technique \citep[e.g., ][]{Bobra08, Savcheva09,yingnaet09,  Su11, Savcheva12a, karnaetal24} when there was no presence of nearby coronal holes. We found our stable models axial flux values were at least 3 times lower than those reported in those studies. However, when  we compared our results with \citet{Yingna09ch}, in which the active region was near a coronal hole, our models' values correspond to the range of values of their best-fit models. Moreover, we find that a magnetic null exists in the corona of the active region prior to eruption. The null lies on the separatrix surface that separates open and closed field lines which are more prone to reconnection. These findings suggest that the proximity of a coronal hole to an active region can influence the magnetic stability, potentially making the system more prone to eruption.

{ The second goal of this study was to assess the influence of coronal hole originated high speed stream on the CME propagation/evolution. From three separate data sets (EUV images (Figures\,\ref{fig:aia193} and \ref{fig:event2}), coronagraph images (Figures\,\ref{fig:sb_STB} and \ref{fig:sb_STA}) and in\,situ observations (Figure\,\ref{fig:dev}) we found that this GS event with two ICMEs embedded in solar wind was influenced by a CH.

 {The remote-sensing data showed two distinct CME structures propagating up to $\sim$20\,R$_\odot$ without signs of interaction. This observation, along with the results from our HD simulations (Figure~\ref{fig:sim0}), supports the notion that the two CMEs evolved independently during their transit.}

 {Our simulations predicted no significant interaction between the transients, yet both arrived at 1,au (Figure~\ref{fig:sim}). This suggests that the CH and its associated HSS played a role in inhibiting interaction between the CMEs, allowing them to propagate more freely along the fast solar wind stream. The in situ signatures also revealed two distinct magnetic flux ropes with an interacting transition region between them, consistent with earlier findings by \citet{Chi2018}. These observations support the idea that the CMEs traveled within the CH-originated HSS, preserving their structure.}

Previously, several researcher \citep[e.g., ][]{Chertoket18,Abuninet20} have noted that CMEs launched close to coronal hole, propagates within the coronal hole originated high-speed stream instead of getting deflected from a radial trajectory of an adjacent coronal hole. \citet{Chertoket18} noticed in their study that ICMEs arrived at Earth earlier than expected as they propagated in the high-speed solar wind similar to our study.

 {Our study provides a comprehensive view connecting the coronal hole–HSS system to filament eruptions and their geo-effectiveness, an aspect that has not been systematically explored before. By combining observations with magnetic and hydrodynamic modeling, we capture multiple components of the CME life cycle: pre-condition, eruption, and propagation. Notably, the modeling approaches used here provide a computationally efficient means to investigate space weather events with sufficient accuracy and physical insight, enabling us to effectively track CME kinematics, arrival times, and interactions with high-speed streams.}

\section{Conclusion}

We presented magnetic field configurations of two filament eruptions observed in EUV that caused a geomagnetic storm. We studied how the presence of a nearby coronal hole impacted the stability of the NLFFF models. The consistency of the model results with clearly observed morphological features in solar observations suggests that we have captured the large-scale magnetic structure of the eruptions. We analyzed STEREO-A/B\,COR2 brightness distributions in space and time to identify both eruptions in coronagraph images and characterize the solar wind speed and density at 10\,R$_{\odot}$.
We also employed analytical models to reconstruct the solar wind speed and density from OMNI (1\,au) back to 10\,R$_{\odot}$. The resulting speed and density time profiles at 10\,R$_{\odot}$, from STEREO observations and by reconstructing the OMNI measurements, were used to perform HD\,1D numerical models to compare with in\,situ data. With the numerical models, we tagged and followed the material from both eruptions propagating up to 1\,au. Further, from remote sensing, we constructed and compared height-time plots near the Sun identifying both CMEs. Our major results are summarized as follows:

\begin{itemize}

\item The geomagnetic storm involved two coronal mass ejections and the source active regions were bordering a large coronal hole. There was a $\sim$8\,h time gap between two eruptions. 

\item In both events magnetic models, two domes with center null points and streamer lines were observed in the SZ\,plane, resembling an x-point null structure. The null  {points} passes in between the two eruptions' locations.

\item The second event had less axial and poloidal flux in comparison to the magnetic models of the first event. The length of the filament associated with CME\,2 is larger than that of CME\,1.

\item At the source region: {\em (a)} both events had positive helicity computed from magnetic models, {\em (b)} CME\,1 had higher free energy and helicity compared to CME\,2.

\item From our magnetic models, we found that the presence of a nearby coronal hole made the flux~rope unstable at relatively low axial flux values.

\item From coronagraph images, we identified two perturbations in the brightness distributions measured by STEREO-B and STEREO-A. 

\item  {{In {\em in-situ} observations: {\em (a)} the two CMEs were embedded in CH-originated high-speed solar wind, {\em (b)} CME\,1 had larger heavy ion ratio compared to CME\,2 and lower alpha/proton ratio, {\em (c)} CME\,1 is a structure with larger magnetic and flow pressure, {\em (d)} Both CMEs have low temperature and high magnetic field strength, {\em (e)} CME\,1 is characterized by larger V and N fluctuations.}}

\item From the V and N second derivative analysis, we captured three solar wind regions: {\em (a)}  characterized by high speed and low density related to the coronal hole, {\em (b)} a region with high speed and high density related with the CMEs path and {\em (c)} a region of ambient solar wind which had low speed and high density. 

\item From the backward OMNI\,reconstruction, we identified the geomagnetic storm related to two CMEs implying that reconstruction from in\,situ data is able to identify the structures near the Sun. 

\item From the HD numerical models, we were able to identify the arrival of material of both CMEs and the interface between them. This is strongly supported by the remote sensing observations where no sign of CME-CME interaction was observed while they propagated towards the Earth. Between the CMEs, fast flows are found indicating the presence of plasma from the CH-HSS. This plasma is comparable in V and N to the CMEs inhibiting CME interaction processes.

\item There is a good correlation between EUV images and the height-time plots from remote sensing (COR2 and SOHO/LASCO\,CME\,Catalog) and in\,situ (from OMNI) observations.

\end{itemize}

\begin{acks}
 The authors would like to thank Alejandro Lara, Katharine Reeves, Edward DeLuca and Kristoff Paulson for all their helpful discussions. 
\end{acks}

 \begin{fundinginformation}
This work is supported in part by the contract SP02H1701R from Lockheed-Martin to the Smithsonian Astrophysical Observatory, SAO-408200 and NSF\,AGS-2201767 grants. NK and TN were supported by NSF\,AGS-2201767 grant.
 \end{fundinginformation}

\begin{dataavailability}
We acknowledge the use of NASA/GSFC's Space Physics Data Facility's OMNIWeb (or CDAWeb or ftp) service, and OMNI data. The D$_{ST}$ (Disturbance Storm-Time) index used in this paper was provided by the WDC for Geomagnetism, Kyoto (\url{http://wdc.kugi.kyoto-u.ac.jp/wdc/Sec3.html}). This CME catalog is generated and maintained at the CDAW Data Center by NASA and The Catholic University of America in cooperation with the Naval Research Laboratory. SOHO is a project of international cooperation between ESA and NASA. The AIA and HMI data are courtesy of NASA/SDO and the AIA and HMI Science Investigation Teams. 
\end{dataavailability}

\begin{authorcontribution}
N.K. computed magnetic models, and and T.N. performed numerical simulations. All authors made massive contributions to the conception of the work and wrote the manuscript.
\end{authorcontribution}

 \begin{ethics}
 \begin{conflict}
The authors declare no competing interests.
 \end{conflict}
\end{ethics}

\bibliographystyle{spr-mp-sola}
\bibliography{gs}

\begin{thebibliography}{96}
\ifx\bisbn     \undefined \def\bisbn  #1{ISBN #1}\fi
\ifx\binits    \undefined \def\binits#1{#1}\fi
\ifx\bauthor   \undefined \def\bauthor#1{#1}\fi
\ifx\batitle   \undefined \def\batitle#1{#1}\fi
\ifx\bjtitle   \undefined \def\bjtitle#1{\textit{#1}}\fi
\ifx\bvolume   \undefined \def\bvolume#1{\textbf{#1}}\fi
\ifx\byear     \undefined \def\byear#1{#1}\fi
\ifx\bissue    \undefined \def\bissue#1{#1}\fi
\ifx\bfpage    \undefined \def\bfpage#1{#1}\fi
\ifx\blpage    \undefined \def\blpage #1{#1}\fi
\ifx\burl      \undefined \def\burl#1{#1}\fi
\ifx\href      \undefined \def\href#1#2{#2}\fi
\ifx\betal     \undefined \def\betal{et al.}\fi
\ifx\bctitle   \undefined \def\bctitle#1{#1}\fi
\ifx\beditor   \undefined \def\beditor#1{#1}\fi
\ifx\bbtitle   \undefined \def\bbtitle#1{\textit{#1}}\fi
\ifx\bedition  \undefined \def\bedition#1{#1}\fi
\ifx\bseriesno \undefined \def\bseriesno#1{\textbf{#1}}\fi
\ifx\blocation \undefined \def\blocation#1{#1}\fi
\ifx\bsertitle \undefined \def\bsertitle#1{\textit{#1}}\fi
\ifx\bsnm      \undefined \def\bsnm#1{#1}\fi
\ifx\bsuffix   \undefined \def\bsuffix#1{#1}\fi
\ifx\bparticle \undefined \def\bparticle#1{#1}\fi
\ifx\barticle  \undefined \def\barticle#1{}\fi
\ifx\binstitute  \undefined \def\binstitute#1{#1}\fi
\ifx\bpublisher  \undefined \def\bpublisher#1{#1}\fi
\ifx\doiurl    \undefined \def\doiurl#1{\href{#1}{DOI}}\fi
\makeatletter
\def\safeHref#1#2#3{\in@{http}{#2}\ifin@\href{#2}{#3}\else\href{#1#2}{#3}\fi}
\makeatother
\ifx\adsurl    \undefined
  \def\adsurl#1{\safeHref{https://ui.adsabs.harvard.edu/abs/}{#1}{ADS}}\fi
\ifx\arxivurl  \undefined
  \def\arxivurl#1{\safeHref{http://arxiv.org/abs/}{#1}{arXiv}}\fi
\ifx\botherref \undefined \def\botherref#1{}\fi
\ifx\url       \undefined \def\url#1{#1}\fi
\ifx\bchapter  \undefined \def\bchapter#1{}\fi
\ifx\bbook     \undefined \def\bbook#1{}\fi
\ifx\bcomment  \undefined \def\bcomment#1{#1}\fi
\ifx\oauthor   \undefined \def\oauthor#1{#1}\fi
\ifx\citeauthoryear \undefined\def \citeauthoryear#1{#1}\fi
\def\endbibitem {}
\ifx\bconflocation  \undefined \def\bconflocation#1{#1} \fi

\bibitem[\protect\citeauthoryear{{Abunin} et~al.}{2020}]{Abuninet20}
\begin{barticle}
\bauthor{\bsnm{{Abunin}}, \binits{A.A.}},
\bauthor{\bsnm{{Abunina}}, \binits{M.A.}},
\bauthor{\bsnm{{Belov}}, \binits{A.V.}},
\bauthor{\bsnm{{Chertok}}, \binits{I.M.}}:
\byear{2020},
\batitle{{Peculiar Solar Sources and Geospace Disturbances on 20-26 August
  2018}}.
\bjtitle{\solphys}
\bvolume{295},
\bfpage{7}.
\doiurl{https://doi.org/10.1007/s11207-019-1574-8}.
\adsurl{2020SoPh..295....7A}.
\end{barticle}
\endbibitem

\bibitem[\protect\citeauthoryear{{Asai} et~al.}{2009}]{Asaietal09}
\begin{barticle}
\bauthor{\bsnm{{Asai}}, \binits{A.}},
\bauthor{\bsnm{{Shibata}}, \binits{K.}},
\bauthor{\bsnm{{Ishii}}, \binits{T.T.}},
\bauthor{\bsnm{{Oka}}, \binits{M.}},
\bauthor{\bsnm{{Kataoka}}, \binits{R.}},
\bauthor{\bsnm{{Fujiki}}, \binits{K.}},
\bauthor{\bsnm{{Gopalswamy}}, \binits{N.}}:
\byear{2009},
\batitle{{Evolution of the anemone AR NOAA 10798 and the related geo-effective
  flares and CMEs}}.
\bjtitle{Journal of Geophysical Research (Space Physics)}
\bvolume{114},
\bfpage{A00A21}.
\doiurl{https://doi.org/10.1029/2008JA013291}.
\adsurl{2009JGRA..114.0A21A}.
\end{barticle}
\endbibitem

\bibitem[\protect\citeauthoryear{{Asgari-Targhi} and {van
  Ballegooijen}}{2012}]{asgari12}
\begin{barticle}
\bauthor{\bsnm{{Asgari-Targhi}}, \binits{M.}},
\bauthor{\bsnm{{van Ballegooijen}}, \binits{A.A.}}:
\byear{2012},
\batitle{{Model for Alfv{\'e}n Wave Turbulence in Solar Coronal Loops: Heating
  Rate Profiles and Temperature Fluctuations}}.
\bjtitle{\apj}
\bvolume{746},
\bfpage{81}.
\doiurl{https://doi.org/10.1088/0004-637X/746/1/81}.
\adsurl{2012ApJ...746...81A}.
\end{barticle}
\endbibitem

\bibitem[\protect\citeauthoryear{ASY/SYM}{1981 - 2023}]{dstlink}
\begin{botherref}
\oauthor{\bsnm{ASY/SYM}}:
1981 - 2023,
\textit{WDC for Geomagnetism at U. Kyoto},
\url{https://wdc.kugi.kyoto-u.ac.jp/aeasy/}.
\end{botherref}
\endbibitem

\bibitem[\protect\citeauthoryear{{Biondo} et~al.}{2021}]{2021JSWSC..11....7B}
\begin{barticle}
\bauthor{\bsnm{{Biondo}}, \binits{R.}},
\bauthor{\bsnm{{Bemporad}}, \binits{A.}},
\bauthor{\bsnm{{Mignone}}, \binits{A.}},
\bauthor{\bsnm{{Reale}}, \binits{F.}}:
\byear{2021},
\batitle{{Reconstruction of the Parker spiral with the Reverse In situ data and
  MHD APproach - RIMAP}}.
\bjtitle{Journal of Space Weather and Space Climate}
\bvolume{11},
\bfpage{7}.
\doiurl{https://doi.org/10.1051/swsc/2020072}.
\adsurl{2021JSWSC..11....7B}.
\end{barticle}
\endbibitem

\bibitem[\protect\citeauthoryear{Bobra, van Ballegooijen, and
  DeLuca}{2008}]{Bobra08}
\begin{barticle}
\bauthor{\bsnm{Bobra}, \binits{M.G.}},
\bauthor{\bparticle{van} \bsnm{Ballegooijen}, \binits{A.A.}},
\bauthor{\bsnm{DeLuca}, \binits{E.E.}}:
\byear{2008},
\batitle{{Modeling Nonpotential Magnetic Fields in Solar Active Regions}}.
\bjtitle{The Astrophysical Journal}
\bvolume{672},
\bfpage{1209}.
\end{barticle}
\endbibitem

\bibitem[\protect\citeauthoryear{Borovsky and Shprits}{2017}]{Borovsky17}
\begin{barticle}
\bauthor{\bsnm{Borovsky}, \binits{J.E.}},
\bauthor{\bsnm{Shprits}, \binits{Y.Y.}}:
\byear{2017},
\batitle{Is the Dst Index Sufficient to Define All Geospace Storms?}
\bjtitle{Journal of Geophysical Research: Space Physics}
\bvolume{122},
\bfpage{11,543}.
\doiurl{https://doi.org/10.1002/2017JA024679}.
\burl{https://agupubs.onlinelibrary.wiley.com/doi/abs/10.1002/2017JA024679}.
\end{barticle}
\endbibitem

\bibitem[\protect\citeauthoryear{{Bravo}, {Cruz-Abeyro}, and
  {Rojas}}{1998}]{bravo98}
\begin{barticle}
\bauthor{\bsnm{{Bravo}}, \binits{S.}},
\bauthor{\bsnm{{Cruz-Abeyro}}, \binits{J.A.L.}},
\bauthor{\bsnm{{Rojas}}, \binits{D.}}:
\byear{1998},
\batitle{{The spatial relationship between active regions and coronal holes and
  the occurrence of intense geomagnetic storms throughout the solar activity
  cycle}}.
\bjtitle{Annales Geophysicae}
\bvolume{16},
\bfpage{49}.
\doiurl{https://doi.org/10.1007/s00585-997-0049-7}.
\adsurl{1998AnGeo..16...49B}.
\end{barticle}
\endbibitem

\bibitem[\protect\citeauthoryear{{Brueckner}
  et~al.}{1995}]{1995SoPh..162..357B}
\begin{barticle}
\bauthor{\bsnm{{Brueckner}}, \binits{G.E.}},
\bauthor{\bsnm{{Howard}}, \binits{R.A.}},
\bauthor{\bsnm{{Koomen}}, \binits{M.J.}},
\bauthor{\bsnm{{Korendyke}}, \binits{C.M.}},
\bauthor{\bsnm{{Michels}}, \binits{D.J.}},
\bauthor{\bsnm{{Moses}}, \binits{J.D.}},
\bauthor{\bsnm{{Socker}}, \binits{D.G.}},
\bauthor{\bsnm{{Dere}}, \binits{K.P.}},
\bauthor{\bsnm{{Lamy}}, \binits{P.L.}},
\bauthor{\bsnm{{Llebaria}}, \binits{A.}},
\bauthor{\bsnm{{Bout}}, \binits{M.V.}},
\bauthor{\bsnm{{Schwenn}}, \binits{R.}},
\bauthor{\bsnm{{Simnett}}, \binits{G.M.}},
\bauthor{\bsnm{{Bedford}}, \binits{D.K.}},
\bauthor{\bsnm{{Eyles}}, \binits{C.J.}}:
\byear{1995},
\batitle{{The Large Angle Spectroscopic Coronagraph (LASCO)}}.
\bjtitle{Sol. Phys.}
\bvolume{162},
\bfpage{357}.
\doiurl{https://doi.org/10.1007/BF00733434}.
\adsurl{1995SoPh..162..357B}.
\end{barticle}
\endbibitem

\bibitem[\protect\citeauthoryear{{Burlaga} et~al.}{1982}]{Burlaga_1982}
\begin{barticle}
\bauthor{\bsnm{{Burlaga}}, \binits{L.F.}},
\bauthor{\bsnm{{Klein}}, \binits{L.}},
\bauthor{\bsnm{{Sheeley}}, \binits{J.} \bsuffix{N.~R.}},
\bauthor{\bsnm{{Michels}}, \binits{D.J.}},
\bauthor{\bsnm{{Howard}}, \binits{R.A.}},
\bauthor{\bsnm{{Koomen}}, \binits{M.J.}},
\bauthor{\bsnm{{Schwenn}}, \binits{R.}},
\bauthor{\bsnm{{Rosenbauer}}, \binits{H.}}:
\byear{1982},
\batitle{{A magnetic cloud and a coronal mass ejection}}.
\bjtitle{GRL}
\bvolume{9},
\bfpage{1317}.
\doiurl{https://doi.org/10.1029/GL009i012p01317}.
\adsurl{1982GeoRL...9.1317B}.
\end{barticle}
\endbibitem

\bibitem[\protect\citeauthoryear{Burt and Smith}{2012}]{Burt2012}
\begin{bchapter}
\bauthor{\bsnm{Burt}, \binits{J.}},
\bauthor{\bsnm{Smith}, \binits{B.}}:
\byear{2012},
\bctitle{Deep Space Climate Observatory: The DSCOVR mission}.
In: \bbtitle{2012 IEEE Aerospace Conference},
\bfpage{1}.
\doiurl{https://doi.org/10.1109/AERO.2012.6187025}.
\end{bchapter}
\endbibitem

\bibitem[\protect\citeauthoryear{{Cane} and
  {Richardson}}{2003}]{2003JGRA..108.1156C}
\begin{barticle}
\bauthor{\bsnm{{Cane}}, \binits{H.V.}},
\bauthor{\bsnm{{Richardson}}, \binits{I.G.}}:
\byear{2003},
\batitle{{Interplanetary coronal mass ejections in the near-Earth solar wind
  during 1996-2002}}.
\bjtitle{Journal of Geophysical Research (Space Physics)}
\bvolume{108},
\bfpage{1156}.
\doiurl{https://doi.org/10.1029/2002JA009817}.
\adsurl{2003JGRA..108.1156C}.
\end{barticle}
\endbibitem

\bibitem[\protect\citeauthoryear{{Chertok}, {Belov}, and
  {Abunin}}{2018}]{Chertoket18}
\begin{barticle}
\bauthor{\bsnm{{Chertok}}, \binits{I.M.}},
\bauthor{\bsnm{{Belov}}, \binits{A.V.}},
\bauthor{\bsnm{{Abunin}}, \binits{A.A.}}:
\byear{2018},
\batitle{{Solar Eruptions, Forbush Decreases, and Geomagnetic Disturbances From
  Outstanding Active Region 12673}}.
\bjtitle{Space Weather}
\bvolume{16},
\bfpage{1549}.
\doiurl{https://doi.org/10.1029/2018SW001899}.
\adsurl{2018SpWea..16.1549C}.
\end{barticle}
\endbibitem

\bibitem[\protect\citeauthoryear{{Chi} et~al.}{2018a}]{Chi2018}
\begin{barticle}
\bauthor{\bsnm{{Chi}}, \binits{Y.}},
\bauthor{\bsnm{{Shen}}, \binits{C.}},
\bauthor{\bsnm{{Luo}}, \binits{B.}},
\bauthor{\bsnm{{Wang}}, \binits{Y.}},
\bauthor{\bsnm{{Xu}}, \binits{M.}}:
\byear{2018}a,
\batitle{{Geoeffectiveness of Stream Interaction Regions From 1995 to 2016}}.
\bjtitle{Space Weather}
\bvolume{16},
\bfpage{1960}.
\doiurl{https://doi.org/10.1029/2018SW001894}.
\adsurl{2018SpWea..16.1960C}.
\end{barticle}
\endbibitem

\bibitem[\protect\citeauthoryear{{Chi} et~al.}{2018b}]{Chi_2018}
\begin{barticle}
\bauthor{\bsnm{{Chi}}, \binits{Y.}},
\bauthor{\bsnm{{Zhang}}, \binits{J.}},
\bauthor{\bsnm{{Shen}}, \binits{C.}},
\bauthor{\bsnm{{Hess}}, \binits{P.}},
\bauthor{\bsnm{{Liu}}, \binits{L.}},
\bauthor{\bsnm{{Mishra}}, \binits{W.}},
\bauthor{\bsnm{{Wang}}, \binits{Y.}}:
\byear{2018}b,
\batitle{{Observational Study of an Earth-affecting Problematic ICME from
  STEREO}}.
\bjtitle{\apj}
\bvolume{863},
\bfpage{108}.
\doiurl{https://doi.org/10.3847/1538-4357/aacf44}.
\adsurl{2018ApJ...863..108C}.
\end{barticle}
\endbibitem

\bibitem[\protect\citeauthoryear{{Colaninno} and
  {Vourlidas}}{2009}]{2009ApJ...698..852C}
\begin{barticle}
\bauthor{\bsnm{{Colaninno}}, \binits{R.C.}},
\bauthor{\bsnm{{Vourlidas}}, \binits{A.}}:
\byear{2009},
\batitle{{First Determination of the True Mass of Coronal Mass Ejections: A
  Novel Approach to Using the Two STEREO Viewpoints}}.
\bjtitle{\apj}
\bvolume{698},
\bfpage{852}.
\doiurl{https://doi.org/10.1088/0004-637X/698/1/852}.
\adsurl{2009ApJ...698..852C}.
\end{barticle}
\endbibitem

\bibitem[\protect\citeauthoryear{{Colaninno}, {Vourlidas}, and
  {Wu}}{2013}]{2013JGRA..118.6866C}
\begin{barticle}
\bauthor{\bsnm{{Colaninno}}, \binits{R.C.}},
\bauthor{\bsnm{{Vourlidas}}, \binits{A.}},
\bauthor{\bsnm{{Wu}}, \binits{C.C.}}:
\byear{2013},
\batitle{{Quantitative comparison of methods for predicting the arrival of
  coronal mass ejections at Earth based on multiview imaging}}.
\bjtitle{Journal of Geophysical Research (Space Physics)}
\bvolume{118},
\bfpage{6866}.
\doiurl{https://doi.org/10.1002/2013JA019205}.
\adsurl{2013JGRA..118.6866C}.
\end{barticle}
\endbibitem

\bibitem[\protect\citeauthoryear{{Cranmer}}{2002}]{2002SSRv..101..229C}
\begin{barticle}
\bauthor{\bsnm{{Cranmer}}, \binits{S.R.}}:
\byear{2002},
\batitle{{Coronal Holes and the High-Speed Solar Wind}}.
\bjtitle{\ssr}
\bvolume{101},
\bfpage{229}.
\doiurl{https://doi.org/10.1023/A:1020840004535}.
\adsurl{2002SSRv..101..229C}.
\end{barticle}
\endbibitem

\bibitem[\protect\citeauthoryear{{Cranmer}}{2009}]{Cranmer09}
\begin{barticle}
\bauthor{\bsnm{{Cranmer}}, \binits{S.R.}}:
\byear{2009},
\batitle{{Coronal Holes}}.
\bjtitle{Living Reviews in Solar Physics}
\bvolume{6},
\bfpage{3}.
\doiurl{https://doi.org/10.12942/lrsp-2009-3}.
\adsurl{2009LRSP....6....3C}.
\end{barticle}
\endbibitem

\bibitem[\protect\citeauthoryear{{Domingo}, {Fleck}, and
  {Poland}}{1995}]{1995SoPh..162....1D}
\begin{barticle}
\bauthor{\bsnm{{Domingo}}, \binits{V.}},
\bauthor{\bsnm{{Fleck}}, \binits{B.}},
\bauthor{\bsnm{{Poland}}, \binits{A.I.}}:
\byear{1995},
\batitle{{The SOHO Mission: an Overview}}.
\bjtitle{\solphys}
\bvolume{162},
\bfpage{1}.
\doiurl{https://doi.org/10.1007/BF00733425}.
\adsurl{1995SoPh..162....1D}.
\end{barticle}
\endbibitem

\bibitem[\protect\citeauthoryear{{Echer} et~al.}{2008}]{Echeret08}
\begin{barticle}
\bauthor{\bsnm{{Echer}}, \binits{E.}},
\bauthor{\bsnm{{Gonzalez}}, \binits{W.D.}},
\bauthor{\bsnm{{Tsurutani}}, \binits{B.T.}},
\bauthor{\bsnm{{Gonzalez}}, \binits{A.L.C.}}:
\byear{2008},
\batitle{{Interplanetary conditions causing intense geomagnetic storms (Dst <=
  -100 nT) during solar cycle 23 (1996-2006)}}.
\bjtitle{Journal of Geophysical Research (Space Physics)}
\bvolume{113},
\bfpage{A05221}.
\doiurl{https://doi.org/10.1029/2007JA012744}.
\adsurl{2008JGRA..113.5221E}.
\end{barticle}
\endbibitem

\bibitem[\protect\citeauthoryear{{Georgieva}, {Kirov}, and
  {Gavruseva}}{2006}]{2006PCE....31...81G}
\begin{barticle}
\bauthor{\bsnm{{Georgieva}}, \binits{K.}},
\bauthor{\bsnm{{Kirov}}, \binits{B.}},
\bauthor{\bsnm{{Gavruseva}}, \binits{E.}}:
\byear{2006},
\batitle{{Geoeffectiveness of different solar drivers, and long-term variations
  of the correlation between sunspot and geomagnetic activity}}.
\bjtitle{Physics and Chemistry of the Earth}
\bvolume{31},
\bfpage{81}.
\doiurl{https://doi.org/10.1016/j.pce.2005.03.003}.
\adsurl{2006PCE....31...81G}.
\end{barticle}
\endbibitem

\bibitem[\protect\citeauthoryear{{Gloeckler} et~al.}{1998}]{Gloeckler98}
\begin{barticle}
\bauthor{\bsnm{{Gloeckler}}, \binits{G.}},
\bauthor{\bsnm{{Cain}}, \binits{J.}},
\bauthor{\bsnm{{Ipavich}}, \binits{F.M.}},
\bauthor{\bsnm{{Tums}}, \binits{E.O.}},
\bauthor{\bsnm{{Bedini}}, \binits{P.}},
\bauthor{\bsnm{{Fisk}}, \binits{L.A.}},
\bauthor{\bsnm{{Zurbuchen}}, \binits{T.H.}},
\bauthor{\bsnm{{Bochsler}}, \binits{P.}},
\bauthor{\bsnm{{Fischer}}, \binits{J.}},
\bauthor{\bsnm{{Wimmer-Schweingruber}}, \binits{R.F.}},
\bauthor{\bsnm{{Geiss}}, \binits{J.}},
\bauthor{\bsnm{{Kallenbach}}, \binits{R.}}:
\byear{1998},
\batitle{{Investigation of the composition of solar and interstellar matter
  using solar wind and pickup ion measurements with SWICS and SWIMS on the ACE
  spacecraft}}.
\bjtitle{\ssr}
\bvolume{86},
\bfpage{497}.
\doiurl{https://doi.org/10.1023/A:1005036131689}.
\adsurl{1998SSRv...86..497G}.
\end{barticle}
\endbibitem

\bibitem[\protect\citeauthoryear{{Gold} et~al.}{1998}]{Gold98}
\begin{barticle}
\bauthor{\bsnm{{Gold}}, \binits{R.E.}},
\bauthor{\bsnm{{Krimigis}}, \binits{S.M.}},
\bauthor{\bsnm{{Hawkins}}, \binits{I.} \bsuffix{S.~E.}},
\bauthor{\bsnm{{Haggerty}}, \binits{D.K.}},
\bauthor{\bsnm{{Lohr}}, \binits{D.A.}},
\bauthor{\bsnm{{Fiore}}, \binits{E.}},
\bauthor{\bsnm{{Armstrong}}, \binits{T.P.}},
\bauthor{\bsnm{{Holland}}, \binits{G.}},
\bauthor{\bsnm{{Lanzerotti}}, \binits{L.J.}}:
\byear{1998},
\batitle{{Electron, Proton, and Alpha Monitor on the Advanced Composition
  Explorer spacecraft}}.
\bjtitle{\ssr}
\bvolume{86},
\bfpage{541}.
\doiurl{https://doi.org/10.1023/A:1005088115759}.
\adsurl{1998SSRv...86..541G}.
\end{barticle}
\endbibitem

\bibitem[\protect\citeauthoryear{{Gonzalez} et~al.}{1994}]{1994JGR....99.5771G}
\begin{barticle}
\bauthor{\bsnm{{Gonzalez}}, \binits{W.D.}},
\bauthor{\bsnm{{Joselyn}}, \binits{J.A.}},
\bauthor{\bsnm{{Kamide}}, \binits{Y.}},
\bauthor{\bsnm{{Kroehl}}, \binits{H.W.}},
\bauthor{\bsnm{{Rostoker}}, \binits{G.}},
\bauthor{\bsnm{{Tsurutani}}, \binits{B.T.}},
\bauthor{\bsnm{{Vasyliunas}}, \binits{V.M.}}:
\byear{1994},
\batitle{{What is a geomagnetic storm?}}
\bjtitle{\jgr}
\bvolume{99},
\bfpage{5771}.
\doiurl{https://doi.org/10.1029/93JA02867}.
\adsurl{1994JGR....99.5771G}.
\end{barticle}
\endbibitem

\bibitem[\protect\citeauthoryear{{Gonzalez} et~al.}{1996}]{Gonzalezet96}
\begin{barticle}
\bauthor{\bsnm{{Gonzalez}}, \binits{W.D.}},
\bauthor{\bsnm{{Tsurutani}}, \binits{B.T.}},
\bauthor{\bsnm{{McIntosh}}, \binits{P.S.}},
\bauthor{\bsnm{{Cl{\'u}a de Gonzalez}}, \binits{A.L.}}:
\byear{1996},
\batitle{{Coronal hole-active region-Current sheet (CHARCS) Association with
  intense interplanetary and geomagnetic activity}}.
\bjtitle{\grl}
\bvolume{23},
\bfpage{2577}.
\doiurl{https://doi.org/10.1029/96GL02393}.
\adsurl{1996GeoRL..23.2577G}.
\end{barticle}
\endbibitem

\bibitem[\protect\citeauthoryear{{Gopalswamy}
  et~al.}{2009a}]{2009JGRA..114.0A22G}
\begin{barticle}
\bauthor{\bsnm{{Gopalswamy}}, \binits{N.}},
\bauthor{\bsnm{{M{\"a}kel{\"a}}}, \binits{P.}},
\bauthor{\bsnm{{Xie}}, \binits{H.}},
\bauthor{\bsnm{{Akiyama}}, \binits{S.}},
\bauthor{\bsnm{{Yashiro}}, \binits{S.}}:
\byear{2009}a,
\batitle{{CME interactions with coronal holes and their interplanetary
  consequences}}.
\bjtitle{Journal of Geophysical Research (Space Physics)}
\bvolume{114},
\bfpage{A00A22}.
\doiurl{https://doi.org/10.1029/2008JA013686}.
\adsurl{2009JGRA..114.0A22G}.
\end{barticle}
\endbibitem

\bibitem[\protect\citeauthoryear{{Gopalswamy}
  et~al.}{2009b}]{2009EM&P..104..295G}
\begin{barticle}
\bauthor{\bsnm{{Gopalswamy}}, \binits{N.}},
\bauthor{\bsnm{{Yashiro}}, \binits{S.}},
\bauthor{\bsnm{{Michalek}}, \binits{G.}},
\bauthor{\bsnm{{Stenborg}}, \binits{G.}},
\bauthor{\bsnm{{Vourlidas}}, \binits{A.}},
\bauthor{\bsnm{{Freeland}}, \binits{S.}},
\bauthor{\bsnm{{Howard}}, \binits{R.}}:
\byear{2009}b,
\batitle{{The SOHO/LASCO CME Catalog}}.
\bjtitle{Earth Moon and Planets}
\bvolume{104},
\bfpage{295}.
\doiurl{https://doi.org/10.1007/s11038-008-9282-7}.
\adsurl{2009EM&P..104..295G}.
\end{barticle}
\endbibitem

\bibitem[\protect\citeauthoryear{{Gopalswamy}
  et~al.}{2010}]{2010SunGe...5....7G}
\begin{barticle}
\bauthor{\bsnm{{Gopalswamy}}, \binits{N.}},
\bauthor{\bsnm{{Yashiro}}, \binits{S.}},
\bauthor{\bsnm{{Michalek}}, \binits{G.}},
\bauthor{\bsnm{{Xie}}, \binits{H.}},
\bauthor{\bsnm{{M{\"a}kel{\"a}}}, \binits{P.}},
\bauthor{\bsnm{{Vourlidas}}, \binits{A.}},
\bauthor{\bsnm{{Howard}}, \binits{R.A.}}:
\byear{2010},
\batitle{{A Catalog of Halo Coronal Mass Ejections from SOHO}}.
\bjtitle{Sun and Geosphere}
\bvolume{5},
\bfpage{7}.
\adsurl{2010SunGe...5....7G}.
\end{barticle}
\endbibitem

\bibitem[\protect\citeauthoryear{{Habbal}, {Scholl}, and
  {McIntosh}}{2008}]{Habbalet08}
\begin{barticle}
\bauthor{\bsnm{{Habbal}}, \binits{S.R.}},
\bauthor{\bsnm{{Scholl}}, \binits{I.F.}},
\bauthor{\bsnm{{McIntosh}}, \binits{S.W.}}:
\byear{2008},
\batitle{{Impact of Active Regions on Coronal Hole Outflows}}.
\bjtitle{\apjl}
\bvolume{683},
\bfpage{L75}.
\doiurl{https://doi.org/10.1086/591315}.
\adsurl{2008ApJ...683L..75H}.
\end{barticle}
\endbibitem

\bibitem[\protect\citeauthoryear{{Heinemann} et~al.}{2019}]{Heinemannetal19}
\begin{barticle}
\bauthor{\bsnm{{Heinemann}}, \binits{S.G.}},
\bauthor{\bsnm{{Temmer}}, \binits{M.}},
\bauthor{\bsnm{{Farrugia}}, \binits{C.J.}},
\bauthor{\bsnm{{Dissauer}}, \binits{K.}},
\bauthor{\bsnm{{Kay}}, \binits{C.}},
\bauthor{\bsnm{{Wiegelmann}}, \binits{T.}},
\bauthor{\bsnm{{Dumbovi{\'c}}}, \binits{M.}},
\bauthor{\bsnm{{Veronig}}, \binits{A.M.}},
\bauthor{\bsnm{{Podladchikova}}, \binits{T.}},
\bauthor{\bsnm{{Hofmeister}}, \binits{S.J.}},
\bauthor{\bsnm{{Lugaz}}, \binits{N.}},
\bauthor{\bsnm{{Carcaboso}}, \binits{F.}}:
\byear{2019},
\batitle{{CME-HSS Interaction and Characteristics Tracked from Sun to Earth}}.
\bjtitle{\solphys}
\bvolume{294},
\bfpage{121}.
\doiurl{https://doi.org/10.1007/s11207-019-1515-6}.
\adsurl{2019SoPh..294..121H}.
\end{barticle}
\endbibitem

\bibitem[\protect\citeauthoryear{{Hinrichs} et~al.}{2021}]{2021JSWSC..11...11H}
\begin{barticle}
\bauthor{\bsnm{{Hinrichs}}, \binits{J.}},
\bauthor{\bsnm{{Davies}}, \binits{J.A.}},
\bauthor{\bsnm{{West}}, \binits{M.J.}},
\bauthor{\bsnm{{Bothmer}}, \binits{V.}},
\bauthor{\bsnm{{Bourgoignie}}, \binits{B.}},
\bauthor{\bsnm{{Eyles}}, \binits{C.J.}},
\bauthor{\bsnm{{Huke}}, \binits{P.}},
\bauthor{\bsnm{{Jiggens}}, \binits{P.}},
\bauthor{\bsnm{{Nicula}}, \binits{B.}},
\bauthor{\bsnm{{Tappin}}, \binits{J.}}:
\byear{2021},
\batitle{{Analysis of signal to noise ratio in coronagraph observations of
  coronal mass ejections}}.
\bjtitle{Journal of Space Weather and Space Climate}
\bvolume{11},
\bfpage{11}.
\doiurl{https://doi.org/10.1051/swsc/2020070}.
\adsurl{2021JSWSC..11...11H}.
\end{barticle}
\endbibitem

\bibitem[\protect\citeauthoryear{{Hood} and {Priest}}{1981}]{hood81}
\begin{barticle}
\bauthor{\bsnm{{Hood}}, \binits{A.W.}},
\bauthor{\bsnm{{Priest}}, \binits{E.R.}}:
\byear{1981},
\batitle{{Critical conditions for magnetic instabilities in force-free coronal
  loops}}.
\bjtitle{Geophysical and Astrophysical Fluid Dynamics}
\bvolume{17},
\bfpage{297}.
\doiurl{https://doi.org/10.1080/03091928108243687}.
\adsurl{1981GApFD..17..297H}.
\end{barticle}
\endbibitem

\bibitem[\protect\citeauthoryear{{Howard} et~al.}{1982}]{1982ApJ...263L.101H}
\begin{barticle}
\bauthor{\bsnm{{Howard}}, \binits{R.A.}},
\bauthor{\bsnm{{Michels}}, \binits{D.J.}},
\bauthor{\bsnm{{Sheeley}}, \binits{J.} \bsuffix{N.~R.}},
\bauthor{\bsnm{{Koomen}}, \binits{M.J.}}:
\byear{1982},
\batitle{{The observation of a coronal transient directed at Earth.}}
\bjtitle{\apjl}
\bvolume{263},
\bfpage{L101}.
\doiurl{https://doi.org/10.1086/183932}.
\adsurl{1982ApJ...263L.101H}.
\end{barticle}
\endbibitem

\bibitem[\protect\citeauthoryear{{Howard} et~al.}{2008}]{2008SSRv..136...67H}
\begin{barticle}
\bauthor{\bsnm{{Howard}}, \binits{R.A.}},
\bauthor{\bsnm{{Moses}}, \binits{J.D.}},
\bauthor{\bsnm{{Vourlidas}}, \binits{A.}},
\bauthor{\bsnm{{Newmark}}, \binits{J.S.}},
\bauthor{\bsnm{{Socker}}, \binits{D.G.}},
\bauthor{\bsnm{{Plunkett}}, \binits{S.P.}},
\bauthor{\bsnm{{Korendyke}}, \binits{C.M.}},
\bauthor{\bsnm{{Cook}}, \binits{J.W.}},
\bauthor{\bsnm{{Hurley}}, \binits{A.}},
\bauthor{\bsnm{{Davila}}, \binits{J.M.}},
\bauthor{\bsnm{{Thompson}}, \binits{W.T.}},
\bauthor{\bsnm{{St Cyr}}, \binits{O.C.}},
\bauthor{\bsnm{{Mentzell}}, \binits{E.}},
\bauthor{\bsnm{{Mehalick}}, \binits{K.}},
\bauthor{\bsnm{{Lemen}}, \binits{J.R.}},
\bauthor{\bsnm{{Wuelser}}, \binits{J.P.}},
\bauthor{\bsnm{{Duncan}}, \binits{D.W.}},
\bauthor{\bsnm{{Tarbell}}, \binits{T.D.}},
\bauthor{\bsnm{{Wolfson}}, \binits{C.J.}},
\bauthor{\bsnm{{Moore}}, \binits{A.}},
\bauthor{\bsnm{{Harrison}}, \binits{R.A.}},
\bauthor{\bsnm{{Waltham}}, \binits{N.R.}},
\bauthor{\bsnm{{Lang}}, \binits{J.}},
\bauthor{\bsnm{{Davis}}, \binits{C.J.}},
\bauthor{\bsnm{{Eyles}}, \binits{C.J.}},
\bauthor{\bsnm{{Mapson-Menard}}, \binits{H.}},
\bauthor{\bsnm{{Simnett}}, \binits{G.M.}},
\bauthor{\bsnm{{Halain}}, \binits{J.P.}},
\bauthor{\bsnm{{Defise}}, \binits{J.M.}},
\bauthor{\bsnm{{Mazy}}, \binits{E.}},
\bauthor{\bsnm{{Rochus}}, \binits{P.}},
\bauthor{\bsnm{{Mercier}}, \binits{R.}},
\bauthor{\bsnm{{Ravet}}, \binits{M.F.}},
\bauthor{\bsnm{{Delmotte}}, \binits{F.}},
\bauthor{\bsnm{{Auchere}}, \binits{F.}},
\bauthor{\bsnm{{Delaboudiniere}}, \binits{J.P.}},
\bauthor{\bsnm{{Bothmer}}, \binits{V.}},
\bauthor{\bsnm{{Deutsch}}, \binits{W.}},
\bauthor{\bsnm{{Wang}}, \binits{D.}},
\bauthor{\bsnm{{Rich}}, \binits{N.}},
\bauthor{\bsnm{{Cooper}}, \binits{S.}},
\bauthor{\bsnm{{Stephens}}, \binits{V.}},
\bauthor{\bsnm{{Maahs}}, \binits{G.}},
\bauthor{\bsnm{{Baugh}}, \binits{R.}},
\bauthor{\bsnm{{McMullin}}, \binits{D.}},
\bauthor{\bsnm{{Carter}}, \binits{T.}}:
\byear{2008},
\batitle{{Sun Earth Connection Coronal and Heliospheric Investigation
  (SECCHI)}}.
\bjtitle{SSR}
\bvolume{136},
\bfpage{67}.
\doiurl{https://doi.org/10.1007/s11214-008-9341-4}.
\adsurl{2008SSRv..136...67H}.
\end{barticle}
\endbibitem

\bibitem[\protect\citeauthoryear{{Kaiser} et~al.}{2008}]{kaiseret08}
\begin{barticle}
\bauthor{\bsnm{{Kaiser}}, \binits{M.L.}},
\bauthor{\bsnm{{Kucera}}, \binits{T.A.}},
\bauthor{\bsnm{{Davila}}, \binits{J.M.}},
\bauthor{\bsnm{{St.~Cyr}}, \binits{O.C.}},
\bauthor{\bsnm{{Guhathakurta}}, \binits{M.}},
\bauthor{\bsnm{{Christian}}, \binits{E.}}:
\byear{2008},
\batitle{{The STEREO Mission: An Introduction}}.
\bjtitle{\ssr}
\bvolume{136},
\bfpage{5}.
\doiurl{https://doi.org/10.1007/s11214-007-9277-0}.
\adsurl{2008SSRv..136....5K}.
\end{barticle}
\endbibitem

\bibitem[\protect\citeauthoryear{{Karna} et~al.}{2020}]{Karnaetal2020}
\begin{barticle}
\bauthor{\bsnm{{Karna}}, \binits{M.L.}},
\bauthor{\bsnm{{Karna}}, \binits{N.}},
\bauthor{\bsnm{{Saar}}, \binits{S.H.}},
\bauthor{\bsnm{{Pesnell}}, \binits{W.D.}},
\bauthor{\bsnm{{DeLuca}}, \binits{E.E.}}:
\byear{2020},
\batitle{{A Study of Equatorial Coronal Holes during the Maximum Phase of Four
  Solar Cycles}}.
\bjtitle{\apj}
\bvolume{901},
\bfpage{124}.
\doiurl{https://doi.org/10.3847/1538-4357/abafae}.
\adsurl{2020ApJ...901..124K}.
\end{barticle}
\endbibitem

\bibitem[\protect\citeauthoryear{Karna et~al.}{2019}]{karnaet2019}
\begin{barticle}
\bauthor{\bsnm{Karna}, \binits{N.}},
\bauthor{\bsnm{Savcheva}, \binits{A.}},
\bauthor{\bsnm{Dalmasse}, \binits{K.}},
\bauthor{\bsnm{Gibson}, \binits{S.}},
\bauthor{\bsnm{Tassev}, \binits{S.}},
\bauthor{\bparticle{de} \bsnm{Toma}, \binits{G.}},
\bauthor{\bsnm{DeLuca}, \binits{E.E.}}:
\byear{2019},
\batitle{Forward Modeling of a Pseudostreamer}.
\bjtitle{The Astrophysical Journal}
\bvolume{883},
\bfpage{74}.
\doiurl{https://doi.org/10.3847/1538-4357/ab3c50}.
\burl{https://doi.org/10.3847\%2F1538-4357\%2Fab3c50}.
\end{barticle}
\endbibitem

\bibitem[\protect\citeauthoryear{{Karna} et~al.}{2021}]{karnaet2021}
\begin{barticle}
\bauthor{\bsnm{{Karna}}, \binits{N.}},
\bauthor{\bsnm{{Savcheva}}, \binits{A.}},
\bauthor{\bsnm{{Gibson}}, \binits{S.}},
\bauthor{\bsnm{{Tassev}}, \binits{S.}},
\bauthor{\bsnm{{Reeves}}, \binits{K.K.}},
\bauthor{\bsnm{{DeLuca}}, \binits{E.E.}},
\bauthor{\bsnm{{Dalmasse}}, \binits{K.}}:
\byear{2021},
\batitle{{Magnetofrictional Modeling of an Erupting Pseudostreamer}}.
\bjtitle{\apj}
\bvolume{913},
\bfpage{47}.
\doiurl{https://doi.org/10.3847/1538-4357/abf2b8}.
\adsurl{2021ApJ...913...47K}.
\end{barticle}
\endbibitem

\bibitem[\protect\citeauthoryear{{Karna} et~al.}{2024}]{karnaetal24}
\begin{barticle}
\bauthor{\bsnm{{Karna}}, \binits{N.}},
\bauthor{\bsnm{{Dhakal}}, \binits{S.}},
\bauthor{\bsnm{{Savcheva}}, \binits{A.}},
\bauthor{\bsnm{{Zhang}}, \binits{J.}},
\bauthor{\bsnm{{Kliem}}, \binits{B.}}:
\byear{2024},
\batitle{{A Double-decker Flux Rope Model for the Solar Eruption on 2012 March
  10}}.
\bjtitle{\apj}
\bvolume{961},
\bfpage{11}.
\doiurl{https://doi.org/10.3847/1538-4357/ad1187}.
\adsurl{2024ApJ...961...11K}.
\end{barticle}
\endbibitem

\bibitem[\protect\citeauthoryear{{Kay}, {Mays}, and
  {Collado-Vega}}{2022}]{Kayetal22}
\begin{barticle}
\bauthor{\bsnm{{Kay}}, \binits{C.}},
\bauthor{\bsnm{{Mays}}, \binits{M.L.}},
\bauthor{\bsnm{{Collado-Vega}}, \binits{Y.M.}}:
\byear{2022},
\batitle{{OSPREI: A Coupled Approach to Modeling CME-Driven Space Weather With
  Automatically Generated, User-Friendly Outputs}}.
\bjtitle{Space Weather}
\bvolume{20},
\bfpage{e2021SW002914}.
\doiurl{https://doi.org/10.1029/2021SW002914}.
\adsurl{2022SpWea..2002914K}.
\end{barticle}
\endbibitem

\bibitem[\protect\citeauthoryear{{Kilpua} et~al.}{2017}]{Kilpuaetal17}
\begin{barticle}
\bauthor{\bsnm{{Kilpua}}, \binits{E.K.J.}},
\bauthor{\bsnm{{Balogh}}, \binits{A.}},
\bauthor{\bsnm{{von Steiger}}, \binits{R.}},
\bauthor{\bsnm{{Liu}}, \binits{Y.D.}}:
\byear{2017},
\batitle{{Geoeffective Properties of Solar Transients and Stream Interaction
  Regions}}.
\bjtitle{\ssr}
\bvolume{212},
\bfpage{1271}.
\doiurl{https://doi.org/10.1007/s11214-017-0411-3}.
\adsurl{2017SSRv..212.1271K}.
\end{barticle}
\endbibitem

\bibitem[\protect\citeauthoryear{{King} and
  {Papitashvili}}{2005}]{2005JGRA..110.2104K}
\begin{barticle}
\bauthor{\bsnm{{King}}, \binits{J.H.}},
\bauthor{\bsnm{{Papitashvili}}, \binits{N.E.}}:
\byear{2005},
\batitle{{Solar wind spatial scales in and comparisons of hourly Wind and ACE
  plasma and magnetic field data}}.
\bjtitle{Journal of Geophysical Research (Space Physics)}
\bvolume{110},
\bfpage{A02104}.
\doiurl{https://doi.org/10.1029/2004JA010649}.
\adsurl{2005JGRA..110.2104K}.
\end{barticle}
\endbibitem

\bibitem[\protect\citeauthoryear{{Lakhina} and
  {Tsurutani}}{2016}]{2016GSL.....3....5L}
\begin{barticle}
\bauthor{\bsnm{{Lakhina}}, \binits{G.S.}},
\bauthor{\bsnm{{Tsurutani}}, \binits{B.T.}}:
\byear{2016},
\batitle{{Geomagnetic storms: historical perspective to modern view}}.
\bjtitle{Geoscience Letters}
\bvolume{3},
\bfpage{5}.
\doiurl{https://doi.org/10.1186/s40562-016-0037-4}.
\adsurl{2016GSL.....3....5L}.
\end{barticle}
\endbibitem

\bibitem[\protect\citeauthoryear{{Landau} and
  {Lifshitz}}{1987}]{1987flme.book.....L}
\begin{bbook}
\bauthor{\bsnm{{Landau}}, \binits{L.D.}},
\bauthor{\bsnm{{Lifshitz}}, \binits{E.M.}}:
\byear{1987},
\bbtitle{{Fluid Mechanics}},
\bpublisher{Butterworth-Heinemann, Elsevier}.
\adsurl{1987flme.book.....L}.
\end{bbook}
\endbibitem

\bibitem[\protect\citeauthoryear{{Lara}, {Gonzalez-Esparza}, and
  {Gopalswamy}}{2004}]{Lara2004}
\begin{botherref}
\oauthor{\bsnm{{Lara}}, \binits{A.}},
\oauthor{\bsnm{{Gonzalez-Esparza}}, \binits{J.A.}},
\oauthor{\bsnm{{Gopalswamy}}, \binits{N.}}:
2004,
{Characteristics of coronal mass ejections in the near Sun interplanetary
  space}.
\textit{Geofis. Int.}
\textbf{43}.
\end{botherref}
\endbibitem

\bibitem[\protect\citeauthoryear{{Lara} et~al.}{2020}]{2020A&A...635A.112L}
\begin{barticle}
\bauthor{\bsnm{{Lara}}, \binits{A.}},
\bauthor{\bsnm{{Gopalswamy}}, \binits{N.}},
\bauthor{\bsnm{{Niembro}}, \binits{T.}},
\bauthor{\bsnm{{P{\'e}rez-Enr{\'\i}quez}}, \binits{R.}},
\bauthor{\bsnm{{Yashiro}}, \binits{S.}}:
\byear{2020},
\batitle{{Space, time and velocity association of successive coronal mass
  ejections}}.
\bjtitle{\aap}
\bvolume{635},
\bfpage{A112}.
\doiurl{https://doi.org/10.1051/0004-6361/201936016}.
\adsurl{2020A&A...635A.112L}.
\end{barticle}
\endbibitem

\bibitem[\protect\citeauthoryear{{Lemen} et~al.}{2012}]{lemenet12}
\begin{barticle}
\bauthor{\bsnm{{Lemen}}, \binits{J.R.}},
\bauthor{\bsnm{{Title}}, \binits{A.M.}},
\bauthor{\bsnm{{Akin}}, \binits{D.J.}},
\bauthor{\bsnm{{Boerner}}, \binits{P.F.}},
\bauthor{\bsnm{{Chou}}, \binits{C.}},
\bauthor{\bsnm{{Drake}}, \binits{J.F.}},
\bauthor{\bsnm{{Duncan}}, \binits{D.W.}},
\bauthor{\bsnm{{Edwards}}, \binits{C.G.}},
\bauthor{\bsnm{{Friedlaender}}, \binits{F.M.}},
\bauthor{\bsnm{{Heyman}}, \binits{G.F.}},
\bauthor{\bsnm{{Hurlburt}}, \binits{N.E.}},
\bauthor{\bsnm{{Katz}}, \binits{N.L.}},
\bauthor{\bsnm{{Kushner}}, \binits{G.D.}},
\bauthor{\bsnm{{Levay}}, \binits{M.}},
\bauthor{\bsnm{{Lindgren}}, \binits{R.W.}},
\bauthor{\bsnm{{Mathur}}, \binits{D.P.}},
\bauthor{\bsnm{{McFeaters}}, \binits{E.L.}},
\bauthor{\bsnm{{Mitchell}}, \binits{S.}},
\bauthor{\bsnm{{Rehse}}, \binits{R.A.}},
\bauthor{\bsnm{{Schrijver}}, \binits{C.J.}},
\bauthor{\bsnm{{Springer}}, \binits{L.A.}},
\bauthor{\bsnm{{Stern}}, \binits{R.A.}},
\bauthor{\bsnm{{Tarbell}}, \binits{T.D.}},
\bauthor{\bsnm{{Wuelser}}, \binits{J.-P.}},
\bauthor{\bsnm{{Wolfson}}, \binits{C.J.}},
\bauthor{\bsnm{{Yanari}}, \binits{C.}},
\bauthor{\bsnm{{Bookbinder}}, \binits{J.A.}},
\bauthor{\bsnm{{Cheimets}}, \binits{P.N.}},
\bauthor{\bsnm{{Caldwell}}, \binits{D.}},
\bauthor{\bsnm{{Deluca}}, \binits{E.E.}},
\bauthor{\bsnm{{Gates}}, \binits{R.}},
\bauthor{\bsnm{{Golub}}, \binits{L.}},
\bauthor{\bsnm{{Park}}, \binits{S.}},
\bauthor{\bsnm{{Podgorski}}, \binits{W.A.}},
\bauthor{\bsnm{{Bush}}, \binits{R.I.}},
\bauthor{\bsnm{{Scherrer}}, \binits{P.H.}},
\bauthor{\bsnm{{Gummin}}, \binits{M.A.}},
\bauthor{\bsnm{{Smith}}, \binits{P.}},
\bauthor{\bsnm{{Auker}}, \binits{G.}},
\bauthor{\bsnm{{Jerram}}, \binits{P.}},
\bauthor{\bsnm{{Pool}}, \binits{P.}},
\bauthor{\bsnm{{Soufli}}, \binits{R.}},
\bauthor{\bsnm{{Windt}}, \binits{D.L.}},
\bauthor{\bsnm{{Beardsley}}, \binits{S.}},
\bauthor{\bsnm{{Clapp}}, \binits{M.}},
\bauthor{\bsnm{{Lang}}, \binits{J.}},
\bauthor{\bsnm{{Waltham}}, \binits{N.}}:
\byear{2012},
\batitle{{The Atmospheric Imaging Assembly (AIA) on the Solar Dynamics
  Observatory (SDO)}}.
\bjtitle{\solphys}
\bvolume{275},
\bfpage{17}.
\doiurl{https://doi.org/10.1007/s11207-011-9776-8}.
\adsurl{2012SoPh..275...17L}.
\end{barticle}
\endbibitem

\bibitem[\protect\citeauthoryear{{Liu} et~al.}{2015}]{Liuetal15}
\begin{barticle}
\bauthor{\bsnm{{Liu}}, \binits{Y.D.}},
\bauthor{\bsnm{{Hu}}, \binits{H.}},
\bauthor{\bsnm{{Wang}}, \binits{R.}},
\bauthor{\bsnm{{Yang}}, \binits{Z.}},
\bauthor{\bsnm{{Zhu}}, \binits{B.}},
\bauthor{\bsnm{{Liu}}, \binits{Y.A.}},
\bauthor{\bsnm{{Luhmann}}, \binits{J.G.}},
\bauthor{\bsnm{{Richardson}}, \binits{J.D.}}:
\byear{2015},
\batitle{{Plasma and Magnetic Field Characteristics of Solar Coronal Mass
  Ejections in Relation to Geomagnetic Storm Intensity and Variability}}.
\bjtitle{\apjl}
\bvolume{809},
\bfpage{L34}.
\doiurl{https://doi.org/10.1088/2041-8205/809/2/L34}.
\adsurl{2015ApJ...809L..34L}.
\end{barticle}
\endbibitem

\bibitem[\protect\citeauthoryear{{Lugaz} et~al.}{2011}]{Lugazetal11}
\begin{barticle}
\bauthor{\bsnm{{Lugaz}}, \binits{N.}},
\bauthor{\bsnm{{Downs}}, \binits{C.}},
\bauthor{\bsnm{{Shibata}}, \binits{K.}},
\bauthor{\bsnm{{Roussev}}, \binits{I.I.}},
\bauthor{\bsnm{{Asai}}, \binits{A.}},
\bauthor{\bsnm{{Gombosi}}, \binits{T.I.}}:
\byear{2011},
\batitle{{Numerical Investigation of a Coronal Mass Ejection from an Anemone
  Active Region: Reconnection and Deflection of the 2005 August 22 Eruption}}.
\bjtitle{\apj}
\bvolume{738},
\bfpage{127}.
\doiurl{https://doi.org/10.1088/0004-637X/738/2/127}.
\adsurl{2011ApJ...738..127L}.
\end{barticle}
\endbibitem

\bibitem[\protect\citeauthoryear{{Mays} et~al.}{2015}]{2015ApJ...812..145M}
\begin{barticle}
\bauthor{\bsnm{{Mays}}, \binits{M.L.}},
\bauthor{\bsnm{{Thompson}}, \binits{B.J.}},
\bauthor{\bsnm{{Jian}}, \binits{L.K.}},
\bauthor{\bsnm{{Colaninno}}, \binits{R.C.}},
\bauthor{\bsnm{{Odstrcil}}, \binits{D.}},
\bauthor{\bsnm{{M{\"o}stl}}, \binits{C.}},
\bauthor{\bsnm{{Temmer}}, \binits{M.}},
\bauthor{\bsnm{{Savani}}, \binits{N.P.}},
\bauthor{\bsnm{{Collinson}}, \binits{G.}},
\bauthor{\bsnm{{Taktakishvili}}, \binits{A.}},
\bauthor{\bsnm{{MacNeice}}, \binits{P.J.}},
\bauthor{\bsnm{{Zheng}}, \binits{Y.}}:
\byear{2015},
\batitle{{Propagation of the 7 January 2014 CME and Resulting Geomagnetic
  Non-event}}.
\bjtitle{\apj}
\bvolume{812},
\bfpage{145}.
\doiurl{https://doi.org/10.1088/0004-637X/812/2/145}.
\adsurl{2015ApJ...812..145M}.
\end{barticle}
\endbibitem

\bibitem[\protect\citeauthoryear{{Niembro} et~al.}{2019}]{Niembro2019}
\begin{barticle}
\bauthor{\bsnm{{Niembro}}, \binits{T.}},
\bauthor{\bsnm{{Lara}}, \binits{A.}},
\bauthor{\bsnm{{Gonz{\'a}lez}}, \binits{R.F.}},
\bauthor{\bsnm{{Cant{\'o}}}, \binits{J.}}:
\byear{2019},
\batitle{{Numerical simulations of ICME-ICME interactions}}.
\bjtitle{Journal of Space Weather and Space Climate}
\bvolume{9},
\bfpage{A4}.
\doiurl{https://doi.org/10.1051/swsc/2018049}.
\adsurl{2019JSWSC...9A...4N}.
\end{barticle}
\endbibitem

\bibitem[\protect\citeauthoryear{{Nieves-Chinchilla}
  et~al.}{2012}]{2012JGRA..117.6106N}
\begin{barticle}
\bauthor{\bsnm{{Nieves-Chinchilla}}, \binits{T.}},
\bauthor{\bsnm{{Colaninno}}, \binits{R.}},
\bauthor{\bsnm{{Vourlidas}}, \binits{A.}},
\bauthor{\bsnm{{Szabo}}, \binits{A.}},
\bauthor{\bsnm{{Lepping}}, \binits{R.P.}},
\bauthor{\bsnm{{Boardsen}}, \binits{S.A.}},
\bauthor{\bsnm{{Anderson}}, \binits{B.J.}},
\bauthor{\bsnm{{Korth}}, \binits{H.}}:
\byear{2012},
\batitle{{Remote and in situ observations of an unusual Earth-directed coronal
  mass ejection from multiple viewpoints}}.
\bjtitle{Journal of Geophysical Research (Space Physics)}
\bvolume{117},
\bfpage{A06106}.
\doiurl{https://doi.org/10.1029/2011JA017243}.
\adsurl{2012JGRA..117.6106N}.
\end{barticle}
\endbibitem

\bibitem[\protect\citeauthoryear{{Nieves-Chinchilla} et~al.}{2016}]{Nieves2016}
\begin{barticle}
\bauthor{\bsnm{{Nieves-Chinchilla}}, \binits{T.}},
\bauthor{\bsnm{{Linton}}, \binits{M.G.}},
\bauthor{\bsnm{{Hidalgo}}, \binits{M.A.}},
\bauthor{\bsnm{{Vourlidas}}, \binits{A.}},
\bauthor{\bsnm{{Savani}}, \binits{N.P.}},
\bauthor{\bsnm{{Szabo}}, \binits{A.}},
\bauthor{\bsnm{{Farrugia}}, \binits{C.}},
\bauthor{\bsnm{{Yu}}, \binits{W.}}:
\byear{2016},
\batitle{{A Circular-cylindrical Flux-rope Analytical Model for Magnetic
  Clouds}}.
\bjtitle{\apj}
\bvolume{823},
\bfpage{27}.
\doiurl{https://doi.org/10.3847/0004-637X/823/1/27}.
\adsurl{2016ApJ...823...27N}.
\end{barticle}
\endbibitem

\bibitem[\protect\citeauthoryear{{Nieves-Chinchilla} et~al.}{2018}]{Nieves2018}
\begin{barticle}
\bauthor{\bsnm{{Nieves-Chinchilla}}, \binits{T.}},
\bauthor{\bsnm{{Vourlidas}}, \binits{A.}},
\bauthor{\bsnm{{Raymond}}, \binits{J.C.}},
\bauthor{\bsnm{{Linton}}, \binits{M.G.}},
\bauthor{\bsnm{{Al-haddad}}, \binits{N.}},
\bauthor{\bsnm{{Savani}}, \binits{N.P.}},
\bauthor{\bsnm{{Szabo}}, \binits{A.}},
\bauthor{\bsnm{{Hidalgo}}, \binits{M.A.}}:
\byear{2018},
\batitle{{Understanding the Internal Magnetic Field Configurations of ICMEs
  Using More than 20 Years of Wind Observations}}.
\bjtitle{\solphys}
\bvolume{293},
\bfpage{25}.
\doiurl{https://doi.org/10.1007/s11207-018-1247-z}.
\adsurl{2018SoPh..293...25N}.
\end{barticle}
\endbibitem

\bibitem[\protect\citeauthoryear{{Nieves-Chinchilla} et~al.}{2019}]{Nieves2019}
\begin{barticle}
\bauthor{\bsnm{{Nieves-Chinchilla}}, \binits{T.}},
\bauthor{\bsnm{{Jian}}, \binits{L.K.}},
\bauthor{\bsnm{{Balmaceda}}, \binits{L.}},
\bauthor{\bsnm{{Vourlidas}}, \binits{A.}},
\bauthor{\bsnm{{dos Santos}}, \binits{L.F.G.}},
\bauthor{\bsnm{{Szabo}}, \binits{A.}}:
\byear{2019},
\batitle{{Unraveling the Internal Magnetic Field Structure of the
  Earth-directed Interplanetary Coronal Mass Ejections During 1995 - 2015}}.
\bjtitle{\solphys}
\bvolume{294},
\bfpage{89}.
\doiurl{https://doi.org/10.1007/s11207-019-1477-8}.
\adsurl{2019SoPh..294...89N}.
\end{barticle}
\endbibitem

\bibitem[\protect\citeauthoryear{{Nitta} and {Mulligan}}{2017}]{Nitta_2017}
\begin{barticle}
\bauthor{\bsnm{{Nitta}}, \binits{N.V.}},
\bauthor{\bsnm{{Mulligan}}, \binits{T.}}:
\byear{2017},
\batitle{{Earth-Affecting Coronal Mass Ejections Without Obvious Low Coronal
  Signatures}}.
\bjtitle{\solphys}
\bvolume{292},
\bfpage{125}.
\doiurl{https://doi.org/10.1007/s11207-017-1147-7}.
\adsurl{2017SoPh..292..125N}.
\end{barticle}
\endbibitem

\bibitem[\protect\citeauthoryear{{Nitta} et~al.}{2021a}]{2021SSRv..217...84N}
\begin{botherref}
\oauthor{\bsnm{{Nitta}}, \binits{N.V.}},
\oauthor{\bsnm{{Mulligan}}, \binits{T.}},
\oauthor{\bsnm{{Kilpua}}, \binits{E.K.J.}},
\oauthor{\bsnm{{Lynch}}, \binits{B.J.}},
\oauthor{\bsnm{{Mierla}}, \binits{M.}},
\oauthor{\bsnm{{O'Kane}}, \binits{J.}},
\oauthor{\bsnm{{Pagano}}, \binits{P.}},
\oauthor{\bsnm{{Palmerio}}, \binits{E.}},
\oauthor{\bsnm{{Pomoell}}, \binits{J.}},
\oauthor{\bsnm{{Richardson}}, \binits{I.G.}},
\oauthor{\bsnm{{Rodriguez}}, \binits{L.}},
\oauthor{\bsnm{{Rouillard}}, \binits{A.P.}},
\oauthor{\bsnm{{Sinha}}, \binits{S.}},
\oauthor{\bsnm{{Srivastava}}, \binits{N.}},
\oauthor{\bsnm{{Talpeanu}}, \binits{D.-C.}},
\oauthor{\bsnm{{Yardley}}, \binits{S.L.}},
\oauthor{\bsnm{{Zhukov}}, \binits{A.N.}}:
2021a,
\textit{{Correction to: Understanding the Origins of Problem Geomagnetic Storms
  Associated with ``Stealth'' Coronal Mass Ejections}},
Space Science Reviews, Volume 217, Issue 8, article id.84.
\doiurl{https://doi.org/10.1007/s11214-021-00860-5}.
\adsurl{2021SSRv..217...84N}.
\end{botherref}
\endbibitem

\bibitem[\protect\citeauthoryear{{Nitta} et~al.}{2021b}]{2021SSRv..217...82N}
\begin{barticle}
\bauthor{\bsnm{{Nitta}}, \binits{N.V.}},
\bauthor{\bsnm{{Mulligan}}, \binits{T.}},
\bauthor{\bsnm{{Kilpua}}, \binits{E.K.J.}},
\bauthor{\bsnm{{Lynch}}, \binits{B.J.}},
\bauthor{\bsnm{{Mierla}}, \binits{M.}},
\bauthor{\bsnm{{O'Kane}}, \binits{J.}},
\bauthor{\bsnm{{Pagano}}, \binits{P.}},
\bauthor{\bsnm{{Palmerio}}, \binits{E.}},
\bauthor{\bsnm{{Pomoell}}, \binits{J.}},
\bauthor{\bsnm{{Richardson}}, \binits{I.G.}},
\bauthor{\bsnm{{Rodriguez}}, \binits{L.}},
\bauthor{\bsnm{{Rouillard}}, \binits{A.P.}},
\bauthor{\bsnm{{Sinha}}, \binits{S.}},
\bauthor{\bsnm{{Srivastava}}, \binits{N.}},
\bauthor{\bsnm{{Talpeanu}}, \binits{D.-C.}},
\bauthor{\bsnm{{Yardley}}, \binits{S.L.}},
\bauthor{\bsnm{{Zhukov}}, \binits{A.N.}}:
\byear{2021}b,
\batitle{{Understanding the Origins of Problem Geomagnetic Storms Associated
  with ``Stealth'' Coronal Mass Ejections}}.
\bjtitle{\ssr}
\bvolume{217},
\bfpage{82}.
\doiurl{https://doi.org/10.1007/s11214-021-00857-0}.
\adsurl{2021SSRv..217...82N}.
\end{barticle}
\endbibitem

\bibitem[\protect\citeauthoryear{{Ogilvie} et~al.}{1995}]{Ogilvie1995}
\begin{barticle}
\bauthor{\bsnm{{Ogilvie}}, \binits{K.W.}},
\bauthor{\bsnm{{Chornay}}, \binits{D.J.}},
\bauthor{\bsnm{{Fritzenreiter}}, \binits{R.J.}},
\bauthor{\bsnm{{Hunsaker}}, \binits{F.}},
\bauthor{\bsnm{{Keller}}, \binits{J.}},
\bauthor{\bsnm{{Lobell}}, \binits{J.}},
\bauthor{\bsnm{{Miller}}, \binits{G.}},
\bauthor{\bsnm{{Scudder}}, \binits{J.D.}},
\bauthor{\bsnm{{Sittler}}, \binits{J.} \bsuffix{E.~C.}},
\bauthor{\bsnm{{Torbert}}, \binits{R.B.}},
\bauthor{\bsnm{{Bodet}}, \binits{D.}},
\bauthor{\bsnm{{Needell}}, \binits{G.}},
\bauthor{\bsnm{{Lazarus}}, \binits{A.J.}},
\bauthor{\bsnm{{Steinberg}}, \binits{J.T.}},
\bauthor{\bsnm{{Tappan}}, \binits{J.H.}},
\bauthor{\bsnm{{Mavretic}}, \binits{A.}},
\bauthor{\bsnm{{Gergin}}, \binits{E.}}:
\byear{1995},
\batitle{{SWE, A Comprehensive Plasma Instrument for the Wind Spacecraft}}.
\bjtitle{\ssr}
\bvolume{71},
\bfpage{55}.
\doiurl{https://doi.org/10.1007/BF00751326}.
\adsurl{1995SSRv...71...55O}.
\end{barticle}
\endbibitem

\bibitem[\protect\citeauthoryear{{Parker}}{1958}]{Parker1958}
\begin{barticle}
\bauthor{\bsnm{{Parker}}, \binits{E.N.}}:
\byear{1958},
\batitle{{Dynamics of the Interplanetary Gas and Magnetic Fields.}}
\bjtitle{\apj}
\bvolume{128},
\bfpage{664}.
\doiurl{https://doi.org/10.1086/146579}.
\adsurl{1958ApJ...128..664P}.
\end{barticle}
\endbibitem

\bibitem[\protect\citeauthoryear{{Pesnell}, {Thompson}, and
  {Chamberlin}}{2012}]{Pesnell:2012rr}
\begin{barticle}
\bauthor{\bsnm{{Pesnell}}, \binits{W.D.}},
\bauthor{\bsnm{{Thompson}}, \binits{B.J.}},
\bauthor{\bsnm{{Chamberlin}}, \binits{P.C.}}:
\byear{2012},
\batitle{{The Solar Dynamics Observatory (SDO)}}.
\bjtitle{\solphys}
\bvolume{275},
\bfpage{3}.
\doiurl{https://doi.org/10.1007/s11207-011-9841-3}.
\adsurl{2012SoPh..275....3P}.
\end{barticle}
\endbibitem

\bibitem[\protect\citeauthoryear{{Pizzo}}{1981}]{1981JGR....86.6685P}
\begin{barticle}
\bauthor{\bsnm{{Pizzo}}, \binits{V.J.}}:
\byear{1981},
\batitle{{On the application of numerical models to the inverse mapping of
  solar wind flow structures}}.
\bjtitle{JGR}
\bvolume{86},
\bfpage{6685}.
\doiurl{https://doi.org/10.1029/JA086iA08p06685}.
\adsurl{1981JGR....86.6685P}.
\end{barticle}
\endbibitem

\bibitem[\protect\citeauthoryear{{Prete} et~al.}{2024}]{Preteetal24}
\begin{barticle}
\bauthor{\bsnm{{Prete}}, \binits{G.}},
\bauthor{\bsnm{{Niemela}}, \binits{A.}},
\bauthor{\bsnm{{Schmieder}}, \binits{B.}},
\bauthor{\bsnm{{Al-Haddad}}, \binits{N.}},
\bauthor{\bsnm{{Zhuang}}, \binits{B.}},
\bauthor{\bsnm{{Lepreti}}, \binits{F.}},
\bauthor{\bsnm{{Carbone}}, \binits{V.}},
\bauthor{\bsnm{{Poedts}}, \binits{S.}}:
\byear{2024},
\batitle{{EUHFORIA modelling of the Sun-Earth chain of the magnetic cloud of 28
  June 2013}}.
\bjtitle{\aap}
\bvolume{683},
\bfpage{A28}.
\doiurl{https://doi.org/10.1051/0004-6361/202346906}.
\adsurl{2024A&A...683A..28P}.
\end{barticle}
\endbibitem

\bibitem[\protect\citeauthoryear{{Priest}, {Hood}, and
  {Anzer}}{1989}]{Priestet89}
\begin{barticle}
\bauthor{\bsnm{{Priest}}, \binits{E.R.}},
\bauthor{\bsnm{{Hood}}, \binits{A.W.}},
\bauthor{\bsnm{{Anzer}}, \binits{U.}}:
\byear{1989},
\batitle{{A twisted flux-tube model for solar prominences. I - General
  properties}}.
\bjtitle{\apj}
\bvolume{344},
\bfpage{1010}.
\doiurl{https://doi.org/10.1086/167868}.
\adsurl{1989ApJ...344.1010P}.
\end{barticle}
\endbibitem

\bibitem[\protect\citeauthoryear{{Raga}, {Navarro-Gonz{\'a}lez}, and
  {Villagr{\'a}n-Muniz}}{2000}]{Raga2000}
\begin{barticle}
\bauthor{\bsnm{{Raga}}, \binits{A.C.}},
\bauthor{\bsnm{{Navarro-Gonz{\'a}lez}}, \binits{R.}},
\bauthor{\bsnm{{Villagr{\'a}n-Muniz}}, \binits{M.}}:
\byear{2000},
\batitle{{A New, 3D Adaptive Grid Code for Astrophysical and Geophysical
  Gasdynamics}}.
\bjtitle{RMxAA}
\bvolume{36},
\bfpage{67}.
\adsurl{2000RMxAA..36...67R}.
\end{barticle}
\endbibitem

\bibitem[\protect\citeauthoryear{{Savcheva} and {van
  Ballegooijen}}{2009}]{Savcheva09}
\begin{barticle}
\bauthor{\bsnm{{Savcheva}}, \binits{A.}},
\bauthor{\bsnm{{van Ballegooijen}}, \binits{A.}}:
\byear{2009},
\batitle{{Nonlinear Force-free Modeling of a Long-lasting Coronal Sigmoid}}.
\bjtitle{\apj}
\bvolume{703},
\bfpage{1766}.
\doiurl{https://doi.org/10.1088/0004-637X/703/2/1766}.
\adsurl{2009ApJ...703.1766S}.
\end{barticle}
\endbibitem

\bibitem[\protect\citeauthoryear{Savcheva, Van~Ballegooijen, and
  DeLuca}{2012}]{Savcheva12a}
\begin{barticle}
\bauthor{\bsnm{Savcheva}, \binits{A.}},
\bauthor{\bsnm{Van~Ballegooijen}, \binits{A.A.}},
\bauthor{\bsnm{DeLuca}, \binits{E.E.}}:
\byear{2012},
\batitle{{Field Topology Analysis of a Long-Lasting Coronal Sigmoid}}.
\bjtitle{The Astrophysical Journal}
\bvolume{744},
\bfpage{SBD12}.
\end{barticle}
\endbibitem

\bibitem[\protect\citeauthoryear{{Savcheva} et~al.}{2015}]{Savchevaet15}
\begin{barticle}
\bauthor{\bsnm{{Savcheva}}, \binits{A.}},
\bauthor{\bsnm{{Pariat}}, \binits{E.}},
\bauthor{\bsnm{{McKillop}}, \binits{S.}},
\bauthor{\bsnm{{McCauley}}, \binits{P.}},
\bauthor{\bsnm{{Hanson}}, \binits{E.}},
\bauthor{\bsnm{{Su}}, \binits{Y.}},
\bauthor{\bsnm{{Werner}}, \binits{E.}},
\bauthor{\bsnm{{DeLuca}}, \binits{E.E.}}:
\byear{2015},
\batitle{{The Relation between Solar Eruption Topologies and Observed Flare
  Features. I. Flare Ribbons}}.
\bjtitle{\apj}
\bvolume{810},
\bfpage{96}.
\doiurl{https://doi.org/10.1088/0004-637X/810/2/96}.
\adsurl{2015ApJ...810...96S}.
\end{barticle}
\endbibitem

\bibitem[\protect\citeauthoryear{{Savcheva} et~al.}{2016}]{savchevaet16}
\begin{barticle}
\bauthor{\bsnm{{Savcheva}}, \binits{A.}},
\bauthor{\bsnm{{Pariat}}, \binits{E.}},
\bauthor{\bsnm{{McKillop}}, \binits{S.}},
\bauthor{\bsnm{{McCauley}}, \binits{P.}},
\bauthor{\bsnm{{Hanson}}, \binits{E.}},
\bauthor{\bsnm{{Su}}, \binits{Y.}},
\bauthor{\bsnm{{DeLuca}}, \binits{E.E.}}:
\byear{2016},
\batitle{{The Relation between Solar Eruption Topologies and Observed Flare
  Features. II. Dynamical Evolution}}.
\bjtitle{\apj}
\bvolume{817},
\bfpage{43}.
\doiurl{https://doi.org/10.3847/0004-637X/817/1/43}.
\adsurl{2016ApJ...817...43S}.
\end{barticle}
\endbibitem

\bibitem[\protect\citeauthoryear{{Schwenn}}{1986}]{1986SSRv...44..139S}
\begin{barticle}
\bauthor{\bsnm{{Schwenn}}, \binits{R.}}:
\byear{1986},
\batitle{{Relationship of Coronal Transients to Interplanetary Shocks - 3d
  Aspects}}.
\bjtitle{\ssr}
\bvolume{44},
\bfpage{139}.
\doiurl{https://doi.org/10.1007/BF00227230}.
\adsurl{1986SSRv...44..139S}.
\end{barticle}
\endbibitem

\bibitem[\protect\citeauthoryear{Schwenn}{2000}]{2000AdSpR..26...43S}
\begin{barticle}
\bauthor{\bsnm{Schwenn}, \binits{R.}}:
\byear{2000},
\batitle{{Heliospheric 3d Structure and CME Propagation as Seen from SOHO:
  Recent Lessons for Space Weather Predictions}}.
\bjtitle{Advances in Space Research}
\bvolume{26},
\bfpage{43}.
\doiurl{https://doi.org/10.1016/S0273-1177(99)01025-X}.
\adsurl{2000AdSpR..26...43S}.
\end{barticle}
\endbibitem

\bibitem[\protect\citeauthoryear{{Shen} et~al.}{2013}]{Shenetal13}
\begin{barticle}
\bauthor{\bsnm{{Shen}}, \binits{C.}},
\bauthor{\bsnm{{Li}}, \binits{G.}},
\bauthor{\bsnm{{Kong}}, \binits{X.}},
\bauthor{\bsnm{{Hu}}, \binits{J.}},
\bauthor{\bsnm{{Sun}}, \binits{X.D.}},
\bauthor{\bsnm{{Ding}}, \binits{L.}},
\bauthor{\bsnm{{Chen}}, \binits{Y.}},
\bauthor{\bsnm{{Wang}}, \binits{Y.}},
\bauthor{\bsnm{{Xia}}, \binits{L.}}:
\byear{2013},
\batitle{{Compound Twin Coronal Mass Ejections in the 2012 May 17 GLE Event}}.
\bjtitle{\apj}
\bvolume{763},
\bfpage{114}.
\doiurl{https://doi.org/10.1088/0004-637X/763/2/114}.
\adsurl{2013ApJ...763..114S}.
\end{barticle}
\endbibitem

\bibitem[\protect\citeauthoryear{{Shen} et~al.}{2014}]{Shenetal14}
\begin{barticle}
\bauthor{\bsnm{{Shen}}, \binits{C.}},
\bauthor{\bsnm{{Wang}}, \binits{Y.}},
\bauthor{\bsnm{{Pan}}, \binits{Z.}},
\bauthor{\bsnm{{Miao}}, \binits{B.}},
\bauthor{\bsnm{{Ye}}, \binits{P.}},
\bauthor{\bsnm{{Wang}}, \binits{S.}}:
\byear{2014},
\batitle{{Full-halo coronal mass ejections: Arrival at the Earth}}.
\bjtitle{Journal of Geophysical Research (Space Physics)}
\bvolume{119},
\bfpage{5107}.
\doiurl{https://doi.org/10.1002/2014JA020001}.
\adsurl{2014JGRA..119.5107S}.
\end{barticle}
\endbibitem

\bibitem[\protect\citeauthoryear{{Shi} et~al.}{2022}]{2022PhPl...29l2901S}
\begin{barticle}
\bauthor{\bsnm{{Shi}}, \binits{C.}},
\bauthor{\bsnm{{Velli}}, \binits{M.}},
\bauthor{\bsnm{{Bale}}, \binits{S.D.}},
\bauthor{\bsnm{{R{\'e}ville}}, \binits{V.}},
\bauthor{\bsnm{{Maksimovi{\'c}}}, \binits{M.}},
\bauthor{\bsnm{{Dakeyo}}, \binits{J.-B.}}:
\byear{2022},
\batitle{{Acceleration of polytropic solar wind: Parker Solar Probe observation
  and one-dimensional model}}.
\bjtitle{Physics of Plasmas}
\bvolume{29},
\bfpage{122901}.
\doiurl{https://doi.org/10.1063/5.0124703}.
\adsurl{2022PhPl...29l2901S}.
\end{barticle}
\endbibitem

\bibitem[\protect\citeauthoryear{{Srivastava} and
  {Venkatakrishnan}}{2002}]{2002GeoRL..29.1287S}
\begin{barticle}
\bauthor{\bsnm{{Srivastava}}, \binits{N.}},
\bauthor{\bsnm{{Venkatakrishnan}}, \binits{P.}}:
\byear{2002},
\batitle{{Relationship between CME Speed and Geomagnetic Storm Intensity}}.
\bjtitle{\grl}
\bvolume{29},
\bfpage{1287}.
\doiurl{https://doi.org/10.1029/2001GL013597}.
\adsurl{2002GeoRL..29.1287S}.
\end{barticle}
\endbibitem

\bibitem[\protect\citeauthoryear{{Stone} et~al.}{1998}]{1998SSRv...86....1S}
\begin{barticle}
\bauthor{\bsnm{{Stone}}, \binits{E.C.}},
\bauthor{\bsnm{{Frandsen}}, \binits{A.M.}},
\bauthor{\bsnm{{Mewaldt}}, \binits{R.A.}},
\bauthor{\bsnm{{Christian}}, \binits{E.R.}},
\bauthor{\bsnm{{Margolies}}, \binits{D.}},
\bauthor{\bsnm{{Ormes}}, \binits{J.F.}},
\bauthor{\bsnm{{Snow}}, \binits{F.}}:
\byear{1998},
\batitle{{The Advanced Composition Explorer}}.
\bjtitle{\ssr}
\bvolume{86},
\bfpage{1}.
\doiurl{https://doi.org/10.1023/A:1005082526237}.
\adsurl{1998SSRv...86....1S}.
\end{barticle}
\endbibitem

\bibitem[\protect\citeauthoryear{{Su} and {van
  Ballegooijen}}{2012}]{yingnaet12}
\begin{barticle}
\bauthor{\bsnm{{Su}}, \binits{Y.}},
\bauthor{\bsnm{{van Ballegooijen}}, \binits{A.}}:
\byear{2012},
\batitle{{Observations and Magnetic Field Modeling of a Solar Polar Crown
  Prominence}}.
\bjtitle{\apj}
\bvolume{757},
\bfpage{168}.
\doiurl{https://doi.org/10.1088/0004-637X/757/2/168}.
\adsurl{2012ApJ...757..168S}.
\end{barticle}
\endbibitem

\bibitem[\protect\citeauthoryear{{Su} et~al.}{2009a}]{Yingna09ch}
\begin{barticle}
\bauthor{\bsnm{{Su}}, \binits{Y.}},
\bauthor{\bsnm{{van Ballegooijen}}, \binits{A.}},
\bauthor{\bsnm{{Schmieder}}, \binits{B.}},
\bauthor{\bsnm{{Berlicki}}, \binits{A.}},
\bauthor{\bsnm{{Guo}}, \binits{Y.}},
\bauthor{\bsnm{{Golub}}, \binits{L.}},
\bauthor{\bsnm{{Huang}}, \binits{G.}}:
\byear{2009}a,
\batitle{{Flare Energy Build-up in a Decaying Active Region Near a Coronal
  Hole}}.
\bjtitle{\apj}
\bvolume{704},
\bfpage{341}.
\doiurl{https://doi.org/10.1088/0004-637X/704/1/341}.
\adsurl{2009ApJ...704..341S}.
\end{barticle}
\endbibitem

\bibitem[\protect\citeauthoryear{{Su} et~al.}{2009b}]{yingnaet09}
\begin{barticle}
\bauthor{\bsnm{{Su}}, \binits{Y.}},
\bauthor{\bsnm{{van Ballegooijen}}, \binits{A.}},
\bauthor{\bsnm{{Lites}}, \binits{B.W.}},
\bauthor{\bsnm{{Deluca}}, \binits{E.E.}},
\bauthor{\bsnm{{Golub}}, \binits{L.}},
\bauthor{\bsnm{{Grigis}}, \binits{P.C.}},
\bauthor{\bsnm{{Huang}}, \binits{G.}},
\bauthor{\bsnm{{Ji}}, \binits{H.}}:
\byear{2009}b,
\batitle{{Observations and Nonlinear Force-Free Field Modeling of Active Region
  10953}}.
\bjtitle{\apj}
\bvolume{691},
\bfpage{105}.
\doiurl{https://doi.org/10.1088/0004-637X/691/1/105}.
\adsurl{2009ApJ...691..105S}.
\end{barticle}
\endbibitem

\bibitem[\protect\citeauthoryear{Su et~al.}{2011}]{Su11}
\begin{barticle}
\bauthor{\bsnm{Su}, \binits{Y.}},
\bauthor{\bsnm{Surges}, \binits{V.}},
\bauthor{\bparticle{van} \bsnm{Ballegooijen}, \binits{A.}},
\bauthor{\bsnm{DeLuca}, \binits{E.E.}},
\bauthor{\bsnm{Golub}, \binits{L.}}:
\byear{2011},
\batitle{{Observations and Magnetic Field Modeling of the Flare/coronal Mass
  Ejection Event on 2010 April 8}}.
\bjtitle{The Astrophysical Journal}
\bvolume{734},
\bfpage{53}.
\end{barticle}
\endbibitem

\bibitem[\protect\citeauthoryear{Sugiura}{1964}]{osti_4554034}
\begin{botherref}
\oauthor{\bsnm{Sugiura}, \binits{M.}}:
1964,
HOURLY VALUES OF EQUATORIAL Dst FOR THE IGY.
\textit{Ann. Int. Geophys. Yr.}
\textbf{35}.
\end{botherref}
\endbibitem

\bibitem[\protect\citeauthoryear{{Thernisien}}{2011}]{2011ApJS..194...33T}
\begin{barticle}
\bauthor{\bsnm{{Thernisien}}, \binits{A.}}:
\byear{2011},
\batitle{{Implementation of the Graduated Cylindrical Shell Model for the
  Three-dimensional Reconstruction of Coronal Mass Ejections}}.
\bjtitle{The Astrophysical Journal Supplement Series}
\bvolume{194},
\bfpage{33}.
\doiurl{https://doi.org/10.1088/0067-0049/194/2/33}.
\adsurl{2011ApJS..194...33T}.
\end{barticle}
\endbibitem

\bibitem[\protect\citeauthoryear{{Thernisien}, {Howard}, and
  {Vourlidas}}{2006}]{Thernisien_etal_2006ApJ}
\begin{barticle}
\bauthor{\bsnm{{Thernisien}}, \binits{A.F.R.}},
\bauthor{\bsnm{{Howard}}, \binits{R.A.}},
\bauthor{\bsnm{{Vourlidas}}, \binits{A.}}:
\byear{2006},
\batitle{{Modeling of Flux Rope Coronal Mass Ejections}}.
\bjtitle{\apj}
\bvolume{652},
\bfpage{763}.
\doiurl{https://doi.org/10.1086/508254}.
\adsurl{2006ApJ...652..763T}.
\end{barticle}
\endbibitem

\bibitem[\protect\citeauthoryear{{Thernisien}, {Vourlidas}, and
  {Howard}}{2009}]{Thernisien_etal_2009SoPh}
\begin{barticle}
\bauthor{\bsnm{{Thernisien}}, \binits{A.}},
\bauthor{\bsnm{{Vourlidas}}, \binits{A.}},
\bauthor{\bsnm{{Howard}}, \binits{R.A.}}:
\byear{2009},
\batitle{{Forward Modeling of Coronal Mass Ejections Using STEREO/SECCHI
  Data}}.
\bjtitle{\solphys}
\bvolume{256},
\bfpage{111}.
\doiurl{https://doi.org/10.1007/s11207-009-9346-5}.
\adsurl{2009SoPh..256..111T}.
\end{barticle}
\endbibitem

\bibitem[\protect\citeauthoryear{{Tsurutani}
  et~al.}{1988}]{1988JGR....93.8519T}
\begin{barticle}
\bauthor{\bsnm{{Tsurutani}}, \binits{B.T.}},
\bauthor{\bsnm{{Gonzalez}}, \binits{W.D.}},
\bauthor{\bsnm{{Tang}}, \binits{F.}},
\bauthor{\bsnm{{Akasofu}}, \binits{S.I.}},
\bauthor{\bsnm{{Smith}}, \binits{E.J.}}:
\byear{1988},
\batitle{{Origin of interplanetary southward magnetic fields responsible for
  major magnetic storms near solar maximum (1978-1979)}}.
\bjtitle{\jgr}
\bvolume{93},
\bfpage{8519}.
\doiurl{https://doi.org/10.1029/JA093iA08p08519}.
\adsurl{1988JGR....93.8519T}.
\end{barticle}
\endbibitem

\bibitem[\protect\citeauthoryear{{van Ballegooijen}}{2004}]{vanBallegooijen04}
\begin{barticle}
\bauthor{\bsnm{{van Ballegooijen}}, \binits{A.A.}}:
\byear{2004},
\batitle{{Observations and Modeling of a Filament on the Sun}}.
\bjtitle{\apj}
\bvolume{612},
\bfpage{519}.
\doiurl{https://doi.org/10.1086/422512}.
\adsurl{2004ApJ...612..519V}.
\end{barticle}
\endbibitem

\bibitem[\protect\citeauthoryear{{van Driel-Gesztelyi}
  et~al.}{2012}]{vanet2012}
\begin{barticle}
\bauthor{\bsnm{{van Driel-Gesztelyi}}, \binits{L.}},
\bauthor{\bsnm{{Culhane}}, \binits{J.L.}},
\bauthor{\bsnm{{Baker}}, \binits{D.}},
\bauthor{\bsnm{{D{\'e}moulin}}, \binits{P.}},
\bauthor{\bsnm{{Mandrini}}, \binits{C.H.}},
\bauthor{\bsnm{{DeRosa}}, \binits{M.L.}},
\bauthor{\bsnm{{Rouillard}}, \binits{A.P.}},
\bauthor{\bsnm{{Opitz}}, \binits{A.}},
\bauthor{\bsnm{{Stenborg}}, \binits{G.}},
\bauthor{\bsnm{{Vourlidas}}, \binits{A.}},
\bauthor{\bsnm{{Brooks}}, \binits{D.H.}}:
\byear{2012},
\batitle{{Magnetic Topology of Active Regions and Coronal Holes: Implications
  for Coronal Outflows and the Solar Wind}}.
\bjtitle{\solphys}
\bvolume{281},
\bfpage{237}.
\doiurl{https://doi.org/10.1007/s11207-012-0076-8}.
\adsurl{2012SoPh..281..237V}.
\end{barticle}
\endbibitem

\bibitem[\protect\citeauthoryear{{Waheed}, {Khan}, and
  {Gwal}}{2019}]{2019InJPh..93.1103W}
\begin{barticle}
\bauthor{\bsnm{{Waheed}}, \binits{M.A.}},
\bauthor{\bsnm{{Khan}}, \binits{P.A.}},
\bauthor{\bsnm{{Gwal}}, \binits{A.K.}}:
\byear{2019},
\batitle{{Distribution of intense, moderate and weak geomagnetic storms over
  the solar cycle}}.
\bjtitle{Indian Journal of Physics}
\bvolume{93},
\bfpage{1103}.
\doiurl{https://doi.org/10.1007/s12648-019-01379-w}.
\adsurl{2019InJPh..93.1103W}.
\end{barticle}
\endbibitem

\bibitem[\protect\citeauthoryear{{Wang}, {Hoeksema}, and
  {Liu}}{2020}]{2020JGRA..12527530W}
\begin{barticle}
\bauthor{\bsnm{{Wang}}, \binits{J.}},
\bauthor{\bsnm{{Hoeksema}}, \binits{J.T.}},
\bauthor{\bsnm{{Liu}}, \binits{S.}}:
\byear{2020},
\batitle{{The Deflection of Coronal Mass Ejections by the Ambient Coronal
  Magnetic Field Configuration}}.
\bjtitle{Journal of Geophysical Research (Space Physics)}
\bvolume{125},
\bfpage{e27530}.
\doiurl{https://doi.org/10.1029/2019JA027530}.
\adsurl{2020JGRA..12527530W}.
\end{barticle}
\endbibitem

\bibitem[\protect\citeauthoryear{{Wang} et~al.}{2011}]{2011JGRA..116.4104W}
\begin{barticle}
\bauthor{\bsnm{{Wang}}, \binits{Y.}},
\bauthor{\bsnm{{Chen}}, \binits{C.}},
\bauthor{\bsnm{{Gui}}, \binits{B.}},
\bauthor{\bsnm{{Shen}}, \binits{C.}},
\bauthor{\bsnm{{Ye}}, \binits{P.}},
\bauthor{\bsnm{{Wang}}, \binits{S.}}:
\byear{2011},
\batitle{{Statistical study of coronal mass ejection source locations:
  Understanding CMEs viewed in coronagraphs}}.
\bjtitle{Journal of Geophysical Research (Space Physics)}
\bvolume{116},
\bfpage{A04104}.
\doiurl{https://doi.org/10.1029/2010JA016101}.
\adsurl{2011JGRA..116.4104W}.
\end{barticle}
\endbibitem

\bibitem[\protect\citeauthoryear{{Wanliss} and
  {Showalter}}{2006}]{2006JGRA..111.2202W}
\begin{barticle}
\bauthor{\bsnm{{Wanliss}}, \binits{J.A.}},
\bauthor{\bsnm{{Showalter}}, \binits{K.M.}}:
\byear{2006},
\batitle{{High-resolution global storm index: Dst versus SYM-H}}.
\bjtitle{Journal of Geophysical Research (Space Physics)}
\bvolume{111},
\bfpage{A02202}.
\doiurl{https://doi.org/10.1029/2005JA011034}.
\adsurl{2006JGRA..111.2202W}.
\end{barticle}
\endbibitem

\bibitem[\protect\citeauthoryear{{Watari}, {Nakamizo}, and
  {Ebihara}}{2023}]{2023EP&S...75...90W}
\begin{barticle}
\bauthor{\bsnm{{Watari}}, \binits{S.}},
\bauthor{\bsnm{{Nakamizo}}, \binits{A.}},
\bauthor{\bsnm{{Ebihara}}, \binits{Y.}}:
\byear{2023},
\batitle{{Solar events and solar wind conditions associated with intense
  geomagnetic storms}}.
\bjtitle{Earth, Planets and Space}
\bvolume{75},
\bfpage{90}.
\doiurl{https://doi.org/10.1186/s40623-023-01843-2}.
\adsurl{2023EP&S...75...90W}.
\end{barticle}
\endbibitem

\bibitem[\protect\citeauthoryear{{Wood} et~al.}{2017}]{Wood_2017}
\begin{barticle}
\bauthor{\bsnm{{Wood}}, \binits{B.E.}},
\bauthor{\bsnm{{Wu}}, \binits{C.-C.}},
\bauthor{\bsnm{{Lepping}}, \binits{R.P.}},
\bauthor{\bsnm{{Nieves-Chinchilla}}, \binits{T.}},
\bauthor{\bsnm{{Howard}}, \binits{R.A.}},
\bauthor{\bsnm{{Linton}}, \binits{M.G.}},
\bauthor{\bsnm{{Socker}}, \binits{D.G.}}:
\byear{2017},
\batitle{{A STEREO Survey of Magnetic Cloud Coronal Mass Ejections Observed at
  Earth in 2008-2012}}.
\bjtitle{\apjs}
\bvolume{229},
\bfpage{29}.
\doiurl{https://doi.org/10.3847/1538-4365/229/2/29}.
\adsurl{2017ApJS..229...29W}.
\end{barticle}
\endbibitem

\bibitem[\protect\citeauthoryear{{Yashiro} et~al.}{2004}]{2004JGRA..109.7105Y}
\begin{barticle}
\bauthor{\bsnm{{Yashiro}}, \binits{S.}},
\bauthor{\bsnm{{Gopalswamy}}, \binits{N.}},
\bauthor{\bsnm{{Michalek}}, \binits{G.}},
\bauthor{\bsnm{{St. Cyr}}, \binits{O.C.}},
\bauthor{\bsnm{{Plunkett}}, \binits{S.P.}},
\bauthor{\bsnm{{Rich}}, \binits{N.B.}},
\bauthor{\bsnm{{Howard}}, \binits{R.A.}}:
\byear{2004},
\batitle{{A catalog of white light coronal mass ejections observed by the SOHO
  spacecraft}}.
\bjtitle{Journal of Geophysical Research (Space Physics)}
\bvolume{109},
\bfpage{A07105}.
\doiurl{https://doi.org/10.1029/2003JA010282}.
\adsurl{2004JGRA..109.7105Y}.
\end{barticle}
\endbibitem

\bibitem[\protect\citeauthoryear{{Zhang} et~al.}{2007}]{Zhanget07}
\begin{barticle}
\bauthor{\bsnm{{Zhang}}, \binits{J.}},
\bauthor{\bsnm{{Richardson}}, \binits{I.G.}},
\bauthor{\bsnm{{Webb}}, \binits{D.F.}},
\bauthor{\bsnm{{Gopalswamy}}, \binits{N.}},
\bauthor{\bsnm{{Huttunen}}, \binits{E.}},
\bauthor{\bsnm{{Kasper}}, \binits{J.C.}},
\bauthor{\bsnm{{Nitta}}, \binits{N.V.}},
\bauthor{\bsnm{{Poomvises}}, \binits{W.}},
\bauthor{\bsnm{{Thompson}}, \binits{B.J.}},
\bauthor{\bsnm{{Wu}}, \binits{C.-C.}},
\bauthor{\bsnm{{Yashiro}}, \binits{S.}},
\bauthor{\bsnm{{Zhukov}}, \binits{A.N.}}:
\byear{2007},
\batitle{{Solar and interplanetary sources of major geomagnetic storms (Dst <=
  -100 nT) during 1996-2005}}.
\bjtitle{Journal of Geophysical Research (Space Physics)}
\bvolume{112},
\bfpage{A10102}.
\doiurl{https://doi.org/10.1029/2007JA012321}.
\adsurl{2007JGRA..11210102Z}.
\end{barticle}
\endbibitem

\end{thebibliography}




\end{document}